\documentclass{IEEEtran}
\IEEEoverridecommandlockouts
\usepackage{amsmath, amsthm, amssymb}
\usepackage{verbatim}
\usepackage{graphicx}
\usepackage{cite}
\usepackage{subfig}
\usepackage[subnum]{cases}

\usepackage[usenames]{color}
\definecolor{plum}  {rgb}{.4,0,.4}
\definecolor{BrickRed} {rgb}{0.6,0,0}

\usepackage[plainpages=false,pdfpagelabels,colorlinks=true,linkcolor=BrickRed,citecolor=plum]{hyperref}

\newtheorem{theorem}{Theorem}
\newtheorem{lemma}{Lemma}

\newtheorem{corollary}{Corollary}

\newtheorem{example}{Example}
\newcommand{\R}{\mathbb{R}}

\newcommand{\E}{\mathbb{E}}

\newcommand{\Var}{{\rm Var}}
\newcommand{\I}{\mathbf{1}}

\def\cA{{\mathcal A}}
\def\cB{{\mathcal B}}

\def\cD{{\mathcal D}}
\def\cE{{\mathcal E}}

\def\cL{{\mathcal L}}
\def\cN{{\mathcal N}}
\def\cP{{\mathcal P}}
\def\cS{{\mathcal S}}
\def\cV{{\mathcal V}}

\def\sW{{\mathsf W}}
\def\sX{{\mathsf X}}
\def\sY{{\mathsf Y}}
\def\sZ{{\mathsf Z}}

\def\BB{{\mathbb B}}
\def\PP{{\mathbb P}}
\def\SS{{\mathbb S}}
\def\QQ{{\mathbb Q}}
\def\N{{\mathbb N}}

\def\deq{\triangleq}
\def\wh#1{{\widehat{#1}}}
\def\lepartial{\overset{{}_{\leftarrow}}{\partial}}
\def\lephi{\overset{{}_{\leftarrow}}{\varphi}}
\def\riphi{\overset{{}_{\rightarrow}}{\varphi}}
\def\eps{\varepsilon}
\begin{document}
\sloppy

\title{Information-Theoretic Lower Bounds for Distributed Function Computation}

\author{Aolin Xu and Maxim Raginsky,~\IEEEmembership{Senior Member,~IEEE}
\thanks{This work was supported by the NSF under grant CCF-1017564, CAREER award CCF-1254041, by the Center for Science of
Information (CSoI), an NSF Science and Technology Center, under grant
agreement CCF-0939370, and by ONR under grant N00014-12-1-0998. The material in this paper was presented in part at the IEEE International Symposium on Information Theory (ISIT), Honolulu, HI, July 2014.}
\thanks{The authors are with the Department of Electrical and Computer Engineering and the Coordinated Science Laboratory, University of Illinois, Urbana, IL 61801, USA. E-mail: \{aolinxu2,maxim\}@illinois.edu. }
}
\maketitle


\begin{abstract}
We derive information-theoretic converses (i.e., lower bounds) for the minimum time required by any algorithm for distributed function computation over a network of point-to-point channels with finite capacity, where each node of the network initially has a random observation and aims to compute a common function of all observations to a given accuracy with a given confidence by exchanging messages with its neighbors.
We obtain the lower bounds on computation time by examining the conditional mutual information between the actual function value and its estimate at an arbitrary node, given the observations in an arbitrary subset of nodes containing that node.
The main contributions include:
1) A lower bound on the conditional mutual information via so-called small ball probabilities, which captures the dependence of the computation time on the joint distribution of the observations at the nodes, the structure of the function, and the accuracy requirement. For linear functions, the small ball probability can be expressed by L\'evy concentration functions of sums of independent random variables, for which tight estimates are available that lead to strict improvements over existing lower bounds on computation time.
2) An upper bound on the conditional mutual information via strong data processing inequalities, which complements and strengthens existing cutset-capacity upper bounds. 
3) A multi-cutset analysis that quantifies the loss (dissipation) of the information needed for computation as it flows across a succession of cutsets in the network. This analysis is based on reducing a general network to a line network with bidirectional links and self-links, and the results highlight the dependence of the computation time on the diameter of the network, a fundamental parameter that is missing from most of the existing lower bounds on computation time.

\end{abstract}

\section{Introduction and preview of results}

\subsection{Model and problem formulation}

The problem of distributed function computation arises in such applications as inference and learning in networks and consensus or coordination of multiple agents. 
Each node of the network has an initial random observation and aims to compute a common function of the observations of all the nodes by exchanging messages with its neighbors over discrete memoryless point-to-point channels and by performing local computations.
A problem of theoretical and practical interest is to determine the fundamental limits on the \emph{computation time}, i.e., the minimum number of steps needed by any distributed computation algorithm to guarantee that, when the algorithm terminates, each node has an accurate estimate of the function value with high probability. 

Formally, a network consisting of nodes connected by point-to-point channels is represented by a directed graph $G = (\cV,\cE)$, where $\cV$ is a finite set of nodes and $\cE \subseteq \cV\times\cV$ is a set of edges. Node $u$ can send messages to node $v$ only if $(u,v) \in \cE$. Accordingly, to each edge $e \in \cE$ we associate a discrete memoryless channel with finite input alphabet $\sX_e$, finite output alphabet $\sY_e$, and stochastic transition law $K_e$ that specifies the transition probabilities $K_e(y_e|x_e)$ for all $(x_e,y_e) \in \sX_e \times \sY_e$. The channels corresponding to different edges are assumed to be independent. 
Initially, each node $v$ has access to an observation given by a random variable (r.v.) $W_v$ taking values in some space $\sW_v$. We assume that the joint probability law $\PP_W$ of  $W \deq (W_v)_{v \in \cV}$ is known to all the nodes. Given a function $f : \prod_{v \in \cV} \sW_v \to \sZ$, each node aims to estimate the value $Z = f(W)$ via local communication and computation. For example, when $f$ is given by the identity mapping $Z = W$, the goal of each node is to estimate the observations of all other nodes in the network.

The operation of the network is synchronized, and takes place in discrete time. A \textit{$T$-step algorithm} $\cA$ is a collection of deterministic encoders $(\varphi_{v,t})$ and estimators $(\psi_v)$, for all $v \in \cV$ and $t \in \{1,\ldots,T\}$, given by mappings
\begin{align*}
	\varphi_{v,t} : \sW_v \times \sY^{t-1}_{v\leftarrow} \to \sX_{v\rightarrow}, \quad \psi_v : \sW_v \times \sY^T_{v\leftarrow} \to \sZ,
\end{align*}
where $\sX_{v\rightarrow} = \prod_{u\in\cN_{v\rightarrow}}\sX_{(v,u)}$ and $\sY_{v\leftarrow} = \prod_{u\in\cN_{v\leftarrow}} \sY_{(u,v)}$. 
Here, $\cN_{v\leftarrow} \deq \{u\in\cV:(u,v)\in\cE\}$ and  $\cN_{v\rightarrow} \deq \{u\in\cV: (v,u) \in \cE\}$ are, respectively, the in-neighborhood and the out-neighborhood of node $v$.  
The algorithm operates as follows: at each step $t$, each node $v$ computes $X_{v,t} \deq (X_{(v,u),t})_{u \in \cN_{v\rightarrow}} = \varphi_{v,t}\big(W_v,Y^{t-1}_v\big) \in \sX_{v\rightarrow}$, and then transmits each message $X_{(v,u),t}$ along the edge $(v,u) \in \cE$. 
For each $(u,v)\in\cE$, the received message $Y_{(u,v),t}$ at each $t$ is related to the transmitted message $X_{(u,v),t}$ via the stochastic transition law $K_{(u,v)}$.
At step $T$, each node $v$ computes $\wh{Z}_v = \psi_v(W_v,Y^T_v)$ as an estimate of $Z$, where $Y_{v,t} \deq (Y_{(u,v),t})_{u\in\cN_{v\leftarrow}} \in \sY_{v\leftarrow}$ for $t\in\{1,\ldots,T\}$.

Given a nonnegative \textit{distortion function} $d : \sZ \times \sZ \to \R^+$, we use the excess distortion probability $\PP \big[d(Z,\wh{Z}_v) > \eps\big]$ to quantify the computation fidelity of the algorithm at node $v$. A key fundamental limit of distributed function computation is the \textit{$(\eps,\delta)$-computation time}:
\begin{align}\label{eq:eps_delta_comp_time}
	T(\eps,\delta) \deq \inf \Big\{ T \in \N: & \exists \text{ a $T$-step algorithm $\cA$ such that } \nonumber \\
& \max_{v \in \cV} \PP\big[d(Z,\wh Z_v) > \eps \big] \le \delta \Big\}.
\end{align}
If an algorithm $\cA$ has the property that
$$
\max_{v \in \cV} \PP\big[d(Z,\wh Z_v) > \eps \big] \le \delta,
$$
then we say that it \textit{achieves accuracy} $\eps$ with \textit{confidence} $1-\delta$. Thus, $T(\eps,\delta)$ is the minimum number of time steps needed by any algorithm to achieve accuracy $\eps$ with confidence $1-\delta$.
The objective of this paper is to derive general lower bounds on $T(\eps,\delta)$ for arbitrary network topologies, {  discrete memoryless} channel models, continuous or discrete observations, and functions $f$.

Previously, this problem (for real-valued functions and quadratic distortion) has been studied by Ayaso {et al.}~\cite{Ayaso_etal} and by Como and Dahleh~\cite{Como_Dahleh} using information-theoretic techniques.
This problem is also related to the study of communication complexity of distributed computing over noisy channels. In that context, Goyal \textit{et al.}~\cite{GKS_noisybc08} studied the problem of computing Boolean functions in complete graphs, where each pair of nodes communicates over a pair of independent binary symmetric channels (BSCs),
and obtained tight lower bounds on the number of serial broadcasts using an approach tailored to that special problem.
The technique used in \cite{GKS_noisybc08} has been extended to random planar networks by Dutta \textit{et al.}~\cite{noisy_bc_Dutta08}.
Other related, but differently formulated, problems include
communication complexity and information complexity in distributed computing over noiseless channels, surveyed in \cite{Braverman_interactiveinformation};
minimum communication rates for distributed computing \cite{OR01,KM79_mod2sum,WTV08_GaussSum}, compression, or estimation based on infinite sequences of observations, surveyed in \cite[Chap.~21]{NITbook11};
and distributed computing in wireless networks, surveyed in \cite{Giridhar_Kumar06}.
Some achievability results for specific distributed function computation problems can be found in \cite{Gallager88, Schulman96_iter_comm, Rajagopalan94_dist_comp,  Ayaso_etal, Carli_etal_erasures_consensus, Kar_Moura, Noorshams_Wainwright,Ying_comp06,Deb_Medard}.

\subsection{Method of analysis and summary of main results}
\label{ssec:summary}

Our analysis builds upon the information-theoretic framework proposed by Ayaso \textit{et al.}~\cite{Ayaso_etal} and Como and Dahleh~\cite{Como_Dahleh}. The underlying idea is rather natural and exploits a fundamental trade-off between the minimal amount of information any good algorithm must \textit{necessarily} extract about the function value $Z$ when it terminates and the maximal amount of information any algorithm is able to obtain due to time and communication constraints. To be more precise, given any set of nodes $\cS \subseteq \cV$, let $W_\cS \deq (W_v)_{v \in \cS}$ denote the vector of observations at all the nodes in $\cS$. The quantity that plays a key role in the analysis is the conditional mutual information $I(Z; \wh Z_v |W_\cS)$ between the actual function value $Z$ and the estimate $\wh Z_v$ at an arbitrary node $v$, given the observations in an arbitrary subset of nodes $\cS$ containing $v$.

 Consider an arbitrary $T$-step algorithm $\cA$ that achieves accuracy $\eps$ with confidence $1-\delta$. Then, as we show in Lemma~\ref{lm:mi_lb_gen} of Sec.~\ref{sec:lb_cond_mi}, this mutual information can be lower-bounded by
\begin{align}\label{eq:smallball_intro}
I(Z; \wh Z_v|W_\cS) &\ge (1-\delta)\log \frac{1}{\E[L(W_\cS,\eps)]} - h_2(\delta),
\end{align}
where $h_2(\delta) \deq -\delta \log \delta -(1-\delta)\log(1-\delta)$ is the binary entropy function, and
\begin{align*}
L(w_\cS,\eps) &\deq \sup_{z \in \sZ} \PP[d(Z,z) \le \eps  | W_\cS = w_\cS] \\
&= \sup_{z \in \sZ} \PP[d(f(W),z) \le \eps | W_\cS = w_\cS]
\end{align*}
is the \textit{conditional small ball probability} of $Z = f(W)$ given $W_\cS=w_\cS$. The conditional small ball probability quantifies the difficulty of localizing the value of $Z = f(W)$ in a ``distortion ball'' of size $\eps$ given partial knowledge about the value of $W$, namely $W_\cS = w_\cS$.  For example, as discussed in Sec.~\ref{sec:comp_linear_fn}, when $f$ is a linear function of the observations $W$, the conditional small ball probability can be expressed in terms of so-called \textit{L\'evy concentration functions} \cite{Petrov_book}, for which tight estimates are available under various regularity conditions. 

\begin{figure}
	\centering
	\includegraphics[scale = 1.0]{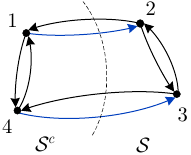}
	\caption{\label{fig:cutset} A four-node network with a cut defined by $\cS = \{2,3\}$ and $\cS^c = \{1,4\}$. The cutset $\cE_\cS$ consists of edges $(1,2)$ and $(4,3)$, marked in blue.}
\end{figure}
On the other hand, if $\cA$ is a $T$-step algorithm, then the amount of information any node $v$ has about $Z$ once $\cA$ terminates can be upper-bounded by a quantity that increases with $T$ and also depends on the network topology and on the information transmission capabilities of the channels connecting the nodes. To quantify this amount of information, we consider a \textit{cut} of the network, i.e., a partition of the set of nodes $\cV$ into two disjoint subsets $\cS$ and $\cS^c \deq \cV \backslash \cS$, such that $v \in \cS$. 
The underlying intuition is that any information that nodes in $\cS$ receive about $W_{\cS^c}$ must flow across the edges from nodes in $\cS^c$ to nodes in $\cS$. The set of these edges, denoted by $\cE_\cS$, is referred to as the \textit{cutset} induced by $\cS$. Figure~\ref{fig:cutset} illustrates these concepts on a simple four-node network.
We then have the following upper bound  \cite{Ayaso_etal,Como_Dahleh} (see also Lemma~\ref{lm:mi_ub_cutset} in Sec.~\ref{sec:ub_mi_cutset_cap}):
\begin{align}\label{eq:cutset_intro}
I(Z;\wh{Z}_{v}| W_{\cS} ) \le T C_{\cS} .
\end{align}
The quantity $C_\cS$, referred to as the \textit{cutset capacity}, is the sum of the Shannon capacities of all the channels located on the edges in the cutset $\cE_\cS$. Thus, if there exists a cut $(\cS,\cS^c)$ with a small value of $C_\cS$, then the amount of information gained by the nodes in $\cS$ about $Z$ will also be small. Note that the cutset upper bound grows linearly with $T$. However, when the initial observations $W$ are discrete, we also know that
$$
I(Z; \wh Z_v |W_{\cS}) \le I(W_{\cS^c}; \wh Z_v |W_{\cS}) \le H(W_{\cS^c}|W_\cS)
$$
where $H(W_{\cS^c}|W_\cS)$ is the conditional entropy of $W_{\cS^c}$ given $W_{\cS}$, which does not depend on $T$.
In fact, we sharpen this bound by showing in Lemma~\ref{lm:mi_ub_SDPI_gen} in Sec.~\ref{sec:ub_mi_SDPI} that
\begin{align}\label{eq:SDPI_intro}
I(Z;\wh{Z}_{v}| W_{\cS} ) & \le  \big(1-(1-\eta_v)^{T}\big)H(W_{\cS^c}|W_{\cS}) .
\end{align}
Here, $\eta_v$ is defined as
$$
\eta_v = \sup \frac{I(U; Y_v)}{I(U;X_v)}
$$ 
where the supremum is over all triples $(U,X_v,Y_v)$ of r.v.'s, such that $U$ takes values in an arbitrary alphabet, $U \to X_v \to Y_v$ is a Markov chain, $X_v$ takes values in $\sX_{v \leftarrow}$, $Y_v$ takes values in $\sY_{v \leftarrow}$, and the conditional probability law $\PP_{Y_v|X_v}$ is equal to the product of all the channels entering $v$.  As we discuss in detail in Sec.~\ref{sec:SDPI_intro}, this constant is related to so-called \textit{strong data processing inequalities} (SDPIs) \cite{MR_SDPI}, and quantifies the information transmission capabilities of the channels entering $v$.  When $\eta_v < 1$, the upper bound \eqref{eq:SDPI_intro} is \textit{strictly smaller} than $H(W_{\cS^c}|W_\cS)$.
With the upper bound \eqref{eq:SDPI_intro}, we can strengthen the cutset bound to the following:
\begin{flalign}\label{eq:cutset_improved_intro}
	I(Z; \wh Z_v | W_\cS) \le 
\min \big\{ TC_\cS, \big(1-(1-\eta_v)^{T}\big)H(W_{\cS^c}|W_{\cS}) \big\}.
\end{flalign}
Combining the bounds in \eqref{eq:smallball_intro} and \eqref{eq:cutset_improved_intro}, we conclude that, if there exists a $T$-step algorithm $\cA$ that achieves accuracy $\eps$ with confidence $1-\delta$, then
\begin{align}\label{eq:lb_intro}
\!&T \ge \max \Bigg\{\frac{1}{C_\cS} \left((1-\delta)\log \frac{1}{\E[L(W_\cS,\eps)]} - h_2(\delta)\right), \nonumber \\
\!\!& \frac{\log\left(1-\frac{1}{H(W_{\cS^c}|W_\cS)}\left((1-\delta)\log \frac{1}{\E[L(W_\cS,\eps)]} - h_2(\delta)\right)\right)}{\log (1-\eta_v)}\Bigg\};
\end{align}
moreover, this inequality holds for all choices of $\cS \subset \cV$ and $v \in \cS$. The precise statements of the resulting lower bounds on $T(\eps,\delta)$ are given in Theorem~\ref{th:gen} and Theorem~\ref{th:SDPI_gen_v}. 

The lower bound in \eqref{eq:lb_intro} accounts for the difficulty of estimating the value of $Z = f(W)$ given only a subset of observations $W_\cS$ through the small ball probability $L(W_\cS,\eps)$, and for the communication bottlenecks in the network through the cutset capacity $C_\cS$ and the constants $\eta_v$. The presence of $L(W_\cS,\eps)$ in the bound ensures the correct scaling of $T(\eps,\delta)$ in the high-accuracy limit $\eps \to 0$. In particular, when the function $f$ is real-valued and the probability distribution of $Z = f(W)$ has a density, it is not hard to see that $L(W_\cS,\eps) = O(\eps)$, and therefore $T(\eps,\delta)$ grows without bound at the rate of $\log (1/\eps)$ as $\eps \to 0$. By contrast, the bounds of Ayaso et al.~\cite{Ayaso_etal} saturate at a finite constant even when no computation error is allowed, i.e., when $\eps=0$. Detailed comparison with existing bounds is given in Sec.~\ref{sec:comp_linear_fn}, where we particularize our lower bounds to the computation of linear functions. Moreover, in certain cases our lower bound on $T(\eps,\delta)$ tends to infinity in the high-confidence regime $\delta \to 0$. By contrast, existing lower bounds that rely on cutset capacity estimates remain bounded regardless of how small we make $\delta$.

Throughout the paper, we provide several concrete examples that illustrate the tightness of the general lower bound in \eqref{eq:lb_intro}. In particular, Example~\ref{ex:2node_mod2+} in Sec.~\ref{sec:lb_comp_time} concerns the problem of computing the mod-$2$ sum of two independent ${\rm Bern}(\tfrac{1}{2})$ random variables in a network of two nodes communicating over binary symmetric channels (BSCs). For that problem, we obtain a lower bound on $T(0,\delta)$ that matches an achievable upper bound within a factor of $2$.
In Example~\ref{ex:dist_W_1:M} in Sec.~\ref{sec:lb_comp_time}, we consider the case where the nodes aim to distribute their discrete observations to all other nodes, and obtain a lower bound on $T(0,\delta)$ that captures the \textit{conductance} of the network, which plays a prominent role in the previously published bounds of Ayaso et al.~\cite{Ayaso_etal}. In Sec.~\ref{sec:compare_ub}, we study two more examples: computing a sum of independent Rademacher random variables in a dumbbell network of BSCs, and distributed averaging of real-valued observations in an arbitrary network of binary erasure channels (BECs). Our lower bound for the former example can precisely capture the dependence of the computation time on the number of nodes in the network, while for the latter example it captures the correct dependence of the computation time on the accuracy parameter $\eps$.

\begin{figure}
\centering
\subfloat[]{
\includegraphics[scale = 1.0]{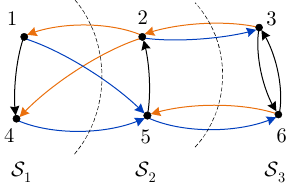}
\label{fg:multi_cutset}
}
\quad\quad\quad\qquad
\centering
\subfloat[]{
\includegraphics[scale = 1.06]{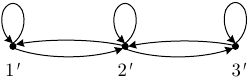}
\label{fg:multi_cutset_chain}
}
\caption{A six-node network partitioned into three sets, $\cS_1 = \{1,4\}$, $\cS_2 = \{2,5\}$, and $\cS_3 = \{3,6\}$. Here, $\cP_1 = \{1,4\}$, $\cP_2 = \{1,2,4,5\}$, and the cutsets $\cE_{\cP_1} = \{(2,1),(2,4)\}$, $\cE_{\cP_2} = \{(3,2),(6,5)\}$, $\cE_{\cP^c_1} = \{(1,5),(4,5)\}$, and $\cE_{\cP^c_2} = \{(2,3),(5,6)\}$ are disjoint. Observe that nodes in $\cS_1$ communicate only with nodes in $\cS_2$ and $\cS_1$, nodes in $\cS_2$ communicate only with nodes in $\cS_1,\cS_2,\cS_3$, and nodes in $\cS_3$ communicate only with nodes in $\cS_2,\cS_3$.
The bidirected chain reduced from the network is shown on the right.}
\end{figure}

A significant limitation of the analysis based on a single cut $(\cS,\cS^c)$ of the network is that it only captures the flow of information across the cutset $\cE_\cS$, but does not account for the time it takes the algorithm to disseminate this information to all the nodes in $\cS$. We address this limitation in Sec.~\ref{sec:multi_cutset} through a multi-cutset analysis. The main idea is to partition the set of nodes $\cV$ into \textit{several} subsets $\cS_1,\ldots,\cS_n$, such that, for all $\cP_i \deq \cS_1 \cup \ldots \cup \cS_i$, the cutsets $\cE_{\cP_1},\ldots,\cE_{\cP_{n-1}}$, $\cE_{\cP^c_1},\ldots,\cE_{\cP^c_{n-1}}$ are disjoint, and to analyze the flow of information across this sequence of cutsets.  Once such a partition is selected, the analysis is based on a network reduction argument (Lemma~\ref{lm:net_red_AA'}), which lumps all the nodes in each $\cS_i$ into a single virtual ``supernode." The construction of the partition ensures that each supernode $i$ only communicates with supernodes $i-1$ and $i+1$, and can also send noisy messages to itself (this is needed to simulate noisy communication among the nodes within $\cS_i$ in the original network). Thus, the reduced network takes the form of a chain with $n$ nodes communicating with their nearest neighbors over bidirectional noisy links and, in addition, sending noisy messages to themselves. We refer to this network as a \textit{bidirected chain} of length $n-1$.
Figure~\ref{fg:multi_cutset} shows the partition of a six-node network, and the bidirected chain reduced from this network is shown in Fig.~\ref{fg:multi_cutset_chain}.

Once this reduction is carried out, we can convert any $T$-step algorithm $\cA$ running on the original network into a \textit{randomized} $T$-step algorithm $\cA'$ running on the reduced network with the same accuracy and confidence guarantees as $\cA$. Consequently, it suffices to analyze distributed function computation in bidirected chains. The key quantitative statement that emerges from this analysis can be informally stated as follows: For any bidirected chain with $n > 3$ nodes, there exists a constant $\eta \in [0,1]$ that plays the same role as $\eta_v$ in \eqref{eq:SDPI_intro} and quantifies the information transmission capabilities of the channels in the chain, such that, for any algorithm $\cA$ that runs on this chain and takes time $T = O(n/\eta)$, the conditional mutual information between the function value $Z$ and its estimate $\wh Z_n$ at the rightmost node $n$ given the observations of nodes $2$ through $n$ is upper-bounded by
\begin{align}\label{eq:dissipation_intro}
I(Z;\wh Z_n|W_{2:n}) = O\left(\frac{C_{(1,2)}n^2}{\eta} e^{-2n\eta^2} \right),
\end{align}
where $C_{(1,2)}$ is the Shannon capacity of the channel from node $1$ to node $2$. The precise statement is given in Lemma~\ref{lm:mi_ub_SDPI_chain_eta} in Sec.~\ref{sec:network_red}. Intuitively, this shows that, unless the algorithm uses $\Omega(n/\eta)$ steps, the information about $W_1$ will \textit{dissipate} at an exponential rate by the time it propagates through the chain from node $1$ to node $n$. Combining \eqref{eq:dissipation_intro} with the lower bound on $I(Z; \wh Z_n |W_{2:n})$ based on small ball probabilities, we can obtain lower bounds on the computation time $T(\eps,\delta)$. The precise statement is given in Theorem~\ref{th:SDPI_red_gen}. Moreover, as we show, it is always possible to reduce an arbitrary network with bidirectional point-to-point channels between the nodes to a bidirected chain whose length is equal to the \textit{diameter} of the original network, which implies that, for networks with sufficiently large diameter, and for sufficiently small values of $\eps,\delta$,
\begin{align}\label{eq:diam_bound}
T(\eps,\delta) = \Omega \left(\frac{{\rm diam}(G)}{\eta}\right),
\end{align}
where ${\rm diam}(G)$ denotes the diameter. This dependence on ${\rm diam}(G)$, which cannot be captured by the single-cutset analysis, is missing in almost all of the existing lower bounds on computation time. An exception is the paper by Rajagopalan and Schulman \cite{Rajagopalan94_dist_comp} that gives an asymptotic lower bound on the time required to broadcast a single bit over a chain of unidirectional BSCs. Our multi-cutset analysis applies to both discrete and continuous observations, and to general network topologies. It can be straightforwardly particularized to specific networks, such as bidirected chains, rings, trees, and grids, as discussed in Sec.~\ref{sec:multi_cut_lb}. We note that techniques involving multiple (though not necessarily disjoint) cutsets have also been proposed in the study of multi-party communication complexity by Tiwari \cite{Tiwari87} and more recently by Chattopadhyay {et al.}~\cite{topology_Cha14}, while our concern is the influence of network topology and channel noise on the computation time.

\subsection{Organization of the paper}

The remainder of the paper is structured as follows. We start with the single-cutset analysis in Sec.~\ref{sec:single_cutset}.
The lower bound on the conditional mutual information via the conditional small ball probability is presented in Sec.~\ref{sec:lb_cond_mi}.
The cutset upper bound and the SDPI upper bound on the conditional mutual information are presented in Sec.~\ref{sec:ub_mi_cutset_cap} and Sec.~\ref{sec:ub_mi_SDPI}. 
An introduction on SDPIs given in Sec.~\ref{sec:SDPI_intro}.
The lower bound on computation time is given in Sec.~\ref{sec:lb_comp_time}, along with two concrete examples.
Sec.~\ref{sec:multi_cutset} is devoted to the multi-cutset analysis, where we first present the network reduction argument in Sec.~\ref{sec:network_red}, then derive general lower bounds on computation time and particularize the results to special networks in Sec.~\ref{sec:multi_cut_lb}.
In Sec.~\ref{sec:comp_linear_fn}, we discuss  lower bounds for computing linear functions, where we relate the conditional small ball probability to L\'evy concentration functions, and evaluate them in a number of special cases. We also make detailed comparisons of our results with existing lower bounds in Sec.~\ref{sec:compare_exist_lb}.
In Sec.~\ref{sec:compare_ub}, we compare the lower bounds on computation time with the achievable upper bounds for two more examples: computing a sum of independent Rademacher random variables in a dumbbell network of BSCs, and distributed averaging of real-valued observations in an arbitrary network of binary erasure channels (BECs).
We conclude this paper and point out future research directions in Sec.~\ref{sec:conclusion}. A couple of lengthy technical proofs are relegated to a series of appendices.

\section{Single-cutset analysis}\label{sec:single_cutset}
We start by deriving information-theoretic lower bounds on the computation time $T(\eps,\delta)$ based on a single cutset in the network. Recall that a \textit{cutset} associated to a partition of $\cV$ into two disjoint sets $\cS$ and $\cS^c \deq \cV \setminus \cS$ consists of all edges that connect a node in $\cS^c$ to a node in $\cS$:
$$
\cE_\cS \deq \big\{ (u,v) \in \cE : u \in \cS^c, v \in \cS \big\} \equiv (\cS^c \times \cS) \cap \cE.
$$
When $\cS$ is a singleton, i.e., $\cS = \{v\}$, we will write $\cE_v$ instead of the more clunky $\cE_{\{v\}}$. As the discussion in Sec.~\ref{ssec:summary} indicates, our analysis revolves around the conditional mutual information $I(Z; \wh Z_v | W_\cS)$ for an arbitrary set of nodes $\cS \subset \cV$ and for an arbitrary node $v \in \cS$.
The lower bound on $I(Z; \wh Z_v | W_\cS)$ expresses quantitatively the intuition that any algorithm that achieves 
\begin{align*}
\max_{v\in\cV} \PP\big[d(Z,\wh Z_v) > \eps \big] \le \delta
\end{align*}
must necessarily extract a sufficient amount of information about the value of $Z = f(W) = f(W_\cS,W_{\cS^c})$. 
On the other hand, the upper bounds on $I(Z; \wh Z_v | W_\cS)$ formalize the idea that this amount cannot be too large, since any information that nodes in $\cS$ receive about $W_{\cS^c}$ must flow across the edges in the cutset $\cE_\cS$ (cf.~\cite[Sec.~15.10]{Cover_book} for a typical illustration of this type of \textit{cutset arguments}). We capture this information limitation in two ways: via channel capacity and via SDPI constants.

The remainder of this section is organized as follows. We first present conditional mutual information lower bounds in Sec.~\ref{sec:lb_cond_mi}. Then we state the upper bound based on cutset capacity in Sec.~\ref{sec:ub_mi_cutset_cap}. After a brief detour to introduce the SDPIs in Sec.~\ref{sec:SDPI_intro}, we state the SDPI-based upper bounds in Sec.~\ref{sec:ub_mi_SDPI}. Finally, we combine the lower and upper bounds to derive lower bounds on $T(\eps,\delta)$ in Sec.~\ref{sec:lb_comp_time}.

\subsection{Lower bound on $I(Z;\wh Z_{v}|W_\cS)$}\label{sec:lb_cond_mi}
For any $\eps \ge 0$, $\cS \subset \cV$, and $w_\cS \in \prod_{v \in \cS} \sW_v$, define the \textit{conditional small ball probability} of $Z$ given $W_\cS = w_\cS$ as
\begin{align}\label{eq:def_cond_smb_prob}
	L(w_\cS,\eps) \deq \sup_{z \in \sZ} \PP[d(Z,z) \le \eps | W_\cS = w_\cS].
\end{align}
This quantity measures how well the conditional distribution of $Z$ given $W_\cS = w_\cS$ concentrates in a small region of size $\eps$ as measured by $d(\cdot,\cdot)$. The following lower bound on $I(Z; \wh Z_v|W_\cS)$ in terms of the conditional small ball probability is essential for proving lower bounds on $T(\eps,\delta)$.
\begin{lemma}\label{lm:mi_lb_gen} 
If an algorithm $\cA$ achieves 
\begin{align}
\max_{v \in \cV} \PP\big[d(Z,\wh{Z}_v) > \eps \big] \le \delta \le 1/2 , \label{eq_excess_distortion_bound}
\end{align}
then for any set $\cS \subset \cV$ and any node $v\in\cS$,
\begin{align}\label{eq:mi_lb_smb}
I(Z; \wh Z_v|W_\cS) &\ge (1-\delta)\log \frac{1}{\E[L(W_\cS,\eps)]} - h_2(\delta) ,
\end{align}
where $h_2(\delta) = -\delta\log\delta - (1-\delta)\log(1-\delta)$ is the binary entropy function.
\end{lemma}
\begin{IEEEproof}
Fix an arbitrary $\cS \subset \cV$ and an arbitrary $v \in \cS$. Consider the probability distributions $\PP = \PP_{W_\cS,Z,\wh{Z}_v}$ and $\QQ = \PP_{W_\cS} \otimes \PP_{Z|W_\cS} \otimes \PP_{\wh{Z}_v|W_\cS}$. Define the indicator random variable $\Upsilon \deq  \I\big\{d(Z,\wh{Z}_v)\le\eps\big\}$. Then from \eqref{eq_excess_distortion_bound} it follows that $\PP[\Upsilon=1] \ge 1-\delta$. On the other hand, since $Z \to W_\cS \to \wh{Z}_v$ form a Markov chain under $\QQ$, by Fubini's theorem,
\begin{align}
&\quad\,\, \QQ[\Upsilon=1] \nonumber \\
&= \int_{\sW_\cS} \int_{\sZ} \int_{\sZ} \I\big\{d(z,\wh{z}_v)\le\eps\big\} \PP({\rm d}z | w_\cS) \PP({\rm d}\wh{z}_v | w_\cS) \PP({\rm d} w_\cS) \nonumber\\
	&= \int_{\sW_\cS} \int_{\sZ}\PP\big[ d(Z,\wh{z}_v) \le \eps \big| W_\cS = w_\cS\big] \PP({\rm d}\wh{z}_v | w_\cS)\PP({\rm d} w_\cS) \nonumber\\
	&\le \int_{\mathcal{\sW}_\cS} \sup_{\wh{z}_v\in \sZ}\PP\big[d(Z,\wh{z}_v) \le \eps \big| W_\cS = w_\cS\big] \PP({\rm d}w_\cS) \nonumber\\
	&= \E[L(W_\cS,\eps)] . \label{eq:Q_bound} 
\end{align}
Consequently,
\begin{align*}
I(Z;\wh{Z}_v|W_\cS) &= D(\PP \| \QQ) \\
&\overset{{\rm (a)}}\ge d_2(\PP[\Upsilon=1] \| \QQ[\Upsilon=1]) \\
&\overset{{\rm (b)}}\ge \PP[\Upsilon=1]\log\frac{1}{\QQ[\Upsilon=1]} - h_2(\PP[\Upsilon=1]) \\
&\overset{{\rm (c)}}\ge (1-\delta)\log\frac{1}{\E[L(W_\cS,\eps)]} - h_2(\delta)
\end{align*}
where 
\begin{enumerate}
  \item[(a)] follows from the data processing inequality for divergence, where $d_2(p \| q) \deq p\log(p/q)+(1-p)\log((1-p)/(1-q))$ is the binary divergence function;
  \item[(b)] follows from the fact that $d_2(p \| q) \ge p \log (1/q) - h_2(p)$;
  \item[(c)] follows from the fact that $\PP[\Upsilon=1]\ge 1-\delta \ge 1/2$ by \eqref{eq_excess_distortion_bound}, and $\QQ[\Upsilon=1] \le \E[L(W_\cS,\eps)]$ by \eqref{eq:Q_bound}.
\end{enumerate}
\end{IEEEproof}
\noindent For a fixed $\eps$, Lemma~\ref{lm:mi_lb_gen} captures the intuition that, the more spread the conditional distribution $\PP_{Z|W_\cS}$ is, the more information we need about $Z$ to achieve the required accuracy; similarly, for a fixed $\PP_{Z|W_\cS}$, the smaller the accuracy parameter $\eps$, the more information is necessary. In Section~\ref{sec:comp_linear_fn}, we provide explicit expressions and upper bounds for the conditional small ball probability $L(\eps,w_\cS)$ in the context of computing linear functions of real-valued r.v.'s with absolutely continuous probability distributions. We show that, in such cases, $L(\eps,w_\cS) = O(\eps)$, which implies that the lower bound of Lemma~\ref{lm:mi_lb_gen} grows at least as fast as $\log (1/\eps)$ in the high-accuracy limit $\eps \to 0$.

\subsection{Upper bound on $I(Z;\wh Z_{v}|W_\cS)$ via cutset capacity}\label{sec:ub_mi_cutset_cap}
Our first upper bound involves the \textit{cutset capacity $C_\cS$}, defined as
\begin{align*}
C_\cS \deq \sum_{e \in \cE_\cS} C_e .
\end{align*}
Here, $C_e$ denotes the Shannon capacity of the channel $K_e$. 
\begin{lemma}\label{lm:mi_ub_cutset} 
For any set $\cS \subset \cV$, let $\wh Z_\cS \deq (\wh Z_v)_{v\in\cS}$. Then, for any $T$-step algorithm $\cA$ and for any $v \in \cS$,
\begin{align*}
I(Z; \wh{Z}_v | W_{\cS}) \le I(Z;\wh{Z}_{\cS}| W_{\cS} ) \le T C_{\cS} .
\end{align*}
\end{lemma}
\begin{IEEEproof}
The first inequality follows from the data processing lemma for mutual information. The second inequality has been obtained in \cite{Ayaso_etal} and \cite{Como_Dahleh} as well, but the proof in \cite{Ayaso_etal} relies heavily on differential entropy. Our proof is more general, as it only uses the properties of mutual information.

For a set of nodes $\cS\subset\cV$, let $X_{\cS,t} \deq (X_{v,t})_{v\in\cS}$ and $Y_{\cS,t} \deq (Y_{v,t})_{v\in\cS}$.
For two subsets $\cS_1$ and $\cS_2$ of $\cV$, define $X_{(\cS_1, \cS_2),t} \deq \big(X_{(u,v),t} : u \in \cS_1, v \in \cS_2, (v,u) \in \cE \big)$ as the messages sent from nodes  in $\cS_1$ to nodes in $\cS_2$ at step $t$, and $Y_{(\cS_1, \cS_2),t} \deq \big(Y_{(u,v),t} : u \in \cS_1, v \in \cS_2, (u,v) \in \cE \big)$ as the messages received by nodes in $\cS_2$ from nodes in $\cS_1$ at step $t$. We will be using this notation in the proofs that follow, as well.

If $T=0$, then for any $v\in\cS$, $\wh Z_v = \psi_v(W_v)$, hence $I(Z;\wh Z_{\cS} | W_{\cS}) \le I(Z;W_\cS | W_{\cS}) = 0$. For $T\ge 1$, we start with the following chain of inequalities:
\begin{align}
I(Z;\wh{Z}_{\cS}| W_{\cS})  
&\overset{\rm (a)}{\le} I(W_{\cS},W_{\cS^c};W_{\cS},Y_{\cS}^T| W_{\cS}) \nonumber \\
&= I(W_{\cS^c};Y_{\cS}^T| W_{\cS}) \nonumber \\
&= \sum\limits_{t=1}^{T} I(W_{\cS^c};Y_{\cS,t}| W_{\cS}, Y_{\cS}^{t-1}) \nonumber \\
&\overset{\rm (b)} = \sum\limits_{t=1}^{T} I(W_{\cS^c};Y_{\cS,t}| W_{\cS}, Y_{\cS}^{t-1}, X_{\cS,t}) \nonumber \\
&\le \sum\limits_{t=1}^{T} I(W_{\cS^c}, X_{\cS^c,t};Y_{\cS,t}| W_{\cS}, Y_{\cS}^{t-1}, X_{\cS,t}) \nonumber \\
&= \sum\limits_{t=1}^{T} \Big( I(X_{\cS^c,t};Y_{\cS,t}| W_{\cS}, Y_{\cS}^{t-1}, X_{\cS,t}) \nonumber \\
&\qquad +  I(W_{\cS^c};Y_{\cS,t}| W_{\cS}, Y_{\cS}^{t-1}, X_{\cS,t}, X_{\cS^c,t}) \Big) \nonumber \\
&\overset{\rm (c)}= \sum\limits_{t=1}^{T} I(X_{\cS^c,t};Y_{\cS,t}| W_{\cS}, Y_{\cS}^{t-1}, X_{\cS,t}) \nonumber \\
&\overset{\rm (d)}{\le} \sum\limits_{t=1}^{T} I(X_{\cS^c,t};Y_{\cS,t}| X_{\cS,t}) \label{eq:pf_lm_mi_ub_1}
\end{align}
where
\begin{enumerate}
  \item[(a)] follows from data processing inequality, and the fact that $Z = f(W_\cS,W_{\cS^c})$ and $\wh{Z}_v = \psi_{v}(W_v,Y_v^T)$;
  \item[(b)] follows from the fact that $X_{v,t} = \varphi_{v,t}(W_v, Y_{v}^{t-1})$;
  \item[(c)] follows from the memorylessness of the channels, hence the Markov chain $W_{\cS^c},W_{\cS}, Y_{\cS}^{t-1} \rightarrow  X_{\cS,t}, X_{\cS^c,t} \rightarrow Y_{\cS,t}$, and the weak union property of conditional independence \cite[p. 25]{PGM_book};
  \item[(d)] follows from the Markov chain 
$$W_{\cS}, Y_{\cS}^{t-1} \rightarrow X_{\cS,t}, X_{\cS^c,t} \rightarrow Y_{\cS,t},$$
together with the fact that, if $X \rightarrow A,B \rightarrow C$ form a Markov chain, then
$$
I(A;C| X,B) \le I(A;C | B).
$$
To prove this, we expand $I\left(\left.A,X;C\right| B \right) $ in two ways to get
\begin{align*}
  I\left(A,X;C | B \right) &= I(X;C | B) + I\left(\left.A;C\right| X,B \right)  \\
&= I(A;C | B) + I\left(\left.X;C\right| A,B \right) .
\end{align*}
The claim follows because $I\left(\left.X;C\right| A,B \right) = 0$.
\end{enumerate}
From now on we drop the step index $t$ and denote $X_{(\cS_1,\cS_2),t}$ as $X_{\cS_1 \cS_2}$ to simplify the notation.
Note that $X_{\cS} = (X_{\cS \cS},X_{\cS \cS^c})$ and $Y_{\cS} = (Y_{\cS \cS}, Y_{\cS^c \cS})$. We have 
\begin{align}
I(X_{\cS^c} ; Y_{\cS}| X_{\cS}) &= I(X_{\cS^c}; Y_{\cS^c \cS}, Y_{\cS \cS}| X_\cS) \nonumber \\
&= I( X_{\cS^c}; Y_{\cS^c \cS} | X_\cS ) + I( X_{\cS^c};  Y_{\cS \cS} | X_\cS,Y_{\cS^c \cS} ) \nonumber \\
&\overset{\rm (a)}{=} I( X_{\cS^c \cS},X_{\cS^c \cS^c}; Y_{\cS^c \cS} | X_\cS ) \nonumber \\
&= I( X_{\cS^c \cS}; Y_{\cS^c \cS} | X_\cS ) 
\nonumber \\
&\quad + I( X_{\cS^c \cS^c};  Y_{\cS^c \cS} | X_\cS,X_{\cS^c \cS} ) \nonumber \\
&\overset{\rm (b)}{\le} I( X_{\cS^c \cS}; Y_{\cS^c \cS} ) \nonumber \\
&\overset{\rm (c)}{\le} \sum_{e \in \cE_{\cS}} C_{e} \label{eq:pf_lm_mi_ub_2}
\end{align}
where
\begin{enumerate}
  \item[(a)] follows from the Markov chain $X_{\cS^c},Y_{\cS^c \cS} \rightarrow  X_\cS \rightarrow Y_{\cS \cS}$ and the weak union property of conditional independence;
  \item[(b)] follows from the Markov chains $X_\cS \rightarrow  X_{\cS^c \cS} \rightarrow Y_{\cS^c \cS}$ and $X_{\cS^c \cS^c},X_\cS \rightarrow X_{\cS^c \cS} \rightarrow Y_{\cS^c \cS}$, and the weak union property of conditional independence;
  \item[(c)] follows from the fact that the channels associated with $\cE_\cS$ are independent, and the fact that the capacity of a product channel is at most the sum of the capacities of the constituent channels \cite{PolWu_IT_lectures}.
\end{enumerate}
Then the statement of Lemma \ref{lm:mi_ub_cutset} follows from (\ref{eq:pf_lm_mi_ub_1}) and (\ref{eq:pf_lm_mi_ub_2}). \end{IEEEproof}

\subsection{Preliminaries on strong data processing inequalities}\label{sec:SDPI_intro}
In Sec.~\ref{sec:ub_mi_SDPI}, we will upper-bound $I(Z; \wh Z_v|W_\cS)$ using so-called \textit{strong data processing inequalities} (SDPI's) for discrete channels (cf.~\cite{MR_SDPI} and references therein). 
Here we provide the necessary background.
A discrete memoryless channel is specified by a triple $(\sX,\sY,K)$, where $\sX$ is the input alphabet, $\sY$ is the output alphabet, and $K = \big( K(y|x) \big)_{(x,y) \in \sX \times \sY}$ is the stochastic transition law. We say that the channel $(\sX,\sY,K)$ \textit{satisfies an SDPI at input distribution $\PP_X$ with constant $c\in[0,1)$} if $D(\QQ_Y \| \PP_Y) \le c D(\QQ_X \| \PP_X)$ for any other input distribution $\QQ_X$.
Here $\PP_Y$ and $\QQ_Y$ denote the marginal distribution of the channel output when the input has distribution $\PP_X$ and $\QQ_X$, respectively. 
Define the SDPI constant of $K$ as
\begin{align*}
\eta(K) \deq \sup_{\PP_X} \sup_{\QQ_X\neq \PP_X}\frac{D(\QQ_Y \| \PP_Y)}{D(\QQ_X \| \PP_X)} .
\end{align*}
The SDPI constants of some common discrete channels have closed form expressions. For example, for a binary symmetric channel (BSC) with crossover probability $p$, $\eta({\rm BSC}(p)) = (1-2p)^2$ \cite{Ahlswede_Gacs_hypercont}, and for a binary erasure channel (BEC) with erasure probability $p$, $\eta({\rm BEC}(p)) = 1-p$.
It can be shown that $\eta(K)$ is also the maximum mutual information contraction ratio in a Markov chain $U \rightarrow X \rightarrow Y$ with $\PP_{Y|X} = K$ \cite{VA_SDPI}:
\begin{align*}
\eta(K) = \sup_{\PP_{U,X}} \frac{I(U;Y)}{I(U;X)}
\end{align*}
(see \cite[App.~B]{PolWu_SDPI_avgP} for a proof of this formula in the setting of abstract alphabets). Consequently, for any such Markov chain,
\begin{align*}
I(U;Y) \le \eta(K) I(U;X) .
\end{align*}
This is a stronger result than the ordinary data processing inequality for mutual information, as it quantitatively captures the amount by which the information contracts after passing through a channel. We will also need a conditional version of the SDPI:
\begin{lemma}\label{lm:cond_SDPI}
For any Markov chain $U,V \rightarrow X \rightarrow Y$ with $\PP_{Y|X} = K$, 
\begin{align*}
I(U;Y|V) \le \eta(K) I(U;X|V) .
\end{align*}
\end{lemma}
\noindent For binary channels, this result was first proved
by Evans and Schulman \cite[Corollary~1]{Eva_Sch99}. A proof for the general case is included in \cite[Lemma~2.7]{Xu_thesis}.
Finally, we will need a bound on the SDPI constant of a product channel. The \textit{tensor product} of two channels $(\sX_1,\sY_2,K_1)$ and $(\sX_2,\sY_2,K_2)$ is a channel $(\sX_1 \times \sX_2, \sY_1 \times \sY_2, K_1 \otimes K_2)$ with
$$
K_1 \otimes K_2 (y_1, y_2 | x_1, x_2) \deq K_1(y_1|x_1) K_2(y_2|x_2)
$$
for all $(x_1,x_2) \in \sX_1 \times \sX_2,\, (y_1,y_2) \in \sY_1 \times \sY_2$.
The extension to more than two channels is obvious. The following lemma is a special case of Corollary 2 of Polyanskiy and Wu \cite{PolWu_ES}, obtained using the method of Evans and Schulman \cite{Eva_Sch99}.  We give the proof, since we adapt the underlying technique at several points in this paper.
\begin{lemma}\label{lm:SDPI_prod_ch}
For a product channel $K = \bigotimes_{i=1}^m K_i$, if the constituent channels satisfy $\eta(K_i) \le \eta$ for $i\in\{1,\ldots,m\}$, then
\begin{align*}
\eta(K) \le 1 - (1-{\eta})^m .
\end{align*}
\end{lemma}
\noindent
\begin{IEEEproof}
	Let $X^m$ and $Y^m$ be the input and output of the product channel $K = K_1 \otimes \ldots \otimes K_m$. Let $U$ be an arbitrary random variable, such that $U \rightarrow X^m \rightarrow Y^m$ form a Markov chain. 
	It suffices to show that
	\begin{align}
	I(U;Y^m) \le \big(1-(1-\eta)^m\big)I(U;X^m) .\label{eq:I_UX_UY}
	\end{align}
	From the chain rule,
	\begin{align*}
	I(U;Y^m) = I(U;Y^{m-1}) + I(U;Y_m|Y^{m-1}) .
	\end{align*}
	Since $U,Y^{m-1} \rightarrow X_m \rightarrow Y_m$ form a Markov chain, and $\PP_{Y_m|X_m} = K_m$, Lemma~\ref{lm:cond_SDPI} gives
	\begin{align*}
	I(U;Y_m|Y^{m-1}) &\le \eta(K_m) I(U;X_m|Y^{m-1}) \\
&\le \eta I(U;X_m|Y^{m-1}) .
	\end{align*}
	It follows that
	\begin{align*}
	I(U;Y^m) &\le I(U;Y^{m-1}) + \eta I(U;X_m|Y^{m-1}) \\
	&= (1-\eta)I(U;Y^{m-1}) + \eta I(U;Y^{m-1},X_m) \\
	&\le (1-\eta)I(U;Y^{m-1}) + \eta I(U;X^m) ,
	\end{align*}
	where the last step follows from the ordinary data processing inequality and the Markov chain $U \rightarrow X^{m} \rightarrow Y^{m-1},X_m$. Unrolling the above recursive upper bound on $I(U;Y^m)$ and noting that $I(U;Y_1) \le \eta I(U;X_1)$, we get
	\begin{align*}
	I(U;Y^m) &\le (1-\eta)^{m-1}\eta I(U;X_1) + \ldots \nonumber \\
&\quad + (1-\eta)\eta I(U;X^{m-1}) + \eta I(U;X^m) \\
	&\le \big((1-\eta)^{m-1}+ \ldots + (1-\eta)+1\big)\eta I(U;X^m) \\
	&= \big(1-(1-\eta)^m\big) I(U;X^m),
	\end{align*}
	which proves (\ref{eq:I_UX_UY}) and hence Lemma~\ref{lm:SDPI_prod_ch}.
\end{IEEEproof}

\subsection{Upper bound on $I(Z;\wh Z_{v}|W_\cS)$ via SDPI}\label{sec:ub_mi_SDPI}

Having the necessary background at hand, we can now state our upper bounds based on SDPI constants. Let $K_v \deq \bigotimes_{e\in\cE_v}K_{e}$ be the overall transition law of the channels across the cutset $\cE_v$. Define
$$
\eta_v \deq \eta(K_v)
$$
as the SDPI constant of $K_v$, and 
$$
\eta^*_v \deq \max_{e \in \cE_v} \eta(K_e)
$$
as the largest SDPI constant among all the channels across $\cE_v$.
Our second upper bound on $I(Z;\wh Z_{v}|W_\cS)$ involves these SDPI constants, and the conditional entropy of $W_{\cS^c}$ given $W_{\cS}$.
\begin{lemma}\label{lm:mi_ub_SDPI_gen} For any set $\cS \subset \cV$, any node $v\in\cS$, and any $T$-step algorithm $\cA$, 
\begin{align*}
I(Z;\wh{Z}_{v}| W_{\cS} ) & \le (1-(1-\eta_v)^{T})H(W_{\cS^c}|W_{\cS}) \\
& \le (1-(1-\eta^*_v)^{|\cE_v|T})H(W_{\cS^c}|W_{\cS}).
\end{align*}
\end{lemma}
\begin{IEEEproof}
We adapt the proof of Lemma~\ref{lm:SDPI_prod_ch}. For any $v$ and $t$, define the shorthand $X_{v\leftarrow, t} \deq X_{(\cN_{v \leftarrow},v),t}$. If $T=0$, then for any $v\in\cS$, $\wh Z_v = \psi_v(W_v)$; hence $I(Z;\wh Z_{v} | W_{\cS}) \le I(Z;W_v | W_{\cS}) = 0$. If $T\ge 1$, then for any $v\in\cS$,
\begin{align}
&\quad\,\, I(Z;\wh Z_{v} | W_{\cS}) \nonumber\\
&\le I(W_\cS,W_{\cS^c};W_v,Y_{v}^T | W_{\cS}) \nonumber\\
&= I(W_{\cS^c};Y_{v}^T | W_{\cS}) \nonumber\\
&= I(W_{\cS^c};Y_{v}^{T-1} | W_{\cS}) + I(W_{\cS^c};Y_{v,T}|W_{\cS},Y_{v}^{T-1}) \nonumber\\
&\overset{\rm (a)}\le I(W_{\cS^c};Y_{v}^{T-1} | W_{\cS}) + \eta_v I(W_{\cS^c};X_{v\leftarrow,T}|W_{\cS},Y_{v}^{T-1}) \nonumber \\
&= (1-\eta_v)I(W_{\cS^c};Y_{v}^{T-1} | W_{\cS}) \nonumber \\
&\quad + \eta_v I(W_{\cS^c};Y_{v}^{T-1}, X_{v\leftarrow,T}|W_\cS) \nonumber 
\end{align}
where (a) follows from the conditional SDPI (Lemma~\ref{lm:cond_SDPI}) and the fact that $W_{\cS^c},W_{\cS},Y_{v}^{t-1} \rightarrow X_{v\leftarrow,t} \rightarrow Y_{v,t}$ form a Markov chain for $t\in\{1,\ldots,T\}$. 
Unrolling the above recursive upper bound on $I(W_{\cS^c};Y_{v}^T | W_{\cS})$, and noting that $I(W_{\cS^c};Y_{{v},1} | W_{\cS}) \le \eta_{v} I(W_{\cS^c};X_{v\leftarrow,1} | W_{\cS})$, we get
\begin{align*}
&\quad\,\, I(W_{\cS^c};Y_{v}^T|W_{\cS}) \\
&\le (1-\eta_{v})^{T-1}\eta_v I(W_{\cS^c};X_{v\leftarrow,1}|W_{\cS}) + 
\ldots  \nonumber \\
& \quad + (1-\eta_v)\eta_v I(W_{\cS^c};Y_{v}^{T-2},X_{v\leftarrow,T-1}|W_{\cS}) \\
&\quad + 
\eta_v I(W_{\cS^c};Y_{v}^{T-1},X_{v\leftarrow,T}|W_{\cS})  \\
&\le \big((1-\eta_v)^{T-1}+ \ldots +(1-\eta_v)+1\big)\eta_v H(W_{\cS^c}|W_{\cS}) \\
&= \big(1-(1-\eta_v)^T\big) H(W_{\cS^c}|W_{\cS}) .
\end{align*}
The weakened upper bound follows from the fact that $\eta_v \le  1-(1-\eta_v^*)^{|\cE_{v}|}$, due to Lemma~\ref{lm:SDPI_prod_ch}.
This completes the proof of Lemma~\ref{lm:mi_ub_SDPI_gen}.\end{IEEEproof}

\begin{figure}[h!]
\centering
\includegraphics[scale = 0.85]{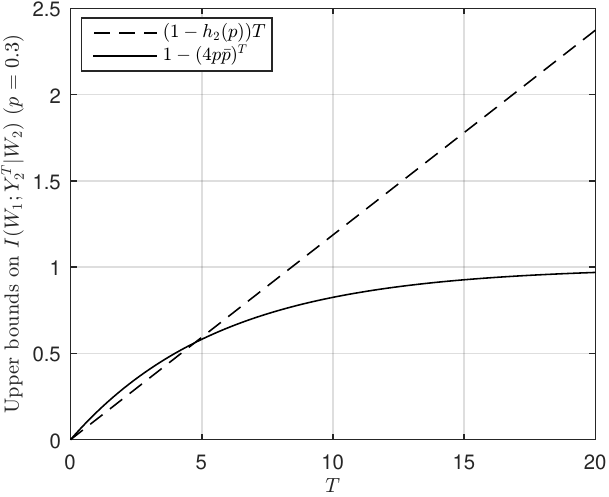}
\caption{Comparison of upper bounds in Lemma~\ref{lm:mi_ub_cutset} and Lemma~\ref{lm:mi_ub_SDPI_gen} for computing mod-$2$ sum in a two-node network.}
\label{fg:cutset_SDPI}
\end{figure}

Comparing Lemma~\ref{lm:mi_ub_cutset} and Lemma~\ref{lm:mi_ub_SDPI_gen}, we note that the upper bound in Lemma~\ref{lm:mi_ub_cutset} captures the communication constraints through the cutset capacity alone, in accordance with the fact that the communication constraints do not depend on $W$ or $Z$. The bound applies when $W$ is either discrete or continuous; however, it grows linearly with $T$.
By contrast, the upper bound in Lemma~\ref{lm:mi_ub_SDPI_gen} builds on the fact that $I(Z;\wh{Z}_{v}| W_{\cS} )$ is upper bounded by $H(W_{\cS^c}|W_{\cS})$, and goes a step further by capturing the communication constraint through a multiplicative contraction of $H(W_{\cS^c}|W_{\cS})$. 
It never exceeds $H(W_{\cS^c}|W_{\cS})$ as $T$ increases.
However, it is useful only when the conditional entropy $H(W_{\cS^c} | W_\cS)$ is well-defined and finite (e.g., when $W$ is discrete).
We give an explicit comparison of Lemma~\ref{lm:mi_ub_cutset} and Lemma~\ref{lm:mi_ub_SDPI_gen} in the following example:
\begin{example}\label{ex:2node_mod2+}
Consider a two-node network, where the nodes are connected by BSCs. The problem is for the two nodes to compute the mod-$2$ sum of their one-bit observations. Formally, we have $G = (\cV,\cE)$ with $\cV = \{1,2\}$, $\cE = \{(1,2),(2,1)\}$, $K_{(1,2)} = K_{(2,1)} = {\rm BSC}(p)$, $W_1$ and $W_2$ are independent ${\rm Bern}(\frac{1}{2})$ r.v.'s, $Z = W_1 \oplus W_2$, and $d(z,\wh z) = \I\{z \neq \wh z\}$.
\end{example}
\noindent Choosing $\cS = \{2\}$, Lemma~\ref{lm:mi_ub_cutset} gives
\begin{align}
I(Z ; \wh Z_2 | W_2) \le (1-h_2(p))T, \label{eq:BSC_cutset}
\end{align}
whereas Lemma~\ref{lm:mi_ub_SDPI_gen}, together with the fact that $\eta({\rm BSC}(p)) = (1-2p)^2$, gives
\begin{align}
I(Z ; \wh Z_2 | W_2) \le 1-(4p\bar p)^T, \label{eq:BSC_SDPI}
\end{align}
where, for $p \in [0,1]$, $\bar{p} \deq 1-p$. For this example, the cutset-capacity upper bound is always tighter for small $T$, as
\begin{align*}
\frac{\partial \big(1-(4p\bar p)^T\big)}{\partial T}\Big|_{T=0} &= \log\frac{1}{4p\bar p} 
\ge 1 - h_2(p) , \quad p \in [0,1].
\end{align*}
Fig.~\ref{fg:cutset_SDPI} shows the two upper bounds with $p = 0.3$: the cutset-capacity upper bound is tighter when $T<5$.

\subsection{Lower bounds on computation time}
\label{sec:lb_comp_time}
We now proceed to derive lower bounds on the computation time $T(\eps,\delta)$ based on the previously derived lower and upper bounds on the conditional mutual information $I(Z;\wh Z_v|W_\cS)$.
Define the shorthand notation
\begin{align}
\ell(\cS,\eps,\delta) \deq (1-\delta)\log \frac{1}{\E[L(W_\cS,\eps)]} - h_2(\delta) ,
\end{align}
which is the lower bound on $I(Z;\wh Z_v|W_\cS)$ in Lemma~\ref{lm:mi_lb_gen} .

\subsubsection{Cutset-capacity bounds}
Combined with the conditional small ball probability lower bound in Lemma~\ref{lm:mi_lb_gen}, the cutset-capacity upper bound in Lemma~\ref{lm:mi_ub_cutset} leads to a lower bound on $T(\eps,\delta) $:
\begin{theorem}\label{th:gen} 
For an arbitrary network, for any $\eps\ge 0$ and $\delta\in[0,1/2]$,
\begin{align*}
T(\eps,\delta) \ge \max_{\cS \subset \cV} \frac{\ell(\cS,\eps,\delta)}{C_\cS}. %
\end{align*}
\end{theorem}

From an operational point of view, the lower bound of Theorem~\ref{th:gen} reflects the fact that the problem of distributed function computation is, in a certain sense, a joint source-channel coding (JSCC) problem { with possibly noisy feedback}. In particular, the lower bound on $I(Z; \wh{Z}_v | W_\cS)$ from Lemma~\ref{lm:mi_lb_gen}, which is used to prove Theorem~\ref{th:gen}, can be interpreted in terms of a reduction of JSCC to generalized list decoding \cite[Sec.~III.B]{Kostina_Verdu_JSCC}. Given any algorithm $\cA$ and any node $v \in \cV$, we may construct a ``list decoder" as follows: given the estimate $\wh{Z}_v$, we generate a ``list'' $\{ z \in \sZ: d(z,\wh{Z}_v) \le \eps \}$. If we fix a set $\cS \subset \cV$ and allow all the nodes in $\cS$ to share their observations $W_\cS$, then $\E[L(W_\cS,\eps)]$ is an upper bound on the $\PP_W$-measure of the list of any node $v \in \cS$.
{ Therefore, $\ell(\cS,\eps,\delta)$ is a lower bound on the total amount of information that is necessary for the JSCC problem.
The complementary cutset upper bound on $I(Z; \wh{Z}_v | W_\cS)$ bounds the amount of information that can be accumulated with each channel use.
The lower bound on $T(\eps,\delta)$ can thus be interpreted as a lower bound on the blocklength of the JSCC problem.}

As we will demonstrate in Section~\ref{sec:comp_linear_fn}, based on Theorem~\ref{th:gen}, it is possible to exploit structural properties of the function $f$ (such as linearity) and of the probability law $\PP_W$ (such as log-concavity) to derive lower bounds on the computation time that are often tighter than existing bounds.

\subsubsection{SDPI bounds}
Combining the lower bound of Lemma~\ref{lm:mi_lb_gen} with the SDPI upper bound of Lemma~\ref{lm:mi_ub_SDPI_gen}, we get the following:
\begin{theorem}\label{th:SDPI_gen_v} 
For an arbitrary network, for any $\eps\ge 0$ and $\delta\in[0,1/2]$,
\begin{align}\label{eq:SDPI_gen_v} 
T(\eps,\delta) \ge \max_{\cS \subset \cV} \max_{v \in \cS} \frac{\log\big(1-\frac{\ell(\cS,\eps,\delta)}{H(W_{\cS^c}|W_\cS)}\big)^{-1}}{|\cE_v|\log(1-\eta^*_v)^{-1}}
\end{align}
where $\eta^*_v \deq \max_{e \in \cE_v} \eta(K_e)$.
\end{theorem}
\noindent The lower bounds in Theorem~\ref{th:gen} and Theorem~\ref{th:SDPI_gen_v} can behave quite differently. To illustrate this, we compare them in two cases:

When $H(W_{\cS^c}|W_\cS) \gg \log \frac{1}{\E[L(W_\cS,\eps)]}$, Theorem~\ref{th:SDPI_gen_v} gives
\begin{align*}
T(\eps,\delta) &\ge \max_{\cS \subset \cV} \max_{v \in \cS} \frac{\log\big(1-\frac{\ell(\cS,\eps,\delta)}{H(W_{\cS^c}|W_\cS)}\big)^{-1}}{|\cE_v|\log(1-\eta^*_v)^{-1}} \\
&\approx \max_{\cS \subset \cV} \max_{v \in \cS} \frac{\ell(\cS,\eps,\delta)\log e}{H(W_{\cS^c}|W_\cS)|\cE_v|\log(1-\eta^*_v)^{-1}} ,
\end{align*}
which has essentially the same dependence on $\ell(\cS,\eps,\delta)$ as the lower bound given by Theorem~\ref{th:gen}.
In this case, Theorem~\ref{th:gen} gives more useful lower bounds as long as $C_\cS \ll H(W_{\cS^c}|W_\cS)$, especially when $W$ is continuous.

When $H(W_{\cS^c}|W_\cS) \approx \log \frac{1}{\E[L(W_\cS,\eps)]}$ and $\delta$ is small, $H(W_{\cS^c}|W_\cS)$ serves as a sharp proxy of $\ell(\cS,\eps,\delta)$.
Theorem~\ref{th:gen} in this case gives
\begin{align*}
T(\eps,\delta) \ge \max_{\cS \subset \cV} \frac{\ell(\cS,\eps,\delta)}{C_\cS} 
\approx \max_{\cS \subset \cV} \frac{H(W_{\cS^c}|W_\cS)}{C_\cS} , 
\end{align*} 
while Theorem~\ref{th:SDPI_gen_v} gives
\begin{align*}
T(\eps,\delta) &\ge \max_{\cS \subset \cV} \max_{v \in \cS} \frac{\log\big(1-\frac{\ell(\cS,\eps,\delta)}{H(W_{\cS^c}|W_\cS)}\big)^{-1}}{|\cE_v|\log(1-\eta^*_v)^{-1}} \\
&\approx \max_{\cS \subset \cV} \max_{v \in \cS} \frac{\log {H(W_{\cS^c}|W_\cS)} + \log\frac{1}{h_2(\delta)}}{|\cE_v|\log(1-\eta^*_v)^{-1}} 
\end{align*}
where in the last step we have used the fact that $\log\left(\delta+\frac{h_2(\delta)}{H(W_{\cS^c}|W_{\cS})}\right) \sim \log\left(\frac{h_2(\delta)}{H(W_{\cS^c}|W_\cS)}\right)$ as $\delta \rightarrow 0$.
Theorem~\ref{th:gen} in this case is sharper in capturing the dependence of $T(\eps,\delta)$ on the amount of information contained in $Z$, in that the lower bound is proportional to $H(W_{\cS^c}|W_\cS)$, whereas the lower bound given by Theorem~\ref{th:SDPI_gen_v} depends on $H(W_{\cS^c}|W_\cS)$ only through $\log H(W_{\cS^c}|W_\cS)$. On the other hand, Theorem~\ref{th:SDPI_gen_v} in this case is much sharper in capturing the dependence of $T(\eps,\delta)$ on the confidence parameter $\delta$, since $\log h_2(\delta)$ grows without bound as $\delta \rightarrow 0$, while the lower bound given by Theorem~\ref{th:gen} remains bounded.
We consider two examples for this case.

The first is Example~\ref{ex:2node_mod2+} in Section~\ref{sec:ub_mi_SDPI}, for the two-node mod-$2$ sum problem. We have $L(w_2,\eps) = \max_{z \in \{0,1\}} \PP[W_1 \oplus W_2 = z |W_2 = w_2] = \frac{1}{2}$, and $\ell(\cS,0,\delta) =   1-\delta - h_2(\delta)$. Theorems~\ref{th:gen} and \ref{th:SDPI_gen_v} imply the following:
\begin{corollary}\label{co:2node_mod2+_Tlb}
For the problem in Example~\ref{ex:2node_mod2+}, for $\delta\in[0,1/2]$, the $(0,\delta)$-computation time satisfies
\begin{align}\label{eq:2node_mod2+_Tlb}
T(0,\delta)  &\ge \max\Big\{\frac{1-\delta - h_2(\delta)}{1-h_2(p)},
\frac{\log (\delta + h_2(\delta))^{-1}}{\log (4p\bar p)^{-1}} \Big\},
\end{align}
where the first lower bound is given by Theorem~\ref{th:gen}, and the second one is given by Theorem~\ref{th:SDPI_gen_v}.
\end{corollary}
{ 
\noindent To obtain an achievable upper bound on $T(0,\delta)$ in Example~\ref{ex:2node_mod2+}, we consider the algorithm where each node uses a length-$T$ repetition code to send its one-bit observation to the other node.
Using the Chernoff bound, as in \cite{Gallager_ITbook}, it can be shown that the probability of decoding error at each node is upper-bounded by $(4p\bar p)^{T/2}$, and therefore this algorithm achieves accuracy $\eps = 0$ with confidence parameter $\delta \le (4p\bar p)^{T/2}$. This gives the upper bound
\begin{align}\label{eq:2node_mod2+_Tub}
T(0,\delta)  &\le \frac{2\log \delta^{-1}}{\log(4p\bar p)^{-1}} .
\end{align}
Comparing \eqref{eq:2node_mod2+_Tub} with the second lower bound in \eqref{eq:2node_mod2+_Tlb}, we see that they asymptotically differ only by a factor of $2$ as $\delta\rightarrow 0$, as $\lim_{\delta\rightarrow 0} \log(\delta+h_2(\delta)) / \log(\delta) = 1$.
Thus, for the problem in Example~\ref{ex:2node_mod2+}, the converse lower bound on $T(0,\delta)$ obtained from the SDPI closely matches the achievable upper bound on $T(0,\delta)$.

The second example concerns the problem of disseminating all of the observations through an arbitrary network:

\begin{example}\label{ex:dist_W_1:M}
Consider the problem where $W_v$'s are i.i.d.\ samples from the uniformly distribution over $\{1,\ldots,M\}$, $Z = W$, and $d(z,\wh z) = \I\{z\neq \wh z\}$. 
In other words, the goal of the nodes is to distribute their observations to all other nodes.
\end{example}
\noindent In this example, $H(W_{\cS^c}|W_{\cS}) = |\cS^c|\log M$,
and $\ell(\cS,0,\delta) = (1-\delta)|\cS^c|\log M - h_2(\delta)$. Following Ayaso et al.~\cite[Def.~III.4]{Ayaso_etal}, we define the \textit{conductance} of the network $G$ as
$$
\Phi(G) \deq \min_{\cS\in\cV: |\cV|/2 < |\cS| < |\cV|} \frac{C_{\cS}}{|\cS^c|} .
$$
Then we have the following corollary:
\begin{corollary}
For the problem in Example~\ref{ex:dist_W_1:M},
Theorem~\ref{th:gen} gives
\begin{align}\label{eq:dist_W_1:M_Tlbcc}
T(0,\delta) 
&\ge \max_{\cS \subset \cV} \frac{(1-\delta)|\cS^c|\log M - h_2(\delta)}{C_\cS} \\
&\gtrsim \frac{\log M}{\Phi(G)} \qquad\text{as $\delta \rightarrow 0$} ,
\end{align}
whereas Theorem~\ref{th:SDPI_gen_v} gives
\begin{align}\label{eq:dist_W_1:M_Tlbsdpi}
T(0,\delta) 
&\gtrsim \max_{\cS \subset \cV} \max_{v\in\cS} \frac{\log\big(|\cS^c|\log M\big) + \log h_2(\delta)^{-1}}{|\cE_v|\log(1-\eta^*_v)^{-1}} 
\end{align}
as $\delta \rightarrow 0$.
\end{corollary}
\noindent Again, we see that the lower bound obtained from SDPI is much sharper for capturing the dependence of $T(0,\delta)$ on $\delta$, since $\log h_2(\delta)^{-1} \to + \infty$ as $ \delta \to 0$. On the other hand, the lower bound obtained from the cutset capacity upper bound is tighter in its dependence on $M$, and can also capture the dependence on the conductance of the network.
}

Finally, we point out that Theorem~\ref{th:gen} gives the correct lower bound $T(\eps,\delta) = + \infty$ when the network graph $G$ is disconnected (assuming $f$ depends on the observations of all nodes): If $\cV$ consists of two disconnected components $\cS$ and $\cS^c$, then $C_\cS = 0$, which results in $T(\eps,\delta) = + \infty$. 
{  Despite the sharp dependence of the lower bounds of Theorems~\ref{th:gen} and \ref{th:SDPI_gen_v} on $\eps$ and $\delta$, they have the same limitation as all previously known bounds obtained via single-cutset arguments: they examine only the flow of information across a cutset $\cE_\cS$, but not within $\cS$; hence they cannot capture the dependence of computation time on the diameter of the network.} We address this limitation in the following section.

\section{Multi-cutset analysis}\label{sec:multi_cutset}
We now extend the techniques of Section~\ref{sec:single_cutset} to a multi-cutset analysis, to address the limitation of the results obtained from the single-cutset analysis. 
In particular, the new results are able to quantify the dissipation of information as it flows across a succession of cutsets in the network.
As briefly sketched in Sec.~\ref{ssec:summary}, we accomplish this by partitioning a general network using multiple disjoint cutsets, such that the operation of any algorithm on the network can be simulated by another algorithm running on a chain of bidirectional noisy links. We then derive tight mutual information upper bounds for such chains, which in turn can be used to lower-bound the computation time for the original network.

\subsection{Network reduction}\label{sec:network_red}

\begin{figure}[h!]
\centering
\subfloat[]{
\includegraphics[scale = 1.12]{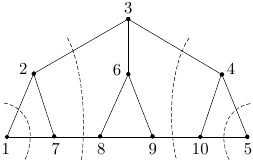}
\label{fg:red_gen1}
}
\quad\quad\quad
\subfloat[]{
\includegraphics[scale = 1.12]{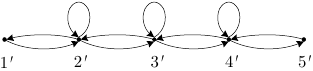}
\label{fg:bichain5_slp}
}
\caption{A successive partition of a network and the chain reduced according to it.}
\label{}
\end{figure}

Consider an arbitrary network $G = (\cV,\cE)$. If there exists a collection of nested subsets $\cP_1 \subset \ldots \subset \cP_{n-1}$ of $\cV$, such that the associated cutsets $\cE_{\cP_1}, \ldots, \cE_{\cP_{n-1}}$ are disjoint, and the cutsets $\cE_{\cP_1^c}, \ldots, \cE_{\cP_{n-1}^c}$ are also disjoint, then we say that $G$ is \textit{successively partitioned} according to $\cP_1,\ldots,\cP_{n-1}$ into $n$ subsets $\cS_1,\ldots,\cS_n$, where $\cS_i = \cP_{i} \setminus \cP_{i-1}$, with $\cP_0 \deq \varnothing$ and $\cP_n \deq \cV$.
For $i\in\{2,\ldots,n\}$, a node in $\cS_i$ is called a left-bound node of $\cS_i$ if there is an edge from it to a node in $\cS_{i-1}$. The set of left-bound nodes of $\cS_i$ is denoted by $\lepartial \cS_i$. For $\cS_1$, define $\lepartial\cS_1 = \{v\}$ for an arbitrary $v \in \cS_1$.
In addition, for $i\in\{2,\ldots,n\}$, let 
\begin{align}\label{eq:def_di}
d_i \deq |\cE_{\cP_{i-1}^c}| + |\cE_{\cP_{i}}| + |\{ \cE \cap (\cS_i \times \lepartial\cS_i)\}| 
\end{align}
be the number of edges entering $\cS_i$ from its neighbors $\cS_{i-1}$ and $\cS_{i+1}$, plus the number of edges entering $\lepartial \cS_i$ from $\cS_i$ itself. 
For example, Fig.~\ref{fg:multi_cutset} in Sec.~\ref{ssec:summary} illustrates a successive partition of a six-node network into three subsets $\cS_1 = \{1,4\}$, $\cS_2 = \{2,5\}$ and $\cS_3 = \{3,6\}$, with $\lepartial\cS_1 = \{4\}$, $\lepartial\cS_2 = \{2\}$ and $\lepartial\cS_3 = \{3,6\}$. In addition, $d_2 = 5$ and $d_3 = 4$.
As another example, the network in Fig.~\ref{fg:red_gen1}, where each undirected edge represents a pair of channels with opposite directions, can be successively partitioned into $\cS_1 = \{1\}$, $\cS_2 = \{2,7\}$, $\cS_3 = \{3,6,8,9\}$, $\cS_4 = \{4,10\}$, and $\cS_5 = \{5\}$, with $\lepartial\cS_1 = \{1\}$, $\lepartial\cS_1 = \{2,7\}$, $\lepartial\cS_3 = \{3,8\}$, $\lepartial\cS_4 = \{4,10\}$, and $\lepartial\cS_i = \{5\}$.
In addition, $d_2 = 6$, $d_3 = 7$, $d_4 = 6$, and $d_5 = 2$.

\begin{figure}[h!]
\centering
\subfloat[]{
\includegraphics[scale = 1.12]{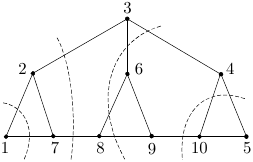}
\label{fg:gen_net_dm}
}
\quad\quad\quad
\subfloat[]{
\includegraphics[scale = 1.12]{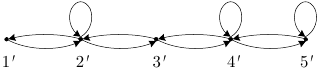}
\label{fg:bichain5_slp2}
}
\caption{Another successive partition (using the construction in the proof of Lemma~\ref{lm:net_red_diameter}) and the chain reduced according to it.}
\label{}
\end{figure}

Formally, a network $G$ has \textit{bidirectional links} if, for any pair of nodes $u,v \in \cV$, $(u,v) \in \cE$ if and only if $(v,u) \in \cE$.  A \textit{path} between $u$ and $v$ is a sequence of edges $\{(v_i,v_{i+1})\}^{k-1}_{i=1}$, such that $v_1 = u$ and $v_k = v$ (if $G$ is connected, there is at least one path between any pair of nodes). The \textit{graph distance} between $u$ and $v$, denoted by $d_G(u,v)$, is the length of a shortest path between $u$ and $v$ (shortest paths are not necessarily unique). The \textit{diameter} of $G$ is then defined by
$$
{\rm diam}(G) \deq \max_{u \in \cV}\max_{v \in \cV} d_G(u,v).
$$
The following lemma states that any such network $G$ can be successively partitioned into $n = {\rm diam}(G)+1$ subsets:

\begin{lemma}\label{lm:net_red_diameter}
Any network $G = (\cV,\cE)$ with bidirectional links (i.e., $(u,v) \in \cE$ if and only if $(v,u) \in \cE$) admits a successive partition  into subsets $\cS_1,\ldots,\cS_n$ with $n = {\rm diam}(G)+1$.
\end{lemma}
\begin{IEEEproof}For any $v \in \cV$ and any $r \in \{0 : {\rm diam}(G)\}$, we define the sets
\begin{align*}
	\BB_G(v,r) \deq \left\{ u \in \cV: d_G(v,u) \le r \right\}
\end{align*}
and
\begin{align*}
	\SS_G(v,r) \deq \left\{u \in \cV: d_G(v,u) = r \right\},
\end{align*}
i.e., the ball and the sphere of radius $r$ centered at $v$. In particular, $\BB_G(v,r) = \BB_G(v,r-1) \cup \SS_G(v,r)$.

We now construct the desired successive partition. Let $n = {\rm diam}(G) + 1$, and pick any pair of nodes $v_0,v_1 \in \cV$ that achieve the maximum in the definition of ${\rm diam}(G)$. With this, we take
$$
\cP_i = \BB_G(v_0,i-1), \qquad i = 1,\ldots,n.
$$
Clearly, $\cP_1 = \{v_0\}\subset \cP_2 \subset \ldots \subset \cP_n = \cV$, and moreover
$$
\cS_i = \SS_G(v_0,i-1), \qquad i = 1,\ldots,n.
$$
From this construction, we see that
$$
\cE_{\cP_i} = \left\{ (u,v) \in \cE: u \in \cS_{i+1},\, v \in \cS_{i} \right\}
$$
and
$$
\cE_{\cP^c_i} = \left\{ (u,v) \in \cV : u \in \cS_{i},\, v \in \cS_{i+1}\right\}.
$$
The pairwise disjointness of the cutsets $\cE_{\cP_i}$, as well as of the cutsets $\cE_{\cP^c_i}$, is immediate.
\end{IEEEproof}
\noindent{\bf Remarks}:
\begin{itemize}
\item
Using the construction underlying the proof, we can also show that, for any two nodes in $G$, we can successively partition $G$ into $n = d_G(u,v)+1$ subsets.
\item
For the successive partition constructed in the proof, all nodes in $\cS_i$ are left-bound nodes, and $d_i$ is the sum of the in-degrees of the nodes in $\cS_i$.
\end{itemize}
As an example, Fig.~\ref{fg:gen_net_dm} shows the successive partition of the network in Fig.~\ref{fg:red_gen1} using the construction in the proof, where $\cS_1 = \{1\}$, $\cS_2 = \{2,7\}$, $\cS_3 = \{3,8\}$, $\cS_4 = \{4,6,9\}$, $\cS_5 = \{5,10\}$, with $\lepartial\cS_i = \cS_i$, $i \in \{1,\ldots,5\}$, and $d_2 = 6$, $d_3 = 6$, $d_4 = 9$, and $d_5 = 5$.

The successive partition of $G$ ensures that nodes in $\cS_i$ only communicate with nodes in $\cS_{i-1}$ and $\cS_{i+1}$, as well as among themselves.
Indeed, suppose that the network graph $G$ includes an edge $e=(u,v) \in \cE$ with $u \in \cS_i$ and $v \in \cS_j$, where $i > j + 1$. By construction of the successive partition, $u \in \cP^c_{j+1} \subset \cP^c_{j}$ and $v \in \cP_j \subset \cP_{j+1}$. Therefore, $e$ belongs to both $\cE_{\cP_j}$ and $\cE_{\cP_{j+1}}$. However, the cutsets $\cE_{\cP_j}$ and $\cE_{\cP_{j+1}}$ are disjoint, so we arrive at a contradiction. Likewise, we can use the disjointness of the cutsets $\cE_{\cP^c_i}$ and $\cE_{\cP^c_j}$ to show that the network graph contains no edges $(u,v)$ with $u \in \cS_i$, $v \in \cS_j$, and $j > i + 1$.

In view of this, we can associate to the partition $\{\cS_i\}$ a \textit{bidirected chain} $G' = (\cV',\cE')$, i.e., a network with vertex set $\cV' = \{1',\ldots,n'\}$, edge set
\begin{align*}
\cE' = \big\{(i',(i-1)')\big\}_{i=2}^{n} \cup  \big\{(i',(i+1)')\big\}_{i=1}^{n-1} \cup \big\{(i',i')\big\}_{i=1}^{n},
\end{align*}
and channel transition laws
\begin{align}
K_{(i',(i-1)')} &= \bigotimes_{(u,v)\in\cE: u\in \cS_i,v\in \cS_{i-1}} K_{(u,v)} \label{eq:chain_Kii-1} \\
K_{(i',(i+1)')} &= \bigotimes_{(u,v)\in\cE: u\in \cS_i,v\in \cS_{i+1}} K_{(u,v)} \label{eq:chain_Kii+1} \\
K_{(i',i')} &= \bigotimes_{(u,v)\in\cE: u\in \cS_i,v\in \lepartial\cS_{i}} K_{(u,v)} ,\label{eq:chain_Kii} 
\end{align}
where node $i'$ in $G'$ observes
\begin{align*}
W_{i'} = W_{\cS_i} . 
\end{align*}
In other words, the subset $\cS_i$ in $G$ is reduced to node $i'$ in $G'$; the channels across the subsets in $G$ are reduced to the channels between the nodes in $G'$; and the channels from $\cS_i$ to $\lepartial \cS_i$ in $G$ are reduced to a self-loop at node $i'$ in $G'$.
The channels from $\cS_i$ to $\cS_i\setminus\lepartial\cS_i$ in $G$ are not included in $G'$, and will be simulated by node $i'$ using private randomness.
For the network in Fig.~\ref{fg:multi_cutset} in Sec.~\ref{ssec:summary}, according to the illustrated partition, it can be reduced to a $3$-node bidirected chain in Fig.~\ref{fg:multi_cutset_chain}, with $K_{(1',1')} = K_{(1,4)}$, $K_{(2',2')} = K_{(5,2)}$, and $K_{(3',3')} = K_{(3,6)} \otimes K_{(6,3)}$.
For the network in Fig.~\ref{fg:red_gen1}, according to the illustrated partition, it can be reduced to a $5$-node bidirected chain in Fig.~\ref{fg:bichain5_slp}, with $K_{(2',2')} = K_{(2,7)} \otimes K_{(7,2)}$, $K_{(3',3')} = K_{(6,3)} \otimes K_{(6,8)} \otimes K_{(9,8)}$, and $K_{(4',4')} = K_{(4,10)} \otimes K_{(10,4)}$.
According to the partition in Fig.~\ref{fg:gen_net_dm}, the same network can be reduced to a $5$-node bidirected chain in Fig.~\ref{fg:bichain5_slp2}, with $K_{(2',2')} = K_{(2,7)} \otimes K_{(7,2)}$, $K_{(4',4')} = K_{(6,9)} \otimes K_{(9,6)}$, and $K_{(5',5')} = K_{(5,10)} \otimes K_{(10,5)}$. 

For the bidirected chain $G'$ reduced from $G$, we consider a class of \textit{randomized} $T$-step  algorithms that run on $G'$ and are of a more general form compared to the deterministic algorithms considered so far. Such a randomized algorithm operates as follows: at step $t\in\{1,\ldots,T\}$, node $i'$ computes the outgoing messages $X_{(i',(i-1)'),t} = \lephi_{i',t}(W_{i'},Y^{t-1}_{i'})$, $X_{(i',(i+1)'),t} = \riphi_{i',t}(W_{i'},Y^{t-1}_{i'},U^{t-1}_{i'})$, and $X_{(i',i'),t} = \mathring\varphi_{i',t}(W_{i'},Y^{t-1}_{i'},U^{t-1}_{i'})$, and computes the private message $U_{i',t} = \vartheta_{i',t}(W_{i'},Y_{i'}^{t-1},U^{t-1}_{i'},R_{i',t})$, where $R_{i',t}$ is the private randomness held by node $i'$, uniformly distributed on $[0,1]$ and independent across $i' \in \cV'$ and $t\in\{1,\ldots,T\}$. 
At step $T$, node $i'$ computes the final estimate $\wh{Z}_{i'} = \psi_{i'}(W_{i'},Y^T_{i'})$ of $Z$. These randomized algorithms have the feature that the message sent to the node on the left and the final estimate of a node are computed solely based on the node's initial observation and received messages, whereas the messages sent to the node on the right and to itself are computed based on the node's initial observation, received messages, as well as private messages, and the computation of the private messages involves the node's private randomness.
Define 
\begin{align}\label{eq:comp_time_rdm}
T'(\eps,\delta) = \inf\Big\{ & T\in\N: \exists \text{ a randomized $T$-step algorithm }\cA'  \nonumber \\
& \text{ such that } \max_{i'\in \cV'} \PP\big[d(Z,\wh Z_{i'}) > \eps\big] \! \le \delta \Big\} 
\end{align}
as the $(\eps,\delta)$-computation time for $Z$ on $G'$ using the randomized algorithms described above.
The following lemma indicates that we can obtain lower bounds on $T(\eps,\delta)$ by lower-bounding $T'(\eps,\delta)$. 
\begin{lemma}\label{lm:net_red_AA'}
Consider an arbitrary network $G$ that can be successively partitioned into $\cS_1,\ldots,\cS_{n}$, such that $\lepartial \cS_i$'s are all nonempty. Let $G' = (\cV',\cE')$ be the bidirected chain constructed from $G$ according to the partition. Then, given any $T$-step algorithm on $G$ that achieves $\max_{v\in\cV} \PP[d(Z,\wh Z_v) > \eps]\le \delta$, we can construct a randomized $T$-step algorithm $\cA'$ on $G'$, such that $\max_{i'\in\cV'} \PP[d(Z,\wh Z_{i'}) > \eps]\le \delta$. 
Consequently, $T(\eps,\delta)$ for computing $Z$ on $G$ is lower bounded by $T'(\eps,\delta)$ defined in \eqref{eq:comp_time_rdm}.
\end{lemma}
\begin{IEEEproof}
Appendix~\ref{sec:pf:lm:net_red_AA'}.
\end{IEEEproof}
\noindent{\bf Remark}:
{  In the network reduction, we can alternatively map all the channels from $\cS_i$ to $\cS_i$ (instead of only mapping the channels from $\cS_i$ to $\lepartial\cS_{i}$) in the original network $G$ to the self-loop at node $i'$ of the reduced chain $G'$.
By doing so, to simulate the operation of an algorithm $\cA$ that runs on $G$, the algorithm $\cA'$ that runs on $G'$ no longer needs to generate private messages using the nodes' private randomness, since all the channels in $G$ are preserved in $G'$.
In other words, under this alternative reduction, any $T$-step algorithm $\cA$ that runs on $G$ can be simulated by a $T$-step algorithm $\cA'$ of the same deterministic type as $\cA$ that runs on $G'$.
However, this alternative reduction increases the information transmission capability of the self-loops in $G'$, and will result in a looser lower bound on $T(\eps,\delta)$, as will be discussed in the remark following Theorem~\ref{th:SDPI_red_gen}.
}

In light of Lemma~\ref{lm:net_red_AA'}, in order to lower-bound $T(\eps,\delta)$ for computing $Z$ on $G$, we just need to lower-bound $T'(\eps,\delta)$ defined in \eqref{eq:comp_time_rdm}. To this end, we derive upper bounds on the conditional mutual information for bidirected chains by extending the techniques behind Lemma~\ref{lm:mi_ub_cutset} and Lemma~\ref{lm:mi_ub_SDPI_gen}:
\begin{lemma}\label{lm:mi_ub_SDPI_chain_eta}
Consider an $n$-node bidirected chain with vertex set $\cV = \{1,\ldots,n\}$ and edge set
\begin{align*}
\cE = \big\{(i,i-1)\big\}_{i=2}^{n} \cup \big\{(i,i+1)\big\}_{i=1}^{n-1} \cup \big\{(i,i)\big\}_{i=1}^{n},
\end{align*}
and an arbitrary randomized $T$-step algorithm $\cA'$ that runs on this chain. 
Let $\eta_i \deq \eta(K_i)$ denote the SDPI constant of the channel
$
K_i \deq \bigotimes_{j:\, (j,i) \in \cE} K_{(j,i)} 
$,
and let $\eta \deq \max_{i=1,\ldots,n}\eta_i$.
If $T\le n-2$, then 
\begin{align*}
I(Z;\wh Z_n | W_{2:n}) = 0 .
\end{align*}
If $T \ge n-1$, then
\begin{flalign*}
 I(Z;\wh Z_n | W_{2:n}) \le  &&
\end{flalign*}
\begin{numcases}
{}
H(W_1|W_{2:n}) \eta \sum_{i = 1}^{T-n+2} \cB(T-i, n-2, \eta) , & \!\!\!\!\!\!\!\!\! $n \ge 2\qquad$ \label{eq:mi_ub_SDPI_chain_opt1} \\
C_{(1,2)} \eta \sum_{i = 1}^{T-n+2}
 \cB(T-i-1, n-3, \eta) i , & \!\!\!\!\!\!\!\!\! $n \ge 3\qquad$ \label{eq:mi_ub_SDPI_chain_opt2}
\end{numcases}
with $\cB(m,k,p) \deq {m \choose k} p^k (1-p)^{m-k}$.
For $n\ge 2$, the above upper bounds can be weakened to
\begin{flalign*}
\quad I(Z;\wh Z_n | W_{2:n}) \le &&
\end{flalign*}
\begin{numcases}
{}
 \!\!\! H(W_1|W_{2:n}) \big(1-(1-\eta)^{T-n+2}\big)^{n-1} , 
\label{eq:mi_ub_SDPI_chain_sub_a} \\
\!\!\! C_{(1,2)} (T-n+2) \big(1-(1-\tilde\eta)^{T-n+2}\big)^{n-2} .\label{eq:mi_ub_SDPI_chain_sub_b} 
\end{numcases}
Moreover, if $n\ge 4$ and 
$$
n-1 \le T \le 2 + \frac{(n-3)\gamma}{\eta} $$
for some $\gamma \in (0,1)$, then 
\begin{align}\label{eq:chain_mi_limit0}
&I(Z;\wh Z_n|W_{2:n}) \le \nonumber \\
&\qquad C_{(1,2)} \frac{(n-3)^2 \gamma^2}{\eta} \exp\left(-2\left(\frac{\eta}{\gamma}-\eta\right)^2 (n-3)\right) .
\end{align}
\end{lemma}
\begin{IEEEproof}
Appendix~\ref{sec:pf:lm:mi_ub_SDPI_chain_eta}.
\end{IEEEproof}
Equation~\eqref{eq:mi_ub_SDPI_chain_opt1} is reminiscent of a result of Rajagopalan and Schulman \cite{Rajagopalan94_dist_comp} on the evolution of mutual information in broadcasting a bit over a unidirectional chain of BSCs. The result in \cite{Rajagopalan94_dist_comp} is obtained by solving a system of recursive inequalities on the mutual information involving suboptimal SDPI constants.
Our results apply to chains of general bidirectional links and to the computation of general functions.
We arrive at a system of inequalities similar to the one in \cite{Rajagopalan94_dist_comp}, which can be solved in a similar manner and gives \eqref{eq:mi_ub_SDPI_chain_opt1} and \eqref{eq:mi_ub_SDPI_chain_opt2}.
We also obtain weakened upper bounds in \eqref{eq:mi_ub_SDPI_chain_sub_a} and \eqref{eq:mi_ub_SDPI_chain_sub_b}, which show that, for a fixed $T$, the conditional mutual information decays at least exponentially fast in $n$.
The upper bound in \eqref{eq:chain_mi_limit0} provides another weakening of \eqref{eq:mi_ub_SDPI_chain_opt1} and \eqref{eq:mi_ub_SDPI_chain_opt2}, and shows explicitly the dependence of the upper bound on $n$.

\begin{figure}[h!]
\centering
\includegraphics[scale = 0.85]{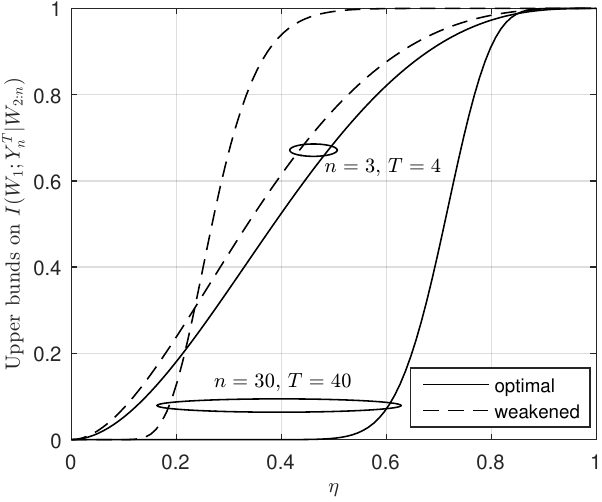}
  \caption{Upper bound in (\ref{eq:mi_ub_SDPI_chain_opt1}) (solid line) vs.\ the weakened one in (\ref{eq:mi_ub_SDPI_chain_sub_a}) (dashed line) for chains.}
  \label{fg:chain_opt}
\end{figure}

Assuming for simplicity that $H(W_1|W_{2:n})=1$, Fig.~\ref{fg:chain_opt} compares (\ref{eq:mi_ub_SDPI_chain_opt1}) with the weakened upper bound in (\ref{eq:mi_ub_SDPI_chain_sub_a}). We can see that the gap can be large when $n$ is large and $T$ is much larger than $n$.
Nevertheless, the weakened upper bounds in \eqref{eq:mi_ub_SDPI_chain_sub_a} and \eqref{eq:mi_ub_SDPI_chain_sub_b} allow us to derive lower bounds on computation time that are non-asymptotic in $n$, and explicit in $\eps$, $\delta$, and channel properties.

\subsection{Lower bounds on computation time}\label{sec:multi_cut_lb}
We now build on the results presented above to obtain lower bounds on the $T(\eps,\delta)$ by reducing the original problem to function computation over bidirected chains. We first provide the result for an arbitrary network, and then particularize it to several specific topologies (namely, chains, rings, grids, and trees).

\subsubsection{Lower bound for an arbitrary network}

Theorem~\ref{th:SDPI_red_gen} below contains general lower bounds on computation time for an arbitrary network. The statement of the theorem is somewhat lengthy, but can be parsed as follows: Given an arbitrary connected network with bidirectional links, any reduction of that network to a bidirected chain gives rise to a system of inequalities that must be satisfied by the computation time $T(\eps,\delta)$. These inequalities, presented in \eqref{eq:T(e,d)_gen_opt}, are nonasymptotic in nature and involve explicitly computable parameters of the network, but cannot be solved in closed form. The first inequality follows from an SDPI-based analysis analogous to Theorem~\ref{th:SDPI_gen_v}, while the second inequality is a cutset bound in the spirit of Theorem~\ref{th:gen}. Explicit but weaker expressions that lower-bound $T(\eps,\delta)$ in terms of network parameters appear below as \eqref{eq:T_multi_gen_SDPI} and \eqref{eq:T_multi_gen_CS}, together with asymptotic expressions for large $n$ (the size of the reduced bidirected chain). Both of these bounds state that $T(\eps,\delta)$ is lower-bounded by the size of the bidirected chain plus a correction term that accounts for the effect of channel noise (via channel capacities and SDPI constants). Finally, \eqref{eq:large_partition} and \eqref{eq:gen_Evans_Schulman} provide the precise version of the bound in \eqref{eq:diam_bound}: asymptotically, the computation time $T(\eps,\delta)$ scales as $\Omega(n/\tilde{\eta})$, where $\tilde{\eta}$ is the worst-case SDPI constant of the reduced network. By Lemma~\ref{lm:net_red_diameter}, it is always possible to reduce the network to a bidirected chain of length ${\rm diam}(G)+1$, so the main message of Theorem~\ref{th:SDPI_red_gen} is that the computation time $T(\eps,\delta)$ scales at least linearly in the network diameter. Thus, the main advantage of the multi-cutset analysis over the usual single-cutset analysis is that it can capture this dependence on the network diameter.

\begin{theorem}\label{th:SDPI_red_gen}
Assume the following:
\begin{itemize}
\item The network graph $G = (\cV,\cE)$ is connected, the capacities of all edge links are upper-bounded by $C$, and the SDPI constants of edge links are upper-bounded by $\eta$.
\item $G$ admits a successive partition into $\cS_1,\ldots,\cS_{n}$, such that $\lepartial \cS_i$'s are all nonempty.
\end{itemize}
Let
$$
\Delta \deq \max_{i \in \{2:n\}} d_i
$$
where 
$$
d_i = |\cE_{\cP_{i-1}^c}| + |\cE_{\cP_{i}}| + |\{ \cE \cap (\cS_i \times \lepartial\cS_i)\}| 
$$
as defined in \eqref{eq:def_di}, and let
$$
\tilde\eta = 1-(1-\eta)^\Delta .
$$
Then for $\eps\ge 0$ and $\delta\in(0,1/2]$, the $(\eps,\delta)$-computation time $T(\eps,\delta)$ must satisfy the inequalities
\begin{align}\label{eq:T(e,d)_gen_opt}
& \ell (\cS_1^c,\eps,\delta) \le \nonumber \\
&\begin{cases}
H(W_{\cS_1}|W_{\cS_1^c}) \tilde\eta \displaystyle\sum\limits_{i = 1}^{T(\eps,\delta)-n+2} \cB(T(\eps,\delta)-i, n-2, \tilde\eta) , & \!\! n \ge 2 \\
C_{\cS_1^c} \tilde\eta \displaystyle\sum\limits_{i = 1}^{T(\eps,\delta)-n+2}
 \cB(T(\eps,\delta)-i-1, n-3, \tilde\eta) i , & \!\! n \ge 3 .
\end{cases} 
\end{align}
The above results can be weakened to
\begin{align}
T(\eps,\delta) 
&\ge \frac{ \log\left(1-\big(\frac{\ell(\cS_1^c,\eps,\delta)}{H(W_{\cS_1}|W_{\cS_1^c})}\big)^{\frac{1}{n-1}}\right)^{-1}}{\Delta\log(1-\eta)^{-1}} + n-2 \label{eq:T_multi_gen_SDPI} \\
&\sim  \frac{\log (n-1) + \log\big(1-\frac{\ell\left(\cS_1^c,\eps,\delta\right)}{H(W_{\cS_1}|W_{\cS_1^c})}\big)^{-1}}{\Delta \log(1-\eta)^{-1}} + n-2  , \nonumber
\end{align}
as $n \rightarrow \infty$, and
\begin{align}\label{eq:T_multi_gen_CS}
T(\eps,\delta) \ge  \frac{\ell\left(\cS_1^c,\eps,\delta\right)}{C_{\cS_1^c}} + n-2 .
\end{align}
Moreover, if the partition size $n$ is large enough, so that $n \ge 4$ and
\begin{align}\label{eq:large_partition}
\frac{C|\cV|^2(n-3)^2}{4 \eta} \exp\left(- 2 \eta^2(n-3) \right) < \ell(\cS^c_1,\eps,\delta),
\end{align}
then 
\begin{align}\label{eq:gen_Evans_Schulman}
T(\eps,\delta) > 2 + \frac{n-3}{2\tilde\eta} 
\ge 2 + \frac{n-3}{2\Delta\eta} .
\end{align}
\end{theorem}
\begin{IEEEproof}
In light of Lemma~\ref{lm:net_red_AA'}, it suffices to show that the lower bounds in Theorem~\ref{th:SDPI_red_gen} need to be satisfied by $T'(\eps,\delta)$ for the bidirected chain $G'$, to which $G$ reduces according to the partition $\{\cS_i\}$. 

Consider any randomized $T$-step algorithm $\cA'$ that achieves $\max_{i'\in \cV'} \PP[d(Z,\wh Z_{i'}) >\eps] \le \delta$ on $G'$.
From Lemma~\ref{lm:mi_lb_gen}, 
$$
I(Z;\wh Z_{n'} | W_{2':n'}) \ge \ell(\{2':n'\},\eps,\delta) .
$$
Then from Lemma~\ref{lm:mi_ub_SDPI_chain_eta} and the fact that
\begin{align}
\eta_{i'} &= \eta(K_{((i-1)',i')} \otimes K_{((i+1)',i')} \otimes K_{i',i'}) \nonumber \\
&\le 1-(1-\eta)^{d_i} \nonumber \\
&\le 1-(1-\eta)^{\Delta} ,  \label{eq:def_tilde_eta}
\end{align}
we have 
\begin{align*}
& \ell(\{2':n'\},\eps,\delta) \le \\
&\begin{cases}
H(W_{1'}|W_{2':n'}) \tilde\eta \displaystyle\sum\limits_{i = 1}^{T-n+2} \cB(T-i, n-2, \tilde\eta) , &\quad n \ge 2 \\
C_{(1',2')} \tilde\eta \displaystyle\sum\limits_{i = 1}^{T-n+2}
 \cB(T-i-1, n-3, \tilde\eta) i , &\quad n \ge 3 ,
\end{cases} 
\end{align*}
and for $n \ge 2$,
\begin{align}\label{eq:chain_T'_sub}
& \ell(\{2':n'\},\eps,\delta) \le \nonumber \\
&\begin{cases}
H(W_{1'}|W_{2':n'}) \displaystyle\prod\limits_{i=2}^{n}\big(1-(1-\eta)^{d_i (T-n+2)} \big) \\
C_{(1',2')} (T-n+2) \displaystyle\prod\limits_{i=3}^{n}\big(1-(1-\eta)^{d_i (T-n+2)}\big) .
\end{cases} 
\end{align}
Since $\ell(\{2':n'\},\eps,\delta) = \ell(\cS_1^c,\eps,\delta)$, $H(W_{1'}|W_{2':n'}) = H(W_{\cS_1}|W_{\cS_1^c})$, and $C_{(1',2')} = C_{\cS_1^c}$, we see that $T'(\eps,\delta)$ must satisfy \eqref{eq:T(e,d)_gen_opt} in Theorem~\ref{th:SDPI_red_gen}.

Using \eqref{eq:def_tilde_eta}, \eqref{eq:chain_T'_sub} can be weakened to
\begin{align}\label{eq:chain_T'_sub_n}
& \ell(\cS_1^c,\eps,\delta) \le \nonumber \\
&\begin{cases}
H(W_{\cS_1}|W_{\cS_1^c}) \big(1-(1-\eta)^{\Delta(T-n+2)}\big)^{n-1}  \\
C_{\cS_1^c} (T-n+2) \big(1-(1-\eta)^{\Delta(T-n+2)}\big)^{n-2}
\end{cases} .
\end{align}
The first line of \eqref{eq:chain_T'_sub_n} leads to
\begin{align*}
T'(\eps,\delta) 
&\ge \frac{ \log\left(1-\big(\frac{\ell(\cS_1^c,\eps,\delta)}{H(W_{\cS_1}|W_{\cS_1^c})}\big)^{\frac{1}{n-1}}\right)^{-1}}{\Delta\log(1-\eta)^{-1}} + n-2 \\
&\sim  \frac{\log (n-1) + \log\big(1-\frac{\ell\left(\cS_1^c,\eps,\delta\right)}{H(W_{\cS_1}|W_{\cS_1^c})}\big)^{-1}}{\Delta \log(1-\eta)^{-1}} + n-2,
\end{align*}
where the last step follows from the fact that $\log\big(1-p^{\frac{1}{n}}\big)^{-1} \sim \log\frac{n}{1-p}$ as $n\rightarrow \infty$ for $p\in(0,1)$.
The second line of \eqref{eq:chain_T'_sub_n} leads to
\begin{align*}
T'(\eps,\delta) \ge  \frac{\ell\left(\cS_1^c,\eps,\delta\right)}{C_{\cS_1^c}} + n-2 .
\end{align*}

Finally, we prove that $T'(\eps,\delta) = \Omega(n/\tilde\eta)$ under the assumption that \eqref{eq:large_partition} holds. Suppose that $T'(\eps,\delta) \le 2 + (n-3)/2\tilde\eta$. Then, from \eqref{eq:chain_mi_limit0} in Lemma~\ref{lm:mi_ub_SDPI_chain_eta}, we have
\begin{align*}
\ell(\cS_1^c,\eps,\delta) \le C_{\cS_1^c} \frac{(n-3)^2}{4 \tilde\eta} \exp\left(-2 {\tilde\eta}^2 (n-3)\right) ,\quad\text{if $n\ge 4$} .
\end{align*}
Note that $\Delta \ge 1$ by the assumption that $G$ is connected, thus $\tilde\eta = 1-(1-\eta)^{\Delta} \ge \eta$. Moreover, $C_{\cS_1^c} \le C|\cE| \le C|\cV|^2$. As a result,
\begin{align*}
\ell(\cS_1^c,\eps,\delta) &\le \frac{C|\cV|^2(n-3)^2}{4 \eta} \exp\left(-2 \eta^2(n-3) \right) ,\quad\text{if $n\ge 4$,}
\end{align*}
which contradicts the assumption that \eqref{eq:large_partition} holds. Thus, 
$$
T'(\eps,\delta) > 2 + \frac{n-3}{2\tilde\eta}
\ge 2 + \frac{n-3}{2\Delta\eta}
$$

Theorem~\ref{th:SDPI_red_gen} then follows from Lemma~\ref{lm:net_red_AA'}.
\end{IEEEproof}
\noindent{\bf Remarks}:
\begin{itemize}
\item
We call a node in $\cS_i$ a \textit{boundary node} if there is an edge (either inward or outward) between it and a node in $\cS_{i-1}$ or $\cS_{i+1}$. 
Denote the set of boundary nodes of $\cS_i$ by $\partial \cS_i$.
The results in Theorem~\ref{th:SDPI_red_gen} can be weakened by replacing $d_i$ with 
\begin{align*}
\partial d_i = \sum_{v \in \partial \cS_i}|\cE_v| ,
\end{align*}
namely the summation of the in-degrees of boundary nodes of $\cS_i$, since $d_i\le \partial d_i$ for $i \in \{2,\ldots,n\}$.

\item
{  
As discussed in the remark following Lemma~\ref{lm:net_red_AA'}, an alternative network reduction is to map all the channels from $\cS_i$ to $\cS_i$ (instead of only mapping the channels from $\cS_i$ to $\lepartial\cS_{i}$) in the original network $G$ to the self-loop at node $i'$ of the reduced chain $G'$.
Using the same proof strategy with this alternative reduction, we can obtain lower bounds on $T(\eps,\delta)$ of the same form as the results in Theorem~\ref{th:SDPI_red_gen}, but with $d_i$'s replaced by
$$
\tilde d_i \deq |\cE_{\cP_{i-1}^c}| + |\cE_{\cP_{i}}| + |\{ \cE \cap (\cS_i \times \cS_i)\}| .
$$
Since $d_i\le \partial d_i \le \tilde d_i$ for $i \in \{2,\ldots,n\}$, the lower bounds on $T(\eps,\delta)$ obtained by this alternative network reduction are weaker than the results in Theorem~\ref{th:SDPI_red_gen}, and are even weaker than the results obtained by replacing $d_i$'s with $\partial d_i$s.
}

\item
Due to Lemma~\ref{lm:net_red_diameter}, for a network $G$ of bidirectional links, we can always find a successive partition of $G$ such that $n$ in Theorem~\ref{th:SDPI_red_gen} is equal to the ${\rm diam}(G)+1$. By contrast, the diameter cannot be captured in general by the theorems in Section~\ref{sec:single_cutset}.

\item
Choosing a successive partition of $G$ with $n=2$ is equivalent to choosing a single cutset. In that case, we see that \eqref{eq:T_multi_gen_CS} recovers Theorem~\ref{th:gen}, while
\eqref{eq:T_multi_gen_SDPI} recovers a weakened version of Theorem~\ref{th:SDPI_gen_v} (in \eqref{eq:T_multi_gen_SDPI}, $\Delta = d_2$ is at least the sum of the in-degrees of the left-bound nodes of $\cS_2$, while Theorem~\ref{th:SDPI_gen_v} involves the in-degree of only one node in $\cS_2$).
\end{itemize}
We now apply Theorem~\ref{th:SDPI_red_gen} to networks with specific topologies. We assume that nodes communicate via bidirectional links. Thus, any such network will be represented by an undirected graph, where each undirected edge represents a pair of channels with opposite directions.
\subsubsection{Chains}
For chains, the proof of Theorem~\ref{th:SDPI_red_gen} already contains lower bounds on $T'(\eps,\delta)$. These lower bounds apply to $T(\eps,\delta)$ as well, since the class of $T$-step algorithms on a chain is a subcollection of randomized $T$-step algorithms on the same chain. We thus have the following corollary.
\begin{corollary}\label{co:SDPI_chain} 
Consider an $n$-node bidirected chain without self-loops, where the SDPI constants of all channels are upper bounded by $\eta$. Then for $\eps\ge 0$ and $\delta\in(0,1/2]$, $T(\eps,\delta)$ must satisfy the inequalities in Theorem~\ref{th:SDPI_red_gen} with $\cS_1 = \{1\}$ and $d_i = 2$ for all $i\in\{1,\ldots,n\}$.
In particular, if all channels are ${\rm BSC}(p)$, then
\begin{align*}
T(\eps, & \delta) \ge \max\bigg\{ \frac{\ell\big(\cV\setminus\{1\},\eps,\delta\big)}{1-h_2(p)}, \\
& \frac{\log (n-1) + \log\big(1-\frac{\ell\big(\cV\setminus\{1\},\eps,\delta\big)}{H(W_{1}|W_{\cV\setminus\{1\}})}\big)^{-1}}{2 \log(4p\bar p)^{-1}} \bigg\} + n-2
\end{align*}
for all sufficiently large $n$.
\end{corollary}
\noindent Here and below, the estimates for a network of bidirectional BSCs are obtained using the bounds \eqref{eq:BSC_cutset} and \eqref{eq:BSC_SDPI}.

\subsubsection{Rings}
\begin{figure}[h!]
\centering
\subfloat[]{
\includegraphics[scale = 1.15]{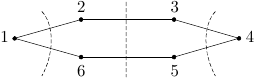}
\label{fg:ring6}
}
\quad\quad\quad
\subfloat[]{
\includegraphics[scale = 1.15]{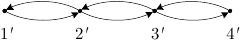}
\label{fg:bichain4}
}
\caption{A ring network and the chain reduced from it.}
\label{}
\end{figure}

Consider a ring with $2n-2$ nodes, where the nodes are labeled clockwise from $1$ to $2n-2$. The diameter is equal to $n-1$. According to the successive partition in the proof of Lemma~\ref{lm:net_red_diameter}, this ring can be partitioned into $\cS_1 = \{1\}$, $\cS_i = \{i,2n-i\}$, $i\in\{2,\ldots,n-1\}$, and $\cS_n = \{n\}$.
As an example, Fig.~\ref{fg:ring6} shows a $6$-node ring and Fig.~\ref{fg:bichain4} shows the chain reduced from it.
With this partition, we can apply Theorem~\ref{th:SDPI_red_gen} and get the following corollary.
\begin{corollary}\label{co:SDPI_ring} 
Consider a $(2n-2)$-node ring, where the SDPI constants of all channels are upper bounded by $\eta$. Then for $\eps\ge 0$ and $\delta\in(0,1/2]$, $T(\eps,\delta)$ must satisfy the inequalities in Theorem~\ref{th:SDPI_red_gen} with $\cS_1 = \{1\}$ and $d_i = 4$ for all $i\in\{1,\ldots,n\}$.
In particular, if all channels are ${\rm BSC}(p)$, then
\begin{align*}
T(\eps,  \delta)& =  \max\bigg\{ \frac{\ell\big(\cV\setminus\{1\},\eps,\delta\big)}{2(1-h_2(p))} ,  \\
& \frac{\log (n-1) + \log\big(1-\frac{\ell\left(\cV\setminus\{1\},\eps,\delta\right)}{H(W_{1}|W_{\cV\setminus\{1\}})}\big)^{-1}}{4 \log(4p\bar p)^{-1}} \bigg\} + n-2
\end{align*}
for all sufficiently large $n$.
\end{corollary}

\subsubsection{Grids}
\begin{figure}[h!]
\centering
\subfloat[]{
\includegraphics[scale = 1.15]{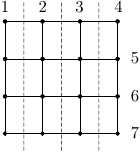}
\label{fg:grid_par1}
}
\quad\quad\quad
\subfloat[]{
\raisebox{1.2mm}
{\includegraphics[scale = 1.15]{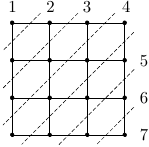}}
\label{fg:grid_par2}
}
\caption{Successive partitions of a $4\times 4$ ($n=7$) grid network.
The length of the labeled path is the diameter of the network.}
\label{}
\end{figure}

Consider an $\frac{n+1}{2} \times \frac{n+1}{2}$ grid (where we assume $n$ is odd), which has diameter $n-1$. Figure~\ref{fg:grid_par1} shows a successive partition of a $\frac{n+1}{2} \times \frac{n+1}{2}$ grid  into $\frac{n+1}{2}$ subsets, with $\Delta = \max_{i \in \{2:n\}} d_i = 2n$.
Figure~\ref{fg:grid_par2} shows the successive partition in the proof of Lemma~\ref{lm:net_red_diameter}, which partitions the network into $n$ subsets, with $\Delta = \max_{i \in \{2:n\}} d_i = 2(n-1)$, thus resulting in strictly tighter lower bounds on computation time compared to the ones obtained from the partition in Fig.~\ref{fg:grid_par1}.
With the latter partition, we get the following corollary.
\begin{corollary}\label{co:grid} 
Consider an $\frac{n+1}{2} \times \frac{n+1}{2}$ grid, where $1-\ldots-n$ is one of the longest paths. Assume that the SDPI constants of all channels are upper bounded by $\eta$. Then for $\eps\ge 0$ and $\delta\in(0,1/2]$, $T(\eps,\delta)$ must satisfy the inequalities in Theorem~\ref{th:SDPI_red_gen} with $\cS_1 = \{1\}$, $d_i = d_{n+1-i} = 4(i-2)+6$, $i\in\{1,\ldots,\frac{n-1}{2}\}$, and $d_{{(n+1)}/{2}} = 2(n-1)$.
In particular, if all channels are ${\rm BSC}(p)$, then
\begin{align*}
T(\eps, & \delta) \ge \max、\bigg\{ \frac{\ell\big(\cV\setminus\{1\},\eps,\delta\big)}{2(1-h_2(p))} ,\\
& \frac{\log (n-1) + \log\big(1-\frac{\ell\left(\cV\setminus\{1\},\eps,\delta\right)}{H(W_{1}|W_{\cV\setminus\{1\}})}\big)^{-1}}{2(n-1) \log(4p\bar p)^{-1}} \bigg\} + n-2
\end{align*}
for all sufficiently large $n$.
\end{corollary}

\subsubsection{Trees} Consider a tree, whose nodes are numbered in such a way that $1 - \ldots - n$ is one of the longest paths. Then the diameter of the tree is $n-1$, and nodes $1$ and $n$ are necessarily leaf nodes. 
The tree can be viewed as being rooted at node $1$. 
Let $\cD_i$ be the union of node $i$ and its descendants in the rooted tree, and
let $\cS_i = \cD_{i} \setminus \cD_{i+1}$, $i\in\{1,\ldots,n\}$.
The tree can then be successively partitioned into $\cS_1,\ldots,\cS_n$. In the $n$-node bidirected chain reduced according to this partition, the edges between nodes $i'$ and $(i+1)'$ are the pair of channels between nodes $i$ and $i+1$ in the tree, and the self-loop of node $i'$, $i\in\{2,\ldots,n-1\}$, is the channel from $\cS_i\setminus\{i\}$ to node $i$ in the tree.
\begin{figure}[h!]
\centering
\subfloat[]{
\includegraphics[scale = 1.15]{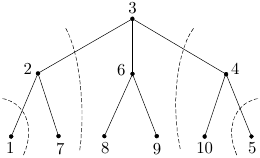}
\label{fg:tree10}
}
\quad\quad\quad
\subfloat[]{
\includegraphics[scale = 1.15]{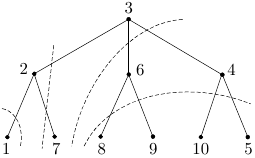}
\label{fg:tree10_par_diameter}
}
\caption{Successive partitions of a tree network.}
\label{}
\end{figure}
As an example, Fig.~\ref{fg:tree10} shows this partition of a tree network, and the chain reduced from it has the same form as the one in Fig.~\ref{fg:bichain5_slp}.
With this partition, we get the following corollary.
\begin{corollary}\label{co:SDPI_tree} 
Consider a $d$-regular tree network where $1-\ldots-n$ is one of the longest paths. Assume that the SDPI constants of all channels are upper bounded by $\eta$.
Then for $\eps\ge 0$ and $\delta\in(0,1/2]$, $T(\eps,\delta)$ must satisfy the inequalities in Theorem~\ref{th:SDPI_red_gen} with $\cS_1 = \{1\}$ and $d_i = d$ for all $i\in\{1,\ldots,n\}$.
In particular, if all channels are ${\rm BSC}(p)$, then
\begin{align*}
T(\eps,& \delta) \ge \max\bigg\{ \frac{\ell\big(\cV\setminus\{1\},\eps,\delta\big)}{1-h_2(p)}, \\
& \frac{\log (n-1) + \log\big(1-\frac{\ell\left(\cV\setminus\{1\},\eps,\delta\right)}{H(W_{1}|W_{\cV\setminus\{1\}})}\big)^{-1}}{d \log(4p\bar p)^{-1}} \bigg\} + n-2
\end{align*}
for all sufficiently large $n$.
\end{corollary}
If we use the successive partition in the proof of Lemma~\ref{lm:net_red_diameter} on a $d$-regular tree with diameter $n-1$, then the tree will be reduced to an $n$-node bidirected chain without self-loops. 
Figure~\ref{fg:tree10_par_diameter} shows such an example.
However, with this partition, $\Delta=\max_{i \in \{2:n\}}d_i$ increases with $n$, which renders the resulting lower bound on computation time looser than the one in Corollary~\ref{co:SDPI_tree}.
It means that, although the partition in the proof of Lemma~\ref{lm:net_red_diameter} always captures the diameter of a network, it may not always give the best lower bound on computation time among all possible successive partitions.

\section{Small ball probability estimates for computation of linear functions}\label{sec:comp_linear_fn}

The bounds stated in the preceding sections involve the conditional small ball probability, defined in \eqref{eq:def_cond_smb_prob}. In this section, we provide estimates for this quantity in the context of a distributed computation problem of wide interest --- the computation of linear functions.  
Specifically, we assume that the observations $W_v, v \in \cV$, are independent real-valued random variables, and the objective is to compute a linear function
\begin{align}\label{eq:lin_func}
Z =	f(W) = \sum_{v \in \cV} a_v W_v
\end{align}
for a fixed vector of coefficients $(a_v)_{v \in \cV} \in \R^{|\cV|}$, subject to the absolute error criterion $d(z,\wh{z}) = |z-\wh{z}|$. We will use the following shorthand notation: for any set $\cS \subset \cV$, let $a_\cS = (a_v)_{v \in \cS}$ and $\langle a_\cS, W_\cS \rangle = \sum_{v \in \cS}a_v W_v$.

The independence of the $W_v$'s and the additive structure of $f$ allow us to express the conditional small ball probability $L(w_\cS,\eps)$ defined in \eqref{eq:def_cond_smb_prob} in terms of so-called \textit{L\'evy concentration functions} of random sums \cite{Petrov_book}. The L\'evy concentration function of a real-valued r.v.\ $U$ (also known as the ``small ball probability") is defined as
\begin{align}
	\cL(U,\rho) = \sup_{u \in \R} \PP\left[ |U-u| \le \rho\right], \qquad \rho > 0.
\end{align}
If we fix a subset $\cS \subset \cV$, and consider a specific realization $W_\cS = w_\cS$ of the observations of the nodes in $\cS$, then
\begin{align}
	L(w_\cS,\eps) &= \sup_{z \in \R} \PP\Bigg[ \Big|\sum_{v \in \cV} a_v W_v - z\Big| \le \eps \Bigg|W_\cS = w_\cS \Bigg] \nonumber \\
	&= \sup_{z \in \R} \PP\left[ \Big|\sum_{v \in \cS^c}a_v W_v + \sum_{v \in \cS}a_v w_v - z\Big| \le \eps\right] \nonumber \\
	&= \sup_{z \in \R} \PP\Bigg[ \Big|\sum_{v \in \cS^c}a_v W_v - z \Big| \le \eps\Bigg] \nonumber \\
	&= \cL\left( \langle a_{\cS^c}, W_{\cS^c} \rangle, \eps\right), \label{eq:LwS_aScWSc}
\end{align}
where in the second line we have used the fact that the $W_v$'s are independent r.v.'s, while in the third line we have used the fact that for any function $g : \R \to \R$ and any $a \in \R$, $\sup_z g(z) = \sup_z g(z+a)$. In other words, for a fixed $\cS$, the quantity $L(w_\cS,\eps)$ is independent of the boundary condition $w_\cS$, and is controlled by the probability law of the random sum $\langle a_{\cS^c}, W_{\cS^c} \rangle$, i.e., the part of the function $f$ that depends on the observations of the nodes in $\cS^c$.

The problem of estimating L\'evy concentration functions of sums of independent random variables has a long history in the theory of probability --- for random variables with densities, some of the first results go back at least to Kolmogorov \cite{Kolmogorov_concent}, while for discrete random variables it is closely related to the so-called Littlewood--Offord problem \cite{Nguyen_Vu}. 
We provide a few examples to illustrate how one can exploit available estimates for L\'evy concentration functions under various regularity conditions to obtain tight lower bounds on the computation time for linear functions. 
The examples are illustrated through Theorem~\ref{th:gen}, as it tightly captures the dependence of computation time on $\ell(\cS,\eps,\delta)$. (However, since the results of Theorems~\ref{th:SDPI_gen_v} and \ref{th:SDPI_red_gen} also involve the quantity $\ell(\cS,\eps,\delta)$, the estimates for L\'evy concentration functions can be applied there as well.)

\subsection{Computing linear functions of continuous observations}
\subsubsection{Gaussian sums}
Suppose that the local observations $W_v$, $v \in \cV$, are i.i.d.\ standard Gaussian random variables. Then, for any $\cS \subseteq \cV$, $\langle a_{\cS}, W_{\cS}\rangle$ is a zero-mean Gaussian r.v.\ with variance $\| a_{\cS} \|^2_2 = \sum_{v \in \cS} a^2_v$ (here, $\| \cdot \|_2$ is the usual Euclidean $\ell_2$ norm). A simple calculation shows that
\begin{align*}
	L(w_\cS,\eps) &= \cL\left( N\left(0,\| a_{\cS^c} \|^2_2\right), \eps\right) 
	\le \sqrt{\frac{2}{\pi}} \frac{\eps}{\| a_{\cS^c} \|_2}.
\end{align*}
Using this in Theorem~\ref{th:gen}, we get the following result.
\begin{corollary}\label{co:Gauss_sum} For the problem of computing a linear function in \eqref{eq:lin_func}, where $(W_v) \overset{{\rm i.i.d.}}\sim N(0,1)$, suppose that the coefficients $a_v$ are all nonzero.
Then for $\eps\ge 0$ and $\delta\in(0,1/2]$, 
	\begin{align*}
		T(\eps,\delta) \ge \max_{\cS \subset \cV} \frac{1}{C_\cS} \left( \frac{1-\delta}{2}\log \frac{\pi \| a_{\cS^c} \|^2_2}{2\eps^2}-h_2(\delta)\right).
	\end{align*}
\end{corollary}
\noindent Thus, the lower bound on the computation time for \eqref{eq:lin_func} depends on the vector of coefficients $a$ only through its $\ell_2$ norm.

\subsubsection{Sums of independent r.v.'s with log-concave distributions}
Another instance in which sharp bounds on the L\'evy concentration function are available is when the observations of the nodes are independent random variables with log-concave distributions (we recall that a real-valued r.v.\ $U$ is said to have a log-concave distribution if it has a density of the form $p_U(u) = e^{-F(u)}$, where $F : \R \to (-\infty,+\infty]$ is a convex function; this includes Gaussian, Laplace, uniform, etc.). The following result was obtained recently by Bobkov and Chistyakov \cite[Theorem~1.1]{Bobkov_Chistyakov_concent}: Let $U_1,\ldots,U_k$ be independent random variables with log-concave distributions, and let $S_k = U_1 + \ldots + U_k$. Then, for any $\rho \ge 0$,
\begin{align}\label{eq:Bobkov_Chistyakov}
\!\!\!\!	\frac{1}{\sqrt{3}} \frac{\rho}{\sqrt{\Var(S_k)+{\rho^2}/{3}}} \le \cL(S_k,\rho) \le \!\frac{2\rho}{\sqrt{\Var(S_k)+{\rho^2}/{3}}}.
\end{align}
\begin{corollary}\label{co:log_concave}
For the problem of computing a linear function in \eqref{eq:lin_func}, where the $W_v$'s are independent random variables with log-concave distributions and with variances at least $\sigma^2$, suppose that the coefficients $a_v$ are all nonzero.
Then for $\eps\ge 0$ and $\delta\in(0,1/2]$, 
	\begin{align*}
\!	T(\eps,\delta) \ge \max_{\cS \subset \cV} \frac{1}{C_\cS} \left( \frac{1-\delta}{2}\log\left(\frac{\sigma^2\|a_{\cS^c}\|^2_2}{4\eps^2} + \frac{1}{12} \right)-h_2(\delta)\right).
	\end{align*}
\end{corollary}
\begin{IEEEproof} For each $v \in \cV$, $a_v W_v$ also has a log-concave distribution, and, for any $\cS \subset \cV$,
\begin{align*}
	\Var(\langle{a_{\cS^c}},W_{\cS^c}\rangle) = \sum_{v \in \cS^c} |a_v|^2 \Var(W_v) \ge \| a_{\cS^c} \|^2_2 \sigma^2.
\end{align*}
The lower bound follows from Theorem~\ref{th:gen} and from \eqref{eq:Bobkov_Chistyakov}.
\end{IEEEproof}

\subsubsection{Sums of independent r.v.'s with bounded third moments}
It is known that random variables with log-concave distributions have bounded moments of any order. Under a much weaker assumption that the local observations $W_v$, $v \in \cV$ have bounded third moments, we can prove the following result.
\begin{corollary} Consider the problem of computing the linear function in \eqref{eq:lin_func}, where the $W_v$'s are independent zero-mean r.v.'s with variances at least $1$ and with third moments bounded by $B$, and the coefficients $a_v$ satisfy the constraint $K_1 \le |a_v| \le K_2$ for some $K_1,K_2 > 0$.
Then for $\eps\ge 0$ and $\delta \in(0,1/2]$,
\begin{align*}
T(\eps,\delta)\ge \max_{\cS \subset \cV} \frac{1}{C_\cS} \Bigg( \frac{1-\delta}{2}\log \frac{|\cV\setminus\cS|}{M^2(\eps)}-h_2(\delta)\Bigg)
	\end{align*}
where $M(\eps) \deq c\big(\eps/K_1 + B(K_2/K_1)^3\big)$ with some absolute constant $c$.
\end{corollary}
\begin{IEEEproof} Under the conditions of the theorem, a small ball estimate due to Rudelson and Vershynin \cite[Corollary~2.10]{Rudelson_Vershynin_LO} can be used to show that, for any $\cS \subset \cV$,
	\begin{align*}
		\cL(\langle a_{\cS}, W_{\cS} \rangle,\eps) \le \frac{M(\eps)}{\sqrt{|\cS|}}.
	\end{align*}
	The desired conclusion follows immediately.
\end{IEEEproof}

\subsection{Linear vector-valued functions}
Similar to the L\'evy concentration function of a real-valued random variable, the L\'evy concentration function of a random vector $U$ taking values in $\R^n$ can be defined as
\begin{align*}
	\cL(U,\rho) = \sup_{u \in \R^n} \PP\left[ \| U-u \|_2 \le \rho\right], \qquad \rho > 0.
\end{align*}
Consider the case where each node observes an independent real-valued random variable $W_v$, and the observations form a $|\cV|\times 1$ vector $W_\cV$. Suppose the nodes wish to compute a linear transform of $W_\cV$,
\begin{align}
Z = AW_\cV \label{eq:lin_vec}
\end{align}
with some fixed $n \times |\cV|$ matrix $A$, subject to the Euclidean-norm distortion criterion $d(z,\wh z) = \|z-\wh z\|_2$. In this case
\begin{align*}
L(w_\cS,\eps) 
&= \sup_{z\in \R^n}\PP[\|AW_\cV - z\|_2 \le \eps |W_\cS = w_\cS] \\
&= \sup_{z\in \R^n}\PP[\|A_{\cS^c}W_{\cS^c} + A_{\cS}w_\cS - z\|_2 \le \eps ] \\
&= \sup_{z\in \R^n}\PP[\|A_{\cS^c}W_{\cS^c} - z\|_2 \le \eps ] \\
&= \cL(A_{\cS^c}W_{\cS^c},\eps) 
\end{align*}
where $A_{\cS^c}$ is the submatrix formed by the columns of $A$ with indices in $\cS^c$. We will need the following result, due to Rudelson and Vershynin \cite{Rudelson_Vershynin_vec}.
Let $s_j(A_{\cS^c})$, $j = 1,\ldots, \min\{n,|\cS^c|\}$, denote the singular values of $A_{\cS^c}$ arranged in non-increasing order, and define the stable rank of $A_{\cS^c}$ by
\begin{align*}
r(A_{\cS^c}) = \Bigg\lfloor \frac{\|A_{\cS^c}\|_{\rm HS}^2}{\|A_{\cS^c}\|^2} \Bigg\rfloor 
\end{align*}
where $\| A_{\cS^c} \|_{\rm HS} = \big(\sum_{j=1}^{\min\{n,|\cS^c|\}}s_j(A_{\cS^c})^2\big)^{1/2} $ is the Hilbert-Schmidt norm of $A_{\cS^c}$, and $\|A_{\cS^c}\| = s_1(A_{\cS^c})$ is the spectral norm of $A_{\cS^c}$. (Note that for any nonzero matrix $A_{\cS^c}$, $1 \le r(A_{\cS^c}) \le {\rm rank}(A_{\cS^c})$.) Then, provided
\begin{align*}
\cL(W_v,\eps/\|A_{\cS^c}\|_{\rm HS}) \le p
\end{align*}
for all $v\in \cS^c$, we will have
\begin{align*}
\cL(A_{\cS^c}W_{\cS^c},\eps) \le (c p)^{0.9r(A_{\cS^c})}
\end{align*}
where $c$ is an absolute constant \cite[Theorem 1.4]{Rudelson_Vershynin_vec}. This result relates the L\'evy concentration function of the linear transform of a vector to the L\'evy concentration function of each coordinate of the vector.
Applying this result in Theorem~\ref{th:gen}, we get a lower bound on $T(\eps,\delta)$ for computing linear vector-valued functions.
\begin{corollary}\label{co:lin_vec}
For the problem of computing a linear transform of the observations defined in (\ref{eq:lin_vec}), where $W_v$'s are independent real-valued r.v.s, suppose the rows of $A$ are nonzero vectors.
Then for $\eps\ge 0$ and $\delta \in(0,1/2]$,
\begin{align*}
T(\eps, \delta)\ge & \max_{\cS \subset \cV} \frac{1}{C_\cS} \Bigg( 0.9(1-\delta)r(A_{\cS^c}) \\
& \log \frac{1}{c \max_{v \in \cS^c}\cL(W_v,\eps/\|A_{\cS^c}\|_{\rm HS})}-h_2(\delta)\Bigg)
\end{align*}
for some absolute constant $c$.
\end{corollary}

\subsection{Linear function of discrete observations}

Finally, we consider a case when the local observations $W_v$ have discrete distributions. Specifically, let the $W_v$'s be i.i.d.\ Rademacher random variables, i.e., each $W_v$ takes values $\pm 1$ with equal probability. 
We still use the absolute distortion function $d(z,\wh z) = |z - \wh z|$ to quantify the estimation error.
In this case, the L\'evy concentration function $\cL(\langle a_\cS, W_\cS \rangle, \eps)$ will be highly sensitive to the \textit{direction} of the vector $a_{\cS}$, rather than just its norm. For example, consider the extreme case when $a_v = |\cV|$ for a single node $v \in \cS$, and all other coefficients are zero. Then $\cL(\langle a_\cS, W_\cS \rangle, 0) = \cL(|\cV|W_v, 0) = 1/2$. On the other hand, if $a_v = 1$ for all $v \in \cV$ and $|\cS|$ is even, then
\begin{align*}
	\cL(\langle a_{\cS},W_{\cS}\rangle, 0) = 2^{-|\cS|}{|\cS| \choose |\cS|/2} \sim \sqrt{\frac{2}{\pi |\cS|}} \quad\text{as $|\cS| \rightarrow \infty$}
\end{align*}
where the last step is due to Stirling's approximation. 
Moreover, a celebrated result due to Littlewood and Offord, improved later by Erd\H{o}s \cite{Erdos_LO}, says that, if $|a_v| \ge 1$ for all $v$, then
\begin{align*}
	\cL(\langle a_{\cS}, W_{\cS} \rangle, 1) \le 2^{-|\cS|} {|\cS| \choose \lfloor |\cS|/2 \rfloor} \sim \sqrt{\frac{2}{\pi |\cS|}} \quad\text{as $|\cS| \rightarrow \infty$,}
\end{align*}
which translates into a lower bound on the $(1,\delta)$-computation time which is of the same order as the lower bound on the zero-error computation time.
\begin{corollary}\label{co:Rademacher_sum}
For the problem of computing the linear function in \eqref{eq:lin_func}, where the $W_v$'s are independent Rademacher random variables, suppose that $|a_v| \ge 1$ for all $v$, and $\delta < {1}/{2}$.
Then 
	\begin{align*}
T(0,\delta) \ge T(1,\delta) \gtrsim \max_{\cS \subset \cV}\frac{1}{C_\cS}\left(\frac{1-\delta}{2}\log\frac{\pi|\cV\setminus\cS|}{2} - h_2(\delta)\right)  
	\end{align*}
as $|\cS| \rightarrow \infty$.
\end{corollary}

\subsection{Comparison with existing results}\label{sec:compare_exist_lb}
We illustrate the utility of the above bounds through comparison with some existing results. For example, Ayaso {et al.}~\cite{Ayaso_etal} derive lower bounds on a related quantity
\begin{align*}
	\tilde T(\eps,\delta) \deq \inf \Big\{ T & \in \N:  \exists \text{ a $T$-step algorithm }\cA  \text{ such that } \\
	& \max_{v \in \cV} \PP\big[ \wh{Z}_v \notin \left[(1-\eps)Z, (1+\eps)Z\right]\big] < \delta \Big\}.
\end{align*}
One of their results is as follows: if $Z = f(W)$ is a linear function of the form \eqref{eq:lin_func} and $\left(W_v\right) \overset{{\rm i.i.d.}}\sim \text{Uniform}([1,1+B])$ for some $B > 0$, then
\begin{align}\label{eq:lb_lin_Ayaso_etal}
	\tilde T(\eps,\delta) \ge \max_{\cS \subset \cV} \frac{|\cS|}{2C_\cS}\log \frac{1}{B\eps^2 + \kappa\delta + (1/B)^{2/|\cV|}}
\end{align}
for all sufficiently small $\eps,\delta > 0$, where $\kappa > 0$ is a fixed constant \cite[Theorem~III.5]{Ayaso_etal}. Let us compare \eqref{eq:lb_lin_Ayaso_etal} with what we can obtain using our techniques. It is not hard to show that
\begin{align}\label{eq:rel_vs_abs}
	\tilde T(\eps,\delta) \ge T\big( \| a \|_1 (1+B)\eps,\delta\big)
\end{align}
where $\|a\|_1 = \sum_{v \in \cV}|a_v|$ is the $\ell_1$ norm of $a$. Moreover, since any r.v.\ uniformly distributed on a bounded interval of the real line has a log-concave distribution, we can use Corollary~\ref{co:log_concave} to lower-bound the right-hand side of \eqref{eq:rel_vs_abs}. This gives
\begin{align}\label{eq:rel_log_concave}
\tilde T(\eps,\delta) 	\ge \max_{\cS \subset \cV} \frac{1}{C_\cS} \left( \frac{1-\delta}{2}\log \frac{B^2\| a_{\cS^c} \|^2_2}{48(B+1)^2 \|a\|^2_1 \eps^2} - h_2(\delta)\right)
\end{align}
for all sufficiently small $\eps,\delta > 0$. We immediately see that this bound is tighter than the one in \eqref{eq:lb_lin_Ayaso_etal}. In particular, the right-hand side of \eqref{eq:lb_lin_Ayaso_etal} remains bounded for vanishingly small $\eps$ and $\delta$, and in the limit of $\eps, \delta \to 0$ tends to
\begin{align*}
	\max_{\cS \subset \cV} \frac{|\cS|}{C_\cS} \frac{\log B}{|\cV|} \le \frac{\log B}{\min_{\cS \subset \cV} C_\cS}.
\end{align*}
By contrast, as $\eps, \delta \to 0$,  the right-hand side of \eqref{eq:rel_log_concave} grows without bound as $\log(1/\eps)$.

Another lower bound on the $(\eps,\delta)$-computation time $T(\eps,\delta)$ was obtained by Como and Dahleh \cite{Como_Dahleh}. Their starting point is the following continuum generalization of Fano's inequality \cite[Lemma~2]{Como_Dahleh} in terms of conditional differential entropy: if $Z, \wh{Z}$ are two jointly distributed real-valued r.v.'s, such that $\E Z^2 < \infty$, then, for any $\eps > 0$,
\begin{align}\label{eq:cond_h_Como_Dahleh}
h(Z|\wh{Z}) \le \PP\big[|Z-\wh{Z}| \le \eps\big]\log\eps + \frac{1}{2}\log\big(16\pi e\E Z^2\big) .
\end{align}
If we use \eqref{eq:cond_h_Como_Dahleh} instead of Lemma~\ref{lm:mi_lb_gen} to lower-bound $I(Z;\wh Z_v|W_\cS)$, then we get
\begin{align}\label{eq:lb_gen_Como_Dahleh}
T(\eps,\delta) \ge \max_{\cS \subset \cV} \frac{1}{C_\cS} \Bigg( & \frac{1-\delta}{2}\log\frac{1}{\eps^2} + h(Z|W_\cS) \nonumber \\
&- \frac{1}{2}\log\big(16\pi e\E Z^2\big) \Bigg) .
\end{align}
Again, let us consider the case when $Z = f(W)$ is a linear function of the form \eqref{eq:lin_func} with all $a_v$ nonzero and with $(W_v) \overset{{\rm i.i.d.}}\sim N(0,1)$. Then \eqref{eq:lb_gen_Como_Dahleh} becomes
\begin{align}\label{eq:lb_Gauss_Como_Dahleh}
	T(\eps,\delta) \ge \max_{\cS \subset \cV} \frac{1}{C_\cS} \left(\frac{1-\delta}{2}\log \frac{1}{\eps^2} + \frac{1}{2}\log \frac{\| a_{\cS^c} \|^2_2}{8\|a\|^2_2}\right). 
\end{align}
The lower bound of our Corollary~\ref{co:Gauss_sum} will be tighter than \eqref{eq:lb_Gauss_Como_Dahleh} for all $\eps >0$ as long as
\begin{align*}
	\frac{1-\delta}{2} \log\frac{\pi \|a_{\cS^c} \|^2_2}{2}-h_2(\delta) \ge \frac{1}{2}\log \frac{\|a_{\cS^c}\|^2_2}{8 \| a \|^2_2}, \quad \forall \cS \subset \cV.
\end{align*}
Note that the quantity on the right-hand side is nonpositive. 
More generally, for observations with log-concave distributions, the result of Lemma~\ref{lm:mi_lb_gen} can be weakened to get a lower bound involving the conditional differential entropy $h(Z|W_\cS)$, which is tighter than similar results obtained in~\cite{Como_Dahleh}.
\begin{corollary}\label{co:sum_logcon_dfen}
If the observations $W_v$, $v\in\cV$, have log-concave distributions, then for computing the sum $Z = \sum_{v \in \cV} W_v$ subject to the absolute error criterion $d(z,\wh{z}) = |z-\wh{z}|$, for $\eps\ge 0$ and $\delta \in(0,1/2]$,
\begin{align*}
T(\eps,\delta) \ge \max_{\cS \subset \cV} \frac{1}{C_\cS} \left( \! (1-\delta)\!\left(\!h(Z|W_\cS) + \log \frac{1}{2e\eps}\! \right)-h_2(\delta)\!\right).
\end{align*}
\end{corollary}
\begin{IEEEproof}
Let $p_\cS(z)$ denote the probability density of $\sum_{v \in \cS^c}W_v$. Then from \eqref{eq:LwS_aScWSc},
	\begin{align}\label{eq:Levy_vs_max_density}
		L(w_\cS,\eps) = \sup_{z \in \R}\int^{z+\eps}_{z-\eps}p_\cS(z){\rm d} z \le 2\eps \| p_\cS \|_\infty 
\end{align}
for all $w_\cS \in \prod_{v\in\cS}\sW_v$, where $\| p_\cS \|_\infty$ is the sup norm of $p_\cS$. By a result of Bobkov and Madiman \cite[Proposition~I.2]{Bobkov_Madiman_logconcave}, if $U$ is a real-valued r.v.\ with a log-concave density $p$, then the differential entropy $h(U)$ is upper-bounded by $\log e + \log \| p \|^{-1}_\infty$. Using this fact together with \eqref{eq:Levy_vs_max_density}, the log-concavity of $p_\cS$, and the fact that the $W_v$'s are mutually independent, we can write
\begin{align*}
	\log \frac{1}{\E[L(W_\cS,\eps)]} &\ge \log \frac{1}{2\eps} + \log \frac{1}{\| p_\cS \|_\infty} \\
	&\ge  \log \frac{1}{2e\eps} + h\Big(\sum_{v \in \cS^c}W_v\Big) \\
	&= \log \frac{1}{2e\eps} + h(Z|W_\cS),
\end{align*}
Using this estimate in Theorem~\ref{th:gen}, we get the desired lower bound on $T(\eps,\delta)$.
\end{IEEEproof}

\section{Comparison with upper bounds on computation time}\label{sec:compare_ub}
For the two-node mod-$2$ sum problem in Example~\ref{ex:2node_mod2+}, we have shown in Corollary~\ref{co:2node_mod2+_Tlb} that the lower bound on computation given by Theorem~\ref{th:SDPI_gen_v} can tightly match the upper bound.
In this section, we provide two more examples in which our lower bounds on computation time are tight.
In the first example, our lower bound precisely captures the dependence of computation time on the number of nodes in the network.
In the second example, our lower bound tightly captures the dependence of computation time on the accuracy parameter $\eps$.

\subsection{Rademacher sum over a dumbbell network}
\begin{example}\label{ex:dumbball_net}
Consider a dumbbell network of bidirectional BSCs with the same crossover probability.
Formally, suppose $|\cV|$ is even, and let the nodes be indexed from $1$ to $|\cV|$. Nodes $1$ to $|\cV|/2$ form a clique (i.e., each pair of nodes are connected by a pair of BSCs), while nodes $|\cV|/2+1$ to $|\cV|$ form another clique. The two cliques are connected by a pair of BSCs between nodes $|\cV|/2$ and $|\cV|/2+1$.
Each node initially observes a ${\rm Bern}(\frac{1}{2})$ (or Rademacher) r.v.
The goal is for the nodes to compute the sum of the observations of all nodes.
The distortion function is $d(z,\wh z) = |z-\wh z|$.
\end{example}
By choosing the cutset as the pair of BSCs that joins the two cliques, our lower bound for random Rademacher sums in Corollary~\ref{co:Rademacher_sum} gives the following lower bound on computation time.
\begin{corollary}
For the problem of in Example~\ref{ex:dumbball_net}, for $\delta\in(0,1/2)$,
\begin{align*}
T(0,\delta) \gtrsim \frac{1}{C}\left(\frac{1-\delta}{2}\log\frac{\pi |\cV|}{4} - h_2(\delta)\right) \quad \text{as $|\cV| \rightarrow \infty$},
\end{align*}
which implies
\begin{align*}
T(0,\delta) = \Omega\left(\log |\cV|\right) .
\end{align*}
\end{corollary}
Now we show that the above lower bound matches the upper bound on the computation time, which turns out to be 
$$
T(0,\delta) = O\left(\log |\cV|\right) .
$$
As shown by Gallager \cite{Gallager88}, for a fixed success probability, nodes $|\cV|/2$ and $|\cV|/2+1$ can learn the partial sum of the observations in their respective cliques in $O\big(\log \log |\cV|\big)$ steps.
These two nodes then exchange their partial sum estimates using binary block codes. Each partial sum can take $|\cV|/2+1$ values, and can be encoded losslessly with $\log(|\cV|/2+1)$ bits.
The blocklength needed for transmission of the encoded partial sums is thus $O\big(\log(|\cV|/2+1)\big)$, where the hidden factor depends on the required success probability and the channel crossover probability, but not on $|\cV|$.
Having learned the partial sum of the other clique, nodes $|\cV|/2$ and $|\cV|/2+1$ continue to broadcast this partial sum to other nodes in their own clique. This takes another $O\big(\log(|\cV|/2+1)\big)$ step.
In total, the computation can be done in $O\big(\log \log |\cV|\big) + 2O\big(\log(|\cV|/2+1)\big) = O(\log |\cV|)$ steps, to have all nodes learn the sum of all observations, for any prescribed success probability.
This shows that $T(0,\delta) = O\left(\log |\cV|\right)$.

\subsection{Distributed averaging over discrete noisy channels}
\begin{example}\label{ex:dist_avg_BEC}
Consider a network where the nodes are connected by binary erasure channels with the same erasure probability.
Each node initially observes a log-concave r.v.
The goal is for the nodes to compute the average of the observations of all nodes.
\end{example}
For this example, Carli et al.~\cite{Carli_etal_erasures_consensus} define the computation time as
\begin{align*}
\tilde T(\eps) &\deq \inf \Big\{ T \in \N: 
	 \frac{1}{|\cV|}\sum_{v \in \cV}\E\big[(Z-\wh{Z}_v(t))^2\big] \le \eps,\,\forall t\ge T \Big\} 
\end{align*}
and show that
\begin{align}\label{eq:dist_avg_BEC_Tub}
\tilde T(\eps) \le c_1 + c_2\frac{\log^3\eps^{-1}}{\log^2\rho^{-1}}
\end{align}
where $\rho$ is the second largest singular value of the consensus matrix adapted to the network, and $c_1$ and $c_2$ are positive constants depending only on channel erasure probability. It can be shown that the above upper bound still holds (with different constants) when channels are BSCs.

{ 
We use Corollary~\ref{co:sum_logcon_dfen} to derive the following lower bound on $\tilde T(\eps)$.
\begin{corollary}
For the problem in Example~\ref{ex:dist_avg_BEC},
\begin{align}\label{eq:dist_avg_BEC_Tlb}
\tilde T(\eps) \ge \max_{\cS \subset \cV} \frac{1}{2C_\cS} \left(\!
 h(Z|W_\cS) + \log\frac{1}{4 e |\cV|} +\frac{1}{2}\log\frac{1}{\eps} - 2\!\right) .
\end{align}
\end{corollary}
\begin{IEEEproof}
Using Jensen's inequality twice, we can write
\begin{align*}
\frac{1}{|\cV|} \sum_{v\in\cV} \E\big[(Z - \wh Z_v(T))^2\big] 
&\ge \frac{1}{|\cV|} \sum_{v\in\cV} \big(\E|Z - \wh Z_v(T)|\big)^2 \\
&\ge \left(\frac{1}{|\cV|} \sum_{v\in\cV} \E|Z - \wh Z_v(T)| \right)^2 .
\end{align*}
Therefore, $|\cV|^{-1} \sum_{v\in\cV} \E\big[(Z - \wh Z_v(T))^2\big] \le \eps$ implies that $\E|Z - \wh Z_v(T)| \le |\cV|\sqrt{\eps}$ for all $v\in|\cV|$, and 
$$
\PP\left[|Z - \wh Z_v(T)| \ge \frac{|\cV|\sqrt{\eps}}{\delta}\right] \le \delta , \qquad\forall v\in\cV, \delta\in(0,1/2]
$$
by Markov's inequality.
Then by Corollary~\ref{co:sum_logcon_dfen},
\begin{align*}
&\tilde T(\eps) \ge T\left(\frac{|\cV|\sqrt{\eps}}{\delta},\delta\right) \\
&\ge \max_{\cS \subset \cV} \frac{1}{C_\cS} \left( (1-\delta)\left(h(Z|W_\cS) + \log \frac{\delta}{2e |\cV|\sqrt{\eps}} \right)-h_2(\delta)\right).
\end{align*}
Choosing $\delta=1/2$, we obtain \eqref{eq:dist_avg_BEC_Tlb}.
\end{IEEEproof}
}
The lower bound given by \eqref{eq:dist_avg_BEC_Tlb} states that $\tilde T(\eps)$ is necessarily logarithmic in $\eps^{-1}$, which tightly matches the poly-logarithmic dependence on $\eps^{-1}$ in the upper bound given by \eqref{eq:dist_avg_BEC_Tub}. 
As pointed out in Carli {et al.} \cite{Carli09_avg_consensus}, it is possible to prove that a computation time logarithmic in $\eps^{-1}$ is achievable by embedding a quantized consensus algorithm for noiseless networks into the simulation framework developed by Rajagopalan and Schulman for noisy networks in \cite{Rajagopalan94_dist_comp}.

\section{Conclusion and future research directions}\label{sec:conclusion}
We have studied the fundamental time limits of distributed function computation from an information-theoretic perspective.
The computation time depends on the amount of information about the function value needed by each node and the rate for the nodes to accumulate such an amount of information.
The small ball probability lower bound on conditional mutual information reveals how much information is necessary, while the cutset-capacity upper bound and the SDPI upper bound capture the bottleneck on the rate for the information to be accumulated.
The multi-cutset analysis provides a more refined characterization of the information dissipation in a network.

Here are some questions that are worthwhile to consider in the future:
\begin{itemize}
\item
In the multi-cutset analysis, the purpose of introducing self-loops when reducing the network to a chain is to establish necessary Markov relations for proving upper bounds on $I(Z ; \wh Z_n | W_{\cS})$ in bidirected chains, and the reason for considering left-bound nodes is to improve the lower bounds on computation time.
We could have included all channels from $\cS_i$ to $\cS_i$ into the self-loop at node $i'$ in $G'$, but this would result in looser lower bounds on computation time (cf.~the remark after Theorem~\ref{th:SDPI_red_gen}).
However, there might be other network reduction methods, e.g., different ways to construct the bidirected chain, that will yield even tighter lower bounds on computation time than our proposed method.

\item
In the first step of the derivation of Lemma~\ref{lm:mi_ub_cutset} and Lemma~\ref{lm:mi_ub_SDPI_gen}, we have upper-bounded $I(Z ; \wh Z_v|W_{\cS})$ using the ordinary data processing inequality as
\begin{align*}
I(Z ; \wh Z_v|W_{\cS}) &\le I(W_{\cS^c} ; \wh Z_v | W_{\cS}) .
\end{align*}
One may wonder whether we can tighten this step by a judicious use of SDPIs. The answer is negative.
It can be shown that
\begin{align*}
I(Z ; & \wh Z_v|W_{\cS}) \le I(W_{\cS^c} ; \wh Z_v| W_{\cS}) \\
&\sup_{w_\cS \in \prod_{v \in \cS}\sW_{v}}\eta\big(\PP_{W_{\cS^c}|W_\cS = w_\cS}, \PP_{Z|W_{\cS^c},W_\cS = w_\cS}\big) 
\end{align*}
where the contraction coefficient depends on the joint distribution of the observations $\PP_W$ and the function $Z = f(W)$.
However, 
\begin{align*}
\eta\big(\PP_{W_{\cS^c}|W_\cS = w_\cS}, \PP_{Z|W_{\cS^c},W_\cS = w_\cS}\big) = 1
\end{align*} 
for both discrete and continuous observations.
For discrete observations, this is a consequence of the fact that $\eta(\PP_X,\PP_{Y|X}) < 1$ if and only if the graph $\big\{(x,y) : \PP_X(x) > 0, \PP_{Y|X}(y|x) > 0\big\}$ is connected \cite{Ahlswede_Gacs_hypercont},
and the fact that, for any $\PP_{Y|X}$ induced by a deterministic function $f: \sX \rightarrow \sY$, this graph is always disconnected. This condition can be extended to continuous alphabets \cite{Witsenhausen_SDPI}.
It would be interesting to see whether \textit{nonlinear} SDPI's, e.g., of the sort recently introduced by Polyanskiy and Wu \cite{PolWu_SDPI_avgP}, can be somehow applied here to tighten the upper bounds.

\item
{ 
If the function to be computed is the identity mapping, i.e., $Z = W$, then the goal of the nodes is to distribute their observations to all other nodes in the network.
In this case, our results on the computation time can provide non-asymptotic lower bounds on the blocklength of the codes for the source-channel coding problems in multi-terminal networks.
In Example~\ref{ex:dist_W_1:M}, we have considered one such case with discrete observations, and obtained lower bounds in Corollary~\ref{ex:dist_W_1:M} based on the single cutset analysis.
It would be interesting to apply the multi-cutset analysis to the source-channel coding problems in multi-terminal, multi-hop networks.
}
\end{itemize}

\section*{Acknowledgment}

The authors would like to thank the Associate Editor Prof.~Chandra Nair and two anonymous referees for numerous constructive suggestions on how to improve the flow and the structure of the paper.

\appendices

\renewcommand{\theequation}{\Alph{section}.\arabic{equation}}
\renewcommand{\thelemma}{\Alph{section}.\arabic{lemma}}
\setcounter{equation}{0}
\setcounter{lemma}{0}

\section{Proof of Lemma~\ref{lm:net_red_AA'}}
\label{sec:pf:lm:net_red_AA'}
\setcounter{equation}{0}
\setcounter{lemma}{0}

{ 
The goal of this proof is to show that, given any $T$-step algorithm $\cA$ running on $G$, we can construct a randomized $T$-step algorithm $\cA'$ running on $G'$ that simulates $\cA$.
}
Fix any $T$-step algorithm $\cA$ that runs on $G$. For each $t$, we can factor the conditional distribution of the messages $X_t \deq (X_{v,t})_{v\in\cV}$ given $W,X^{t-1},Y^{t-1}$ as follows:
\begin{align}
&\quad\,\, \PP_{X_t|W,X^{t-1},Y^{t-1}}(x_t|w,x^{t-1},y^{t-1}) \nonumber \\
&= \prod_{v \in \cV} \PP_{X_{v,t}|W_v,Y^{t-1}_v}(x_{v,t}|w_v,y^{t-1}_v) \nonumber \\
&= \prod^n_{i=1} \prod_{v \in \cS_i} \PP_{X_{v,t} | W_v,Y^{t-1}_v}\Big( x_{v,t} \Big| w_v,y^{t-1}_v \Big) \nonumber \\
&= \prod^n_{i=1} \PP_{X_{\cS_i,t} | W_{\cS_i},Y^{t-1}_{\cS_i}}\Big( x_{\cS_i,t} \Big| w_{\cS_i},y^{t-1}_{\cS_i} \Big).
 \label{eq:partition_factorization_1}
\end{align}
Likewise, the conditional distribution of the received messages $Y_t \deq (Y_{v,t})_{v\in\cV}$ given $W,X^t,Y^{t-1}$ can be factored as
\begin{align}
&\quad\,\,	\PP_{Y_t|W,X^t,Y^{t-1}}(y_t|w,x^t,y^{t-1}) \nonumber \\
&= \prod_{e \in \cE} \PP_{Y_{e,t}|X_{e,t}}(y_{e,t}|x_{e,t}) \nonumber \\
	&= \prod_{e \in \cE} K_e (y_{e,t}|x_{e,t}) \nonumber \\
	&= \prod^n_{i=1} \prod_{u \in \cS_i} \prod_{v \in \cV: \, (u,v) \in \cE} K_{(u,v)}(y_{(u,v),t}|x_{(u,v),t}). \label{eq:partition_factorization_2}
\end{align}
Since the successive partition of $G$ ensures that nodes in $\cS_i$ can communicate with nodes in $\cS_j$ only if $|i-j| \le 1$, the messages originating from $\cS_i$ at step $t$ can be decomposed as
\begin{align*}
X_{\cS_i,t} &= ( X_{(\cS_i,\cS_{i-1}),t}, X_{(\cS_i,\cS_{i+1}),t}, X_{(\cS_i,\cS_{i}),t} ) \\
&= ( X_{(\cS_i,\cS_{i-1}),t}, X_{(\cS_i,\cS_{i+1}),t}, X\!{\raisebox{-3pt}{$\scriptstyle{(\cS_i, \lepartial\cS_i),t}$}}, X\!{\raisebox{-3pt}{$\scriptstyle{(\cS_i,\cS_i \setminus \lepartial\cS_i),t}$}} ) ,
\end{align*}
and the messages received by nodes in $\cS_i$ at step $t$ can be decomposed as
\begin{align}
Y_{\cS_i,t} &= ( Y_{(\cS_{i-1},\cS_{i}),t}, Y_{(\cS_{i+1},\cS_{i}),t}, Y_{(\cS_{i},\cS_{i}),t} ) \nonumber \\
&= ( Y_{(\cS_{i-1},\cS_{i}),t}, Y_{(\cS_{i+1},\cS_{i}),t}, Y\!{\raisebox{-3pt}{$\scriptstyle{(\cS_i, \lepartial\cS_i),t}$}}, Y\!{\raisebox{-3pt}{$\scriptstyle{(\cS_i,\cS_i \setminus \lepartial\cS_i),t}$}} ) . \label{eq:decomp_Y_Si}
\end{align}
According to the operation of algorithm $\cA$, for each $(u,v)\in\cE$ there exists a mapping $\varphi_{(u,v),t}$, such that $X_{(u,v),t} = \varphi_{(u,v),t}(W_u,Y_u^{t-1})$.
By the definition of $\lepartial \cS_i$, we can write
\begin{align*}
X_{(\cS_i,\cS_{i-1}),t} &= \big(\varphi_{(u,v),t}(W_u,Y_u^{t-1}): \\
& \quad (u,v)\in\cE, u\in \lepartial \cS_i, v\in\cS_{i-1}\big) . 
\end{align*}
Thus, there exists a mapping $\lephi_{\cS_i,t}$, such that
\begin{align}
X_{(\cS_i,\cS_{i-1}),t} &= \lephi_{\cS_i,t}(W\hskip-1pt{\raisebox{-4pt}{$\scriptstyle{ \lepartial\cS_i}$}},Y^{t-1}\hskip-17pt{\raisebox{-5pt}{$\scriptstyle{ \lepartial\cS_i}$}})  \label{eq:X_SiSi-1}
\end{align}
where
\begin{align}\label{eq:Y_leSi}
Y\hskip-1pt{\raisebox{-3pt}{$\scriptstyle{ \lepartial\cS_i,t}$}} 
= \big(
Y\hskip-1pt{\raisebox{-3pt}{$\scriptstyle{(\cS_{i-1}, \lepartial\cS_i),t}$}} , 
Y\hskip-1pt{\raisebox{-3pt}{$\scriptstyle{(\cS_{i+1}, \lepartial\cS_i),t}$}} , 
Y\hskip-1pt{\raisebox{-3pt}{$\scriptstyle{(\cS_{i}, \lepartial\cS_i),t}$}}
\big) .
\end{align}
By the same token, there exist mappings $\riphi_{\cS_i,t}$, $\mathring\varphi_{\cS_i,t}$ and $\bar\varphi_{\cS_i,t}$, such that
\begin{align}
X_{(\cS_i,\cS_{i+1}),t} &= \riphi_{\cS_i,t}(W_{\cS_i},Y^{t-1}_{\cS_i}) , \label{eq:X_SiSi+1} \\
X\!{\raisebox{-3pt}{$\scriptstyle{(\cS_{i}, \lepartial\cS_i),t}$}}  
&= \mathring\varphi_{\cS_i,t}(W_{\cS_i}, Y^{t-1}_{\cS_i}), \label{eq:X_SileparSi} \\
X\!{\raisebox{-3pt}{$\scriptstyle{(\cS_{i}, \cS_i \setminus \lepartial\cS_i),t}$}} 
&= \bar\varphi_{\cS_i,t}(W_{\cS_i}, Y^{t-1}_{\cS_i}) . \label{eq:X_SiSi-leparSi} 
\end{align}
Define the random variables
\begin{align*}
W_i &\deq W_{\cS_i}, \\
X_{i,t} &= (X_{(i,i-1),t}, X_{(i,i+1),t}, X_{(i,i),t}) \\
&\deq (X_{(\cS_i,\cS_{i-1}),t}, X_{(\cS_i,\cS_{i+1}),t}, X\!{\raisebox{-3pt}{$\scriptstyle{(\cS_{i}, \lepartial\cS_i),t}$}}) , \\
Y_{i,t} &= (Y_{(i-1,i),t}, Y_{(i+1,i),t}, Y_{(i,i),t}) \\
&\deq (Y_{(\cS_{i-1},\cS_{i}),t}, Y_{(\cS_{i+1},\cS_i),t}, Y\!{\raisebox{-3pt}{$\scriptstyle{(\cS_{i}, \lepartial\cS_i),t}$}}) , 
\\
U_{i,t} 
&\deq (X\!{\raisebox{-3pt}{$\scriptstyle{(\cS_{i}, \cS_i \setminus \lepartial\cS_i),t}$}}, Y\!{\raisebox{-3pt}{$\scriptstyle{(\cS_{i}, \cS_i \setminus \lepartial\cS_i),t}$}}) .
\end{align*}
{ 
From the decomposition of $Y_{\cS_i,t}$ in \eqref{eq:decomp_Y_Si}, we know that $(Y_i^{t-1},U_i^{t-1})$ contains $Y_{\cS_i}^{t-1}$; while from the decomposition of $Y\hskip-1pt{\raisebox{-3pt}{$\scriptstyle{ \lepartial\cS_i,t}$}} $ in \eqref{eq:Y_leSi}, we know that $Y_i^{t-1}$ contains $Y^{t-1}\hskip-17pt{\raisebox{-5pt}{$\scriptstyle{ \lepartial\cS_i}$}}$.
}
Therefore, from Eqs.~\eqref{eq:X_SiSi-1} and \eqref{eq:X_SiSi+1}-\eqref{eq:X_SiSi-leparSi}, we deduce the existence of mappings $\lephi_{i,t}$, $\riphi_{i,t}$,  $\mathring\varphi_{i,t}$, and $\bar\varphi_{i,t}$, such that the messages transmitted by nodes in $\cS_i$ at time $t$ can be generated as
\begin{align}
X_{(i,i-1),t} &= \lephi_{i,t}(W_{i},Y_{i}^{t-1}) , \label{eq:Xi'(i-1)'} \\
X_{(i,i+1),t} &= \riphi_{i,t}(W_{i},Y_{i}^{t-1},U_{i}^{t-1}) , \\
X_{(i,i),t} &= \mathring\varphi_{i,t}(W_{i},Y_{i}^{t-1},U_{i}^{t-1}), \label{eq:Xi'i'} \\
X\!{\raisebox{-3pt}{$\scriptstyle{(\cS_{i}, \cS_i \setminus \lepartial\cS_i),t}$}} &= \bar\varphi_{i,t}(W_i,Y^{t-1}_i,U^{t-1}_i) . \label{eq:Xsi_si-lesi}
\end{align}
{  Note that the computation of $X_{(i,i-1),t}$ does not involve $U_{i}^{t-1}$.}
Next, the messages received by nodes in $\cS_i$ at step $t$ are related to the transmitted messages as
\begin{align*}
X_{(i-1,i),t} &\xrightarrow{K_{(i-1,i)}} Y_{(i-1,i),t}, \\
X_{(i+1,i),t} &\xrightarrow{K_{(i+1,i)}} Y_{(i+1,i),t}  ,\\
X_{(i,i),t} &\xrightarrow{K_{(i,i)}} Y_{(i,i),t} ,
\end{align*}
where the stochastic transition laws have the same form as those in Eqs.~\eqref{eq:chain_Kii-1} to \eqref{eq:chain_Kii}. 
{ 
In addition, since $X\!{\raisebox{-3pt}{$\scriptstyle{(\cS_{i}, \cS_i \setminus \lepartial\cS_i),t}$}}$ and $Y\!{\raisebox{-3pt}{$\scriptstyle{(\cS_{i}, \cS_i \setminus \lepartial\cS_i),t}$}}$ are related through the channels from $\cS_i$ to $\cS_i \setminus \lepartial\cS_i$, there exists a mapping $\kappa_{i,t}$ such that
$Y\!{\raisebox{-3pt}{$\scriptstyle{(\cS_{i}, \cS_i \setminus \lepartial\cS_i),t}$}}$ can be realized as
\begin{align}
Y\!{\raisebox{-3pt}{$\scriptstyle{(\cS_{i}, \cS_i \setminus \lepartial\cS_i),t}$}} &= \kappa_{i,t}(X\!{\raisebox{-3pt}{$\scriptstyle{(\cS_{i}, \cS_i \setminus \lepartial\cS_i),t}$}},R_{i,t}) , \label{eq:kappa_i}
\end{align}
where $R_{i,t}$ can be taken as a random variable uniformly distributed over $[0,1]$ and independent of everything else.
}
From \eqref{eq:Xsi_si-lesi} and \eqref{eq:kappa_i}, we know that
$U_{i,t}$ can be realized by a mapping $\vartheta_{i,t}$ as
\begin{align}
U_{i,t} &= \vartheta_{i,t}(W_{i},Y_{i}^{t-1},U_{i}^{t-1},R_{i,t}) . \label{eq:theta_i}
\end{align}
Taking all of this into account, we can rewrite the factorization \eqref{eq:partition_factorization_1} as follows:
\begin{align}
&\quad\,\, \PP_{X_t|W,X^{t-1},Y^{t-1}}(x_t|w,x^{t-1},y^{t-1}) \nonumber\\
	&= \prod^n_{i=1} \I\big\{ x_{(i-1,i),t} = \lephi_{i,t}(w_{i},y_{i}^{t-1}) \big\} \nonumber \\
& \quad\cdot \I\big\{ x_{(i,i+1),t} = \riphi_{i,t}(w_{i},y_{i}^{t-1},u_{i}^{t-1})\big\} \nonumber\\
	& \quad  \cdot \I\big\{ x_{(i,i),t} = \mathring\varphi_{i,t}(w_{i},y_{i}^{t-1},u_{i}^{t-1})\big\} \nonumber \\
&\quad \cdot \I\big\{ x\!{\raisebox{-3pt}{$\scriptstyle{(\cS_{i}, \cS_i \setminus \lepartial\cS_i),t}$}}  = \bar\varphi_{i,t}(w_i, y^{t-1}_i, u^{t-1}_i) \big\} ,
	 \label{eq:partition_factorization_1R}
\end{align}
and we can rewrite the factorization \eqref{eq:partition_factorization_2} as
\begin{align}
&\quad\,\,	\PP_{Y_t|W,X^t,Y^{t-1}}(y_t|w,x^t,y^{t-1}) \nonumber\\
&= \prod^n_{i=1} K_{(i-1,i)}(y_{(i-1,i),t}|x_{(i-1,i),t}) \nonumber \\
&\quad \cdot K_{(i+1,i)}(y_{(i+1,i),t}|x_{(i+1,i),t}) \cdot K_{(i,i)}(y_{(i,i),t} | x_{(i,i),t}) \nonumber\\
& \quad \cdot \bigotimes_{(u,v) \in \cE:  u \in \cS_i, v \in \cS_i \setminus \lepartial\cS_i} K_{(u,v)} (y\!{\raisebox{-3pt}{$\scriptstyle{(\cS_{i}, \cS_i \setminus \lepartial\cS_i),t}$}} | x\!{\raisebox{-3pt}{$\scriptstyle{(\cS_{i}, \cS_i \setminus \lepartial\cS_i),t}$}}) \label{eq:partition_factorization_2R}
\end{align}
where the channel $\bigotimes_{(u,v) \in \cE: u \in \cS_i, v \in \cS_i \setminus \lepartial\cS_i} K_{(u,v)}$ can be realized by the mapping $\kappa_{i,t}$ with the r.v. $R_{i,t}$.

To summarize: the mappings defined in \eqref{eq:Xi'(i-1)'} to \eqref{eq:Xi'i'} and \eqref{eq:theta_i} specify a randomized $T$-step algorithm $\cA'$ that runs on $G'$ and simulates the $T$-step algorithm $\cA$ that runs on $G$. Specifically, using these mappings, each node $i'$ in $G'$ can generate all the transmitted and received messages of $\cS_i$ in $\cA$ as $(X_{i'}^T,Y_{i'}^T,U_{i'}^T)$. Moreover, from \eqref{eq:partition_factorization_1R} and \eqref{eq:partition_factorization_2R} we see that the random objects
$$
\big(W_{\cS_i},X_{\cS_i}^T,Y_{\cS_i}^T : i \in \{1,\ldots,n\}\big)
$$
and
$$
\big(W_{i'}, X_{i'}^T,Y_{i'}^T,U_{i'}^T : i' \in \{1,\ldots,n\}\big)
$$
have the same joint distribution.

Finally, as we have assumed that $\lepartial\cS_i$'s are all nonempty, we can define
\begin{align*}
\wh Z_{i} \deq \wh Z_v = \psi_v(W_v,Y_v^{T})
\end{align*} 
with an arbitrary $v\in\lepartial\cS_i$.
{  From the definition of $Y_{i,t}$ and the fact that $Y_i^{T}$ contains $Y_v^{T}$}, it follows that there exists a mapping $\psi_{i}$ such that
\begin{align*}
\wh Z_{i} = \psi_{i}(W_{i},Y_{i}^{T}) . 
\end{align*} 
Using this mapping, node $i'$ in $G'$ can generate the final estimate of the chosen $v \in \lepartial\cS_i$ in $\cA$ as $\wh Z_{i'}$, such that $(Z,\wh Z_i:  i \in \{1,\ldots,n\})$ and $(Z,\wh Z_{i'}:  i \in \{1,\ldots,n\})$ have the same joint distribution. This guarantees that
\begin{align*}
\max_{i'\in\cV'}\PP[d(Z,\wh Z_{i'}) > \eps] 
&= \max_{i\in\{1:n\}}\PP[d(Z,\wh Z_{i}) > \eps] \\
&\le \max_{v\in\cV}\PP[d(Z,\wh Z_{v}) > \eps] \\
&\le \delta .
\end{align*}
The claim that $T(\eps,\delta)$ for computing $Z$ on $G$ is lower bounded by $T'(\eps,\delta)$ for computing $Z$ on $G'$ then follows from the definition of $T'(\eps,\delta)$ in \eqref{eq:comp_time_rdm}.
This proves Lemma~\ref{lm:net_red_AA'}.

\section{Proof of Lemma~\ref{lm:mi_ub_SDPI_chain_eta}}
\label{sec:pf:lm:mi_ub_SDPI_chain_eta}
\setcounter{equation}{0}
\setcounter{lemma}{0}

\begin{figure}[h!]
\centering
  \includegraphics[scale = 1.0]{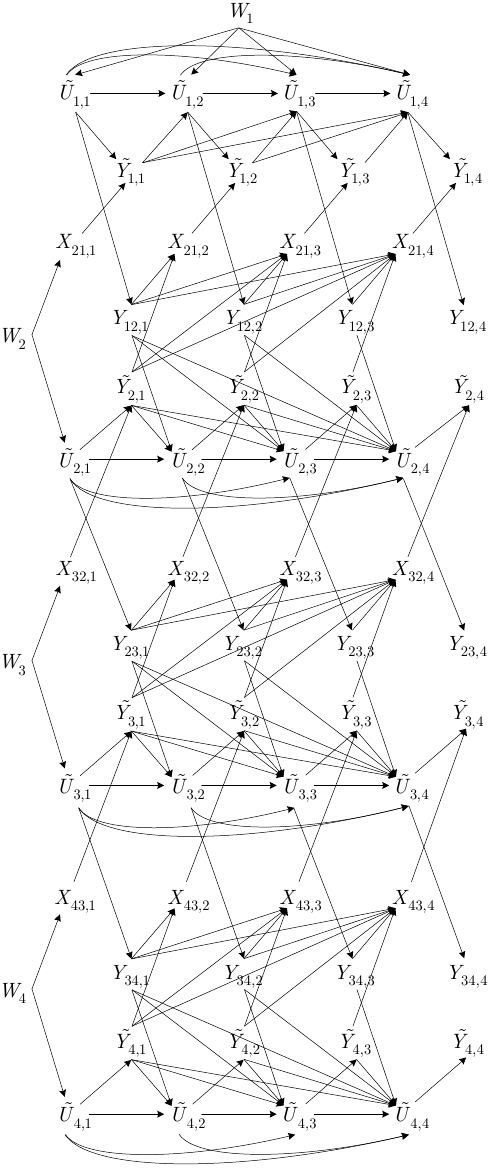}
  \caption{Bayesian network of $(W,X^T,U^T,Y^T)$ for the randomized algorithm $\cA'$ on a $4$-node bidirected chain with $T=4$. ($W_{1:4}$ are arbitrarily correlated, and not all edges emanating from $W_{2:4}$ are shown.)}
  \label{fg:BN_chain4_slp}
\end{figure}

Recall that, for any randomized $T$-step algorithm $\cA'$, at step $t\in\{1,\ldots,T\}$, node $i \in \{1,\ldots,n\}$ computes the outgoing messages $X_{(i,i-1),t} = \lephi_{i,t}(W_{i},Y^{t-1}_{i})$, $X_{(i,i+1),t} = \riphi_{i,t}(W_{i},Y^{t-1}_{i},U^{t-1}_{i})$, and $X_{(i,i),t} = \mathring\varphi_{i,t}(W_{i},Y^{t-1}_{i},U^{t-1}_{i})$, and the private message $U_{i,t} = \vartheta_{i,t}(W_{i},Y_{i}^{t-1},U^{t-1}_{i},R_{i,t})$, where $R_{i,t}$ is the private randomness of node $i$.
At step $T$, node $i$ computes $\wh{Z}_{i} = \psi_{i}(W_{i},Y^T_{i})$.
We will use the Bayesian network formed by all the relevant variables and the  d-separation criterion \cite[Theorem 3.3]{PGM_book} to find conditional independences among these variables.
To simplify the Bayesian network, we merge some of the variables by defining 
$$
\tilde U_{i,t} \deq (X_{(i,i),t},X_{(i,i+1),t},U_{i,t})
$$ 
and 
$$
\tilde Y_{i,t} \deq (Y_{(i,i),t},Y_{(i+1,i),t}) 
$$
for $i\in\{1,\ldots,n\}$. 
The joint distribution of the variables can then be factored as
\begin{align}
&\quad\,\, \PP_{W,X^T,U^T,Y^T}(w,x^T,u^T,y^T)\nonumber\\
&= \PP_W(w) \prod_{t=1}^T \prod_{i=1}^n \I\big\{x_{(i,i-1),t} = \lephi_{i,t}(w_i,y_i^{t-1})\big\} \nonumber \\
&\quad \cdot\PP_{\tilde{U}_{i,t}|W_i,Y_i^{t-1},\tilde{U}_i^{t-1}}(\tilde u_{i,t}|w_i,y_i^{t-1},\tilde{u}_{i}^{t-1}) \nonumber \\ 
& \quad \cdot \prod_{i=1}^n \PP_{Y_{(i-1,i),t} | \tilde U_{i-1,t}}(y_{(i-1,i),t} | \tilde u_{i-1,t}) \nonumber \\
&\quad \cdot \PP_{\tilde Y_{i,t}|\tilde U_{i,t}, X_{(i+1,i),t}}(\tilde y_{i,t}|\tilde u_{i,t}, x_{(i+1,i),t}) .\label{eq:BN_factor}
\end{align}
The Bayesian network corresponding to this factorization for $n=4$ and $T=4$ is shown in Fig.~\ref{fg:BN_chain4_slp}.

If $T=0$, then $\wh Z_n = \psi(W_n)$, hence $I(Z;\wh Z_n | W_{2:n}) \le I(Z;W_n | W_{2:n}) = 0$.
For $T\ge 1$, we prove the upper bounds in the following steps, where we assume $n\ge 4$. The case $n=3$ can be proved by skipping Step~2, and the case $n=2$ can be proved by skipping Step~1 and Step~2.

\smallskip
\noindent{\bf Step 1}:\\
For any $i$ and $t$, define the shorthand $X_{i\leftarrow, t} \deq X_{(\cN_{i \leftarrow},i),t}$, where $\cN_{i \leftarrow}$ is the in-neighborhood of node $i$. From the Markov chain $W, Y_n^{T-1} \rightarrow X_{n \leftarrow,T} \rightarrow Y_{n,T}$ and Lemma~\ref{lm:cond_SDPI}, we follow the same argument as the one used for proving Lemma~\ref{lm:mi_ub_SDPI_gen} to show that
\begin{align*}
I(Z;\wh Z_n | W_{2:n}) &\le I(W_1;Y_n^T|W_{2:n}) \nonumber\\
&\le (1-\eta_n)I(W_1;Y_n^{T-1}|W_{2:n}) \nonumber \\
&\quad  + \eta_n I(W_1; Y_n^{T-1},X_{{n \leftarrow},T}|W_{2:n}). 
\end{align*}
Applying the d-separation criterion to the Bayesian network corresponding to \eqref{eq:BN_factor} (see Fig.~\ref{fg:BN_chain4_slp} for an illustration), we can read off the Markov chain  
\begin{align*}
{%
W_{1} \rightarrow W_{2:n}, Y_{n-1}^{t-1} \rightarrow Y_n^{t-1}, \tilde U_{n-1,t}, \tilde U_{n,t}
}
\end{align*}
for $t\in\{1,\ldots,T\}$, since all trails from $W_{1}$ to $(Y_n^{t-1}, \tilde U_{n-1,t}, \tilde U_{n,t})$ are blocked by $(W_{2:n},Y_{(n-2,n-1)}^{t-1})$, and all trails from $(Y_n^{t-1}, \tilde U_{n-1,t}, \tilde U_{n,t})$ to $W_{1}$ are blocked by $(W_{2:n},\tilde Y_{n-1}^{t-1})$.
This implies the Markov chain $W_{1} \rightarrow W_{2:n}, Y_{n-1}^{T-1} \rightarrow Y_n^{T-1}, X_{n\leftarrow,T}$, since $X_{(n-1,n),T}$ is included in $\tilde U_{n-1,T}$ and $X_{(n,n),T}$ is included in $\tilde U_{n,T}$. Consequently,\footnote{This follows from the ordinary DPI and from the fact that, if $X \to A,B \to C$ is a Markov chain, then $X \to B \to C$ is a Markov chain conditioned on $A=a$.}
\begin{align}
I(W_1;Y_n^T|W_{2:n}) &\le (1-\eta_n)I(W_1;Y_n^{T-1}|W_{2:n}) \nonumber \\
&\quad + \eta_n I(W_1; Y_{n-1}^{T-1}|W_{2:n}) \label{eq:SDPI_chain_rec1_slp} .
\end{align}
Also note that $I(W_1;Y_{n,1}|W_{2:n}) \le I(W_1;X_{n\leftarrow,1}|W_{2:n}) \le I(W_1; W_{\cN_{n\leftarrow}}|W_{2:n})=0$.

\smallskip
\noindent{\bf Step 2}:\\
For $i\in\{1,\ldots,n-3\}$, from the Markov chain
$W, Y_{n-i}^{T-i-1} \rightarrow X_{{{(n-i)}\leftarrow},T-i} \rightarrow Y_{n-i,T-i}$
and Lemma~\ref{lm:cond_SDPI}, 
\begin{align*}
I(W_1;Y_{n-i}^{T-i}| & W_{2:n}) 
\le (1-\eta_{n-i}) I(W_1;Y_{n-i}^{T-i-1}|W_{2:n}) \\
& +
\eta_{n-i} I(W_1;Y_{n-i}^{T-i-1},X_{({n-i)}\leftarrow,T-i}|W_{2:n}) 
\end{align*}
From the Bayesian network corresponding to \eqref{eq:BN_factor}, we can read off the Markov chain
\begin{align*}
W_{1} &\rightarrow W_{2:n}, Y_{n-i-1}^{t-1} \\
&\rightarrow 
 Y_{n-i}^{t-1}, \tilde U_{n-i-1,t}, \tilde U_{n-i,t}, X_{(n-i+1,n-i),t}
\end{align*}
for $t\in\{1,\ldots,T-i\}$, since all trails from $W_{1}$ to 
$$
(Y_{n-i}^{t-1}, \tilde U_{n-i-1,t}, \tilde U_{n-i,t}, X_{(n-i+1,n-i),t})
$$
are blocked by $(W_{2:n}, Y_{(n-i-2,n-i-1)}^{t-1})$, and all trails from 
$$
(Y_{n-i}^{t-1}, \tilde U_{n-i-1,t}, \tilde U_{n-i,t}, X_{(n-i+1,n-i),t})
$$
to $W_{1}$ are blocked by $(W_{2:n},\tilde Y_{n-i-1}^{t-1})$.  
This implies the Markov chain
$$
W_{1} \rightarrow W_{2:n}, Y_{n-i-1}^{T-i-1} \rightarrow Y_{n-i}^{T-i-1}, X_{(n-i)\leftarrow,T-i} ,
$$
since $X_{(n-i-1,n-i),T-i}$ is included in $\tilde U_{n-i-1,T-i}$ and $X_{(n-i,n-i),T-i}$ is included in $\tilde U_{n-i,T-i}$. Therefore,
\begin{align}
I(W_1;Y_{n-i}^{T-i}|W_{2:n}) &\le (1-\eta_{n-i}) I(W_1;Y_{n-i}^{T-i-1}|W_{2:n}) \nonumber \\
&\quad + \eta_{n-i} I(W_1;Y_{n-i-1}^{T-i-1}|W_{2:n})  \label{eq:SDPI_chain_rec2_slp} 
\end{align}
for $i\in \{1,\ldots,n-3\}$.
Also note that 
\begin{align*}
I(W_1;Y_{n-i,1}|W_{2:n}) &\le I(W_1;X_{(n-i)\leftarrow,1}|W_{2:n}) \\
&\le I(W_1 ; W_{\cN_ {(n-i) \leftarrow}} | W_{2:n}) \\
&= 0.
\end{align*}

\smallskip
\noindent{\bf Step 3}:\\
Finally, we upper-bound $I(W_1;Y_{2}^{T-n+2}|W_{2:n})$ for $T\ge n-1$. 
From the Markov chain $W,Y_2^{t-1} \rightarrow X_{2\leftarrow,t} \rightarrow Y_{2,t}$ and Lemma~\ref{lm:cond_SDPI}, 
\begin{align}
I(W_1;Y_{2}^{T-n+2}|W_{2:n})
&\le (1-\eta_2) I(W_1;Y_2^{T-n+1}|W_{2:n}) \nonumber \\
&\quad + \eta_2 H(W_1|W_{2:n}) \label{eq:SDPI_chain_rec3_slp}  .
\end{align}
This upper bound is useful only when $H(W_1|W_{2:n})$ is finite. If the observations are continuous r.v.'s, we can upper bound $I(W_1;Y_{2}^{T-n+2}|W_{2:n})$ in terms of the channel capacity $C_{(1,2)}$:
\begin{align}
&\quad\,\, I(W_1;Y_{2}^{T-n+2}|W_{2:n}) \nonumber \\
&= \sum_{t=1}^{T-n+2} I(W_1;Y_{2,t}|W_{2:n},Y_2^{t-1}) \nonumber \\
&\overset{\rm (a)}= \sum_{t=1}^{T-n+2} \Big( I(W_1;Y_{(1,2),t}|W_{2:n},Y_{2}^{t-1}) \nonumber \\
&\qquad\qquad + I(W_1;\tilde Y_{2,t}|W_{2:n},Y_{2}^{t-1},Y_{(1,2),t}) \Big) \nonumber\\
&\overset{\rm (b)}\le \sum_{t=1}^{T-n+2} I(X_{(1,2),t};Y_{(1,2),t}|W_{2:n},Y_{2}^{t-1}) \nonumber\\
&\overset{\rm (c)}\le \sum_{t=1}^{T-n+2} I(X_{(1,2),t};Y_{(1,2),t}) \nonumber\\
&\le C_{(1,2)}(T-n+2) , \label{eq:SDPI_chain_rec31_slp}
\end{align}
where we have used the Markov chain {$W_{1} \rightarrow W_{2:n}, Y_{2}^{t-1},Y_{(1,2),t} \rightarrow \tilde Y_{2,t}$} for $t\in\{1,\ldots,T-n+2\}$,
which follows by applying the d-separation criterion to the Bayesian network corresponding to the factorization in \eqref{eq:BN_factor},
so that the second term in (a) is zero; the Markov chain $W,Y_{2}^{t-1} \rightarrow X_{(1,2),t} \rightarrow Y_{(1,2),t}$, which also implies the Markov chain $W_1 \rightarrow X_{(1,2),t} ,W_{2:n},Y_{2}^{t-1} \rightarrow Y_{(1,2),t}$ by the weak union property of conditional independence, hence (b) and (c); and the fact that $I(X_{(1,2),t};Y_{(1,2),t}) \le C_{(1,2)}$.

\smallskip
\noindent{\bf Step 4}:\\
Define $I_{i,t} = I(W_1;Y_{i}^{t}|W_{2:i})$ for $i\ge 2$ and $t\ge 1$.
From (\ref{eq:SDPI_chain_rec1_slp}), (\ref{eq:SDPI_chain_rec2_slp}), (\ref{eq:SDPI_chain_rec3_slp}), and (\ref{eq:SDPI_chain_rec31_slp}), we can write, for $n\ge 3$, $T\ge n-1$, and $i\in\{0,\ldots,n-3\}$,
\begin{align}
I_{n-i,T-i} &\le \bar\eta_{n-i} I_{n-i,T-i-1} + \eta_{n-i} I_{n-i-1,T-i-1} \label{eq:SDPI_chain_rec_n} 
\end{align}
where $\bar\eta_{n-i} = 1- \eta_{n-i}$, and $I_{n-i,1} = 0$.
In addition, for $T\ge n-1$,
\begin{align}
I_{2,T-n+2} &\le 
\begin{cases}
\bar\eta_2 I_{2,T-n+1} + \eta_2 H(W_1|W_{2:n}) \\
C_{(1,2)} (T-n+2) 
\end{cases},
\end{align}
and $I_{2,0} = 0$.

An upper bound on $I(W_1;Y_{n}^{T}|W_{2:n})$ can be obtained by solving this set of recursive inequalities with the specified boundary conditions. 
It can be checked by induction that $I(W_1;Y_{n}^{T}|W_{2:n}) = 0$ if $T\le n-2$.
For $T\ge n-1$, if $\eta_i \le \tilde\eta$ for all $i\in\{1,\ldots,n\}$, then the above  inequalities continue to hold with $\eta_i$'s replaced with $\tilde\eta$. The resulting set of inequalities is similar to the one obtained by Rajagopalan and Schulman \cite{Rajagopalan94_dist_comp} for the evolution of mutual information in broadcasting a bit over a unidirectional chain of BSCs. 
With 
$$
\cB(m,k,p) \deq {m \choose k} p^k (1-p)^{m-k} ,
$$
the exact solution is given by
\begin{align*}
&\quad\,\, I(W_1;Y_{n}^{T}|W_{2:n}) \\
&\le H(W_1|W_{2:n}) \tilde\eta \sum_{i = 1}^{T-n+2} \tilde\eta^{n-2} (1-\tilde\eta)^{T-i-n+2} {T-i \choose n-2}  \\
&= H(W_1|W_{2:n}) \eta \sum_{i = 1}^{T-n+2} \cB(T-i, n-2, \eta) 
\end{align*}
for $n\ge 2$, and
\begin{align*}
&\quad\,\,I(W_1;Y_{n}^{T}|W_{2:n}) \\
&\le
C_{(1,2)} \tilde\eta \sum_{i = 1}^{T-n+2} \tilde\eta^{n-3} (1-\tilde\eta)^{T-i-n+2} {T-i-1 \choose n-3} i  \\
&= C_{(1,2)} \eta \sum_{i = 1}^{T-n+2}
 \cB(T-i-1, n-3, \eta) i 
\end{align*}
for $n\ge 3$.
This proves \eqref{eq:mi_ub_SDPI_chain_opt1} and \eqref{eq:mi_ub_SDPI_chain_opt2}.

For general $\eta_i$'s, we obtain a suboptimal upper bound by unrolling the first term in (\ref{eq:SDPI_chain_rec_n}) for each $i$ and using the fact that $I_{n-i,t} = 0$ for $t\le n-i-2$, getting
\begin{align*}
I_{n-i,T-i} &\le \bar\eta_{n-i}^{T-n+1} \eta_{n-i} I_{n-i-1,n-i-2} + \ldots \\
&\quad + \bar\eta_{n-i} \eta_{n-i} I_{n-i-1,T-i-2} + \eta_{n-i} I_{n-i-1,T-i-1} \\
&\le \big(\bar\eta_{n-i}^{T-n+1}  + \ldots + \bar\eta_{n-i}+1\big)\eta_{n-i} I_{n-i-1,T-i-1} \\
&= \big(1-\bar\eta_{n-i}^{T-n+2}\big)I_{n-i-1,T-i-1} .
\end{align*}
Iterating over $i$, and noting that 
\begin{align*}
&\quad\,\, I_{2,T-n+2} \\
&\le \min\big\{H(W_1|W_{2:n})(1-\bar\eta_{2}^{T-n+2}),C_{(1,2)}(T-n+2)\big\} ,
\end{align*}
we get for $n\ge 2$ and $T\ge n-1$,
\begin{align}\label{eq:SDPI_chain_mi_gen}
& I(W_1;Y_n^T | W_{2:n}) \le \nonumber \\
&\begin{cases}
H(W_1|W_{2:n}) \prod_{i=2}^{n}\big(1-(1-\eta_{i})^{T-n+2}\big) \\
C_{(1,2)} (T-n+2) \prod_{i=3}^{n}\big(1-(1-\eta_{i})^{T-n+2}\big) 
\end{cases} .
\end{align}
The weakened upper bounds in \eqref{eq:mi_ub_SDPI_chain_sub_a} and \eqref{eq:mi_ub_SDPI_chain_sub_b} are obtained by replacing $\eta_{i}$ in \eqref{eq:SDPI_chain_mi_gen} with 
$$\eta \deq \max_{i=1,\ldots,n} \eta_i .$$

Finally, we show \eqref{eq:chain_mi_limit0} using an argument similar to the one in \cite{Rajagopalan94_dist_comp}. If $n\ge 4$ and $T \le 2 + (n-3)\gamma/\eta$ for some $\gamma \in (0,1)$, then
\begin{align*}
\eta < \frac{\eta}{\gamma} \le \frac{n-3}{T-2} \le \frac{n-2}{T-1} \le 1 
\end{align*}
where the last inequality follows from the assumption that $T \ge n-1$, since otherwise $I(Z;\wh Z_n|W_{2:n})=0$. 
The upper bounds in (\ref{eq:mi_ub_SDPI_chain_opt1}) and (\ref{eq:mi_ub_SDPI_chain_opt2}) can be weakened to
\begin{align*}
&\quad\,\, I(Z;\wh Z_n | W_{2:n}) \\
&\overset{\rm (a)}\le 
\begin{cases}
H(W_1|W_{2:n}) \eta (T-n+2) \cB(T-1, n-2, \eta)  \\
C_{(1,2)} \eta (T-n+2)^2 \cB(T-2, n-3, \eta)  
\end{cases} \\
&\overset{\rm (b)}\le 
\min\big\{H(W_1|W_{2:n}),C_{(1,2)}\big\} \eta (T-n+2)^2 \cB(T-2, n-3, \eta) \\
&\overset{\rm (c)}\le 
C_{(1,2)} \eta (T-n+2)^2 \exp\left(-2\left(\frac{n-3}{T-2} - \eta\right)^2 (T-2)\right) \\
&\overset{\rm (d)}\le C_{(1,2)} \frac{(n-3)^2 \gamma^2}{\eta} \exp\left(-2\left(\frac{\eta}{\gamma}-\eta\right)^2 (n-3)\right) 
\end{align*}
where 
\begin{enumerate}
  \item[(a)] and (b) follow from monotonicity properties of the binomial distribution;
  \item[(c)] follows from the Chernoff--Hoeffding bound;
  \item[(d)] follows from the fact that the channels associated with $\cE_\cS$ are independent, and the fact that the assumption that $n \ge 4$ and $n-1 \le T \le 2 + (n-3)\gamma/ \eta$.
\end{enumerate}

\begin{thebibliography}{10}
\providecommand{\url}[1]{#1}
\csname url@samestyle\endcsname
\providecommand{\newblock}{\relax}
\providecommand{\bibinfo}[2]{#2}
\providecommand{\BIBentrySTDinterwordspacing}{\spaceskip=0pt\relax}
\providecommand{\BIBentryALTinterwordstretchfactor}{4}
\providecommand{\BIBentryALTinterwordspacing}{\spaceskip=\fontdimen2\font plus
\BIBentryALTinterwordstretchfactor\fontdimen3\font minus
  \fontdimen4\font\relax}
\providecommand{\BIBforeignlanguage}[2]{{%
\expandafter\ifx\csname l@#1\endcsname\relax
\typeout{** WARNING: IEEEtran.bst: No hyphenation pattern has been}%
\typeout{** loaded for the language `#1'. Using the pattern for}%
\typeout{** the default language instead.}%
\else
\language=\csname l@#1\endcsname
\fi
#2}}
\providecommand{\BIBdecl}{\relax}
\BIBdecl

\bibitem{Ayaso_etal}
O.~Ayaso, D.~Shah, and M.~Dahleh, ``Information-theoretic bounds for
  distributed computation over networks of point-to-point channels,''
  \emph{IEEE Trans. Inform. Theory}, vol.~56, no.~12, pp. 6020--6039, 2010.

\bibitem{Como_Dahleh}
G.~Como and M.~Dahleh, ``Lower bounds on the estimation error in problems of
  distributed computation,'' in \emph{Proc. Inform. Theory and Applications
  Workshop}, 2009, pp. 70--76.

\bibitem{GKS_noisybc08}
N.~Goyal, G.~Kindler, and M.~Saks, ``Lower bounds for the noisy broadcast
  problem,'' \emph{SIAM Journal on Computing}, vol.~37, no.~6, pp. 1806--1841,
  2008.

\bibitem{noisy_bc_Dutta08}
C.~Dutta, Y.~Kanoria, D.~Manjunath, and J.~Radhakrishnan, ``A tight lower bound
  for parity in noisy communication networks,'' in \emph{Proc. ACM Symposium on
  Discrete Algorithms ({SODA})}, 2014, pp. 1056--1065.

\bibitem{Braverman_interactiveinformation}
M.~Braverman, ``Interactive information and coding theory,'' in \emph{Proc.
  Int. Congress Math.}, 2014.

\bibitem{OR01}
A.~Orlitsky and J.~Roche, ``Coding for computing,'' \emph{IEEE Trans. Inform.
  Theory}, vol.~47, no.~3, pp. 903--917, 2001.

\bibitem{KM79_mod2sum}
J.~K\"orner and K.~Marton, ``How to encode the modulo-two sum of binary
  sources,'' \emph{IEEE Trans. Inform. Theory}, vol.~25, no.~2, pp. 219--221,
  1979.

\bibitem{WTV08_GaussSum}
A.~B. Wagner, S.~Tavildar, and P.~Viswanath, ``Rate region of the quadratic
  {Gaussian} two-encoder source-coding problem,'' \emph{IEEE Trans. Inform.
  Theory}, vol.~54, no.~5, pp. 1938--1961, 2008.

\bibitem{NITbook11}
A.~{El Gamal} and Y.-H. Kim, \emph{Network Information Theory}.\hskip 1em plus
  0.5em minus 0.4em\relax Cambridge Univ. Press, 2011.

\bibitem{Giridhar_Kumar06}
A.~Giridhar and P.~Kumar, ``Toward a theory of in-network computation in
  wireless sensor networks,'' \emph{IEEE Communications Magazine}, vol.~44,
  no.~4, pp. 98--107, April 2006.

\bibitem{Gallager88}
R.~Gallager, ``Finding parity in a simple broadcast network,'' \emph{IEEE
  Trans. Inform. Theory}, vol.~34, no.~2, pp. 176--180, 1988.

\bibitem{Schulman96_iter_comm}
L.~Schulman, ``Coding for interactive communication,'' \emph{IEEE Trans.
  Inform. Theory}, vol.~42, no.~6, pp. 1745--1756, 1996.

\bibitem{Rajagopalan94_dist_comp}
S.~Rajagopalan and L.~Schulman, ``A coding theorem for distributed
  computation,'' in \emph{ACM Symposium on Theory of Computing}, 1994.

\bibitem{Carli_etal_erasures_consensus}
R.~Carli, G.~Como, P.~Frasca, and F.~Garin, ``Distributed averaging on digital
  erasure networks,'' \emph{Automatica}, vol.~47, no. 115-121, 2011.

\bibitem{Kar_Moura}
S.~Kar and J.~Moura, ``Distributed consensus algorithms in sensor networks with
  imperfect communication: Link failures and channel noise,'' \emph{IEEE Trans.
  Signal Process.}, vol.~57, no.~1, pp. 355--369, 2009.

\bibitem{Noorshams_Wainwright}
N.~Noorshams and M.~Wainwright, ``Non-asymptotic analysis of an optimal
  algorithm for network-constrained averaging with noisy links,'' \emph{IEEE J.
  Sel. Top. Sign. Proces.}, vol.~5, no.~4, pp. 833--844, 2011.

\bibitem{Ying_comp06}
L.~Ying, R.~Srikant, and G.~Dullerud, ``Distributed symmetric function
  computation in noisy wireless sensor networks with binary data,'' in
  \emph{International Symposium on Modeling and Optimization in Mobile, Ad-Hoc
  and Wireless networks (WiOpt)}, 2006.

\bibitem{Deb_Medard}
S.~Deb, M.~Medard, and C.~Choute, ``Algebraic gossip: a network coding approach
  to optimal multiple rumor mongering,'' \emph{IEEE Trans. Inform. Theory},
  vol.~52, no.~6, pp. 2486--2507, 2006.

\bibitem{Petrov_book}
V.~V. Petrov, \emph{Sums of Independent Random Variables}.\hskip 1em plus 0.5em
  minus 0.4em\relax Berlin: Springer-Verlag, 1975.

\bibitem{MR_SDPI}
M.~Raginsky, ``{Strong data processing inequalities and $\Phi$-Sobolev
  inequalities for discrete channels},'' \emph{IEEE Trans. Inform. Theory},
  vol.~62, no.~6, pp. 3355--3389, 2016.

\bibitem{Tiwari87}
P.~Tiwari, ``Lower bounds on communication complexity in distributed computer
  networks,'' \emph{J. ACM}, vol.~34, no.~4, pp. 921--938, Oct. 1987.

\bibitem{topology_Cha14}
A.~Chattopadhyay, J.~Radhakrishnan, and A.~Rudra, ``Topology matters in
  communication,'' in \emph{Proc. IEEE Annu. Symp. on Foundations of Comp. Sci.
  ({FOCS})}, Oct 2014, pp. 631--640.

\bibitem{Cover_book}
T.~Cover and J.~Thomas, \emph{Elements of Information Theory}, 2nd~ed.\hskip
  1em plus 0.5em minus 0.4em\relax New York: Wiley.

\bibitem{PGM_book}
D.~Koller and N.~Friedman, \emph{Probabilistic Graphical Models: Principles and
  Techniques}.\hskip 1em plus 0.5em minus 0.4em\relax MIT Press, 2009.

\bibitem{PolWu_IT_lectures}
\BIBentryALTinterwordspacing
Y.~Polyanskiy and Y.~Wu, ``{Lecture Notes on Information Theory},'' Lecture
  Notes for {ECE563 (UIUC) and 6.441 (MIT)}, 2012-2016. [Online]. Available:
  \url{http://people.lids.mit.edu/yp/homepage/data/itlectures_v4.pdf}
\BIBentrySTDinterwordspacing

\bibitem{Ahlswede_Gacs_hypercont}
R.~Ahlswede and P.~G\'acs, ``Spreading of sets in product spaces and
  hypercontraction of the {M}arkov operator,'' \emph{Ann. Probab.}, vol.~4,
  no.~6, pp. 925--939, 1976.

\bibitem{VA_SDPI}
\BIBentryALTinterwordspacing
V.~Anantharam, A.~Gohari, S.~Kamath, and C.~Nair, ``{On maximal correlation,
  hypercontractivity, and the data processing inequality studied by Erkip and
  Cover},'' \emph{arXiv preprint}, 2013. [Online]. Available:
  \url{http://arxiv.org/abs/1304.6133}
\BIBentrySTDinterwordspacing

\bibitem{PolWu_SDPI_avgP}
Y.~Polyanskiy and Y.~Wu, ``Dissipation of information in channels with input
  constraints,'' \emph{IEEE Trans. Inform. Theory}, vol.~62, no.~1, pp. 35--55,
  2016.

\bibitem{Eva_Sch99}
W.~Evans and L.~Schulman, ``Signal propagation and noisy circuits,'' \emph{IEEE
  Trans. Inform. Theory}, vol.~45, no.~7, pp. 2367--2373, 1999.

\bibitem{Xu_thesis}
A.~Xu, ``Information-theoretic limitations of distributed information
  processing,'' Ph.D. dissertation, University of Illinois at Urbana-Champaign,
  2016.

\bibitem{PolWu_ES}
\BIBentryALTinterwordspacing
Y.~Polyanskiy and Y.~Wu, ``Strong data-processing inequalities for channels and
  {Bayesian} networks,'' \emph{arXiv preprint}, 2015. [Online]. Available:
  \url{http://arxiv.org/abs/1508.06025}
\BIBentrySTDinterwordspacing

\bibitem{Kostina_Verdu_JSCC}
V.~Kostina and S.~Verd\'u, ``Lossy joint source-channel coding in the finite
  blocklength regime,'' \emph{IEEE Trans. Inform. Theory}, vol.~59, no.~5, pp.
  2545--2575, 2013.

\bibitem{Gallager_ITbook}
R.~Gallager, \emph{Information Theory and Reliable Communication}.\hskip 1em
  plus 0.5em minus 0.4em\relax New York: Wiley, 1968.

\bibitem{Kolmogorov_concent}
A.~Kolmogorov, ``Sur les propri\'et\'es des fonctions de concentrations de {M.
  P. L\'evy},'' \emph{Ann. Inst. H. Poincar\'e}, vol.~16, pp. 27--34, 1958.

\bibitem{Nguyen_Vu}
\BIBentryALTinterwordspacing
H.~H. Nguyen and V.~H. Vu, ``Small ball probability, inverse theorems, and
  applications,'' in \emph{Erd\H{o}s Centennial}, ser. Bolyai Society
  Mathematical Studies.\hskip 1em plus 0.5em minus 0.4em\relax Springer, 2013,
  vol.~25. [Online]. Available: \url{http://arxiv.org/abs/1301.0019}
\BIBentrySTDinterwordspacing

\bibitem{Bobkov_Chistyakov_concent}
S.~G. Bobkov and G.~P. Chistyakov, ``On concentration functions of random
  variables,'' \emph{J. Theor. Probab.}, vol.~28, no.~3, pp. 976--988, 2015,
  published online.

\bibitem{Rudelson_Vershynin_LO}
M.~Rudelson and R.~Vershynin, ``The {L}ittlewood--{O}fford problem and
  invertibility of random matrices,'' \emph{Adv. Math.}, vol. 218, pp.
  600--633, 2008.

\bibitem{Rudelson_Vershynin_vec}
\BIBentryALTinterwordspacing
M.~{Rudelson} and R.~{Vershynin}, ``Small ball probabilities for linear images
  of high dimensional distributions,'' \emph{arXiv1402.4492R}, Feb. 2014.
  [Online]. Available: \url{https://arxiv.org/abs/1402.4492}
\BIBentrySTDinterwordspacing

\bibitem{Erdos_LO}
P.~Erd\H{o}s, ``On a lemma of {L}ittlewood and {O}fford,'' \emph{Bull. Amer.
  Math. Soc.}, vol.~51, pp. 898--902, 1945.

\bibitem{Bobkov_Madiman_logconcave}
S.~Bobkov and M.~Madiman, ``The entropy per coordinate of a random vector is
  highly constrained under convexity conditions,'' \emph{IEEE Trans. Inform.
  Theory}, vol.~57, no.~8, pp. 4940--4954, 2011.

\bibitem{Carli09_avg_consensus}
R.~Carli, G.~Como, P.~Frasca, and F.~Garin, ``Average consensus on digital
  noisy networks,'' \emph{1st IFAC Workshop on Estimation and Control of
  Networked Systems}, 2009.

\bibitem{Witsenhausen_SDPI}
H.~S. Witsenhausen, ``On sequences of pairs of dependent random variables,''
  \emph{SIAM J. Appl. Math.}, vol.~28, no.~1, pp. 100--113, Jan. 1975.

\end{thebibliography}


\end{document}